\documentclass[pra,twocolumn,preprintnumbers,amsmath,amssymb,floatfix,longbibliography,superscriptaddress]{revtex4-2}
\usepackage{graphicx}
\usepackage{graphics}
\usepackage{psfrag}
\usepackage{hyperref}
\usepackage{multirow}
\usepackage[sort&compress]{natbib}
\usepackage{amssymb}
\usepackage{amsmath}
\usepackage{amsthm}
\usepackage{bbold}
\usepackage{bbm}
\usepackage{enumerate}
\usepackage{textcomp}

\newcommand{\vare}{\varepsilon}

\newcommand{\rmi}{{\rm i}}

\newcommand{\up}{\uparrow}
\newcommand{\down}{\downarrow}

\usepackage{xcolor}

\begin{document}

\hypersetup{pdftitle={title}}
\title{Dimensional crossover of superfluid $^{3}$He in a magnetic field}

\author{Leyla Saraj}
\affiliation{Department of Physics, University of Alberta, Edmonton, AB, Canada}

\author{Daksh Malhotra}
\affiliation{Department of Physics, University of Alberta, Edmonton, AB, Canada}

\author{Aymar Muhikira }
\affiliation{Department of Physics, University of Alberta, Edmonton, AB, Canada}

\author{Alexander J. Shook}
\affiliation{Department of Physics, University of Alberta, Edmonton, AB, Canada}

\author{John P. Davis}
\affiliation{Department of Physics, University of Alberta, Edmonton, AB, Canada}

\author{Igor Boettcher}
\email{iboettch@ualberta.ca}
\affiliation{Department of Physics, University of Alberta, Edmonton, AB, Canada}
\affiliation{Theoretical Physics Institute, University of Alberta, Edmonton, AB, Canada}
\affiliation{Quantum Horizons Alberta, University of Alberta, Edmonton, AB, Canada}

\begin{abstract}
Motivated by recent experiments on superfluid $^3$He in nanoscale-confined geometries,  we theoretically investigate the associated phase diagram in a slab geometry and perpendicular magnetic field as the size of confinement is varied. Our analysis is based on minimizing the Ginzburg--Landau free energy for the $3\times 3$ matrix superfluid order parameter for three different boundary conditions. We observe a smooth crossover from the phase diagram of the 3D system to the quasi-2D limit for slab heights of several hundred nanometres and magnetic fields of several kilogauss. We illuminate that, despite the apparent complexity of the underlying equations, many precise numerical and even analytical statements can be made about the phase structure for general values of the coefficients of the free energy functional, which can in turn be used to constrain or measure these parameters. To guide future experimental studies, we compute the phase diagram in dependence of pressure, temperature, slab height, and magnetic field.
\end{abstract}

\maketitle

\section{Introduction}\label{SecIntro}

Superfluid $^3$He is a fermionic superfluid exhibiting p-wave, spin-triplet pairing \cite{RevModPhys.47.415,RevModPhys.47.331,volovik2003universe,vollhardt2013superfluid}. The spin and orbital degrees of freedom associated with this pairing state give rise to a large number of potential broken symmetries. In the bulk or three-dimensional (3D) system, two superfluid phases are thermodynamically stable, known as the A- and B-phases. The B-phase is characterized by an isotropic gap and a relative spin-orbit symmetry-breaking. The A-phase features an anisotropic gap with two point nodes and independently breaks the spin and orbital rotational symmetry, resulting in a phase that breaks time-reversal symmetry and carries intrinsic angular momentum. The bulk phase diagram describing the transitions between these phases is well described within the framework of Ginzburg--Landau theory, but only when the so-called strong-coupling corrections have been applied to the parameters \cite{PhysRevB.24.183,PhysRevB.75.174503}. 

The existence of surfaces under spatial confinement, however, plays a dramatic role in modifying the free energy of superfluid $^3$He \cite{PhysRevA.9.2676,PhysRevLett.33.624,BartonMoore,PhysRevB.15.199,Kjaeldman,Fujita,Jacobsen87,FetterUllah,PhysRevB.38.2362,PhysRevB.104.094520}. Physically this can be seen as a consequence of the fact that scattering at the surfaces results in pair breaking, which is anisotropic in nature due to the orientation of the surface \cite{nagato1998rough}. The container walls then set the boundary conditions for the Ginzburg--Landau equations, which modifies the stability of the various potential phases. Geometric considerations therefore become highly important in experimental work on superfluid $^3$He. It has been shown that confining $^3$He to a slab of thickness on the order of microns or smaller causes the A-phase to be stable over a wider region of the phase diagram \cite{science.1233621,levitin2019evidence,PhysRevLett.124.015301,PhysRevLett.132.156001,PhysRevLett.134.136001,shook2024topologically}. Additional phases not present in the bulk system can also be stabilized by the boundaries. For instance, it is known that the polar phase can be stabilized by infusing $^3$He into a nematically ordered aerogel system \cite{thuneberg1998ginzburg,zhelev2016observation}, and the search for inhomogeneous or pair-density-wave states is an active field of research as we discuss below.

\begin{figure}[t!]
\centering
\includegraphics[width=8.6cm]{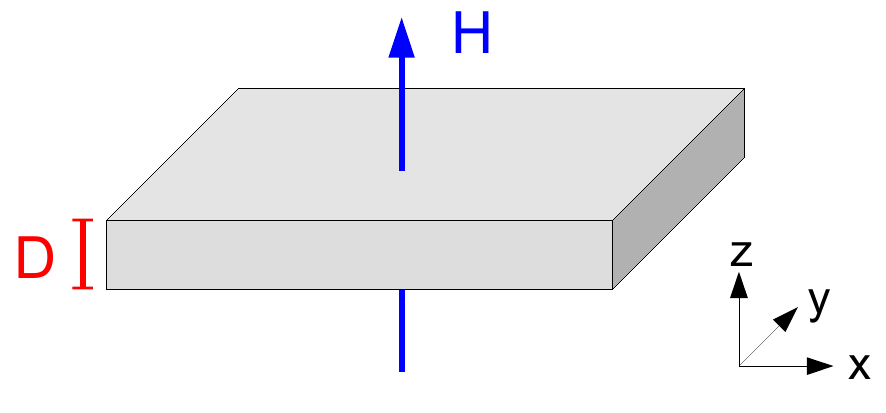}
\caption{In this work, we consider a scenario where superfluid $^3$He is confined to a slab geometry of height $D$ in a perpendicular magnetic field of magnitude $H$. We choose the coordinate system such that the magnetic field points in the z-direction and the superfluid is limited to the region $z\in[0,D]$, with idealized infinite extension in the x- and y-directions. Three different boundary conditions are chosen at the surfaces $z=0$ and $z=D$.}
\label{FigSetting}
\end{figure}

Experimental studies of confined $^3$He-systems motivate the development of numerical approaches, such as finite element solvers, for solving the Ginzburg--Landau equations with arbitrary boundary conditions \cite{wiman2014superfluid,PhysRevB.92.144515,wiman2016strong,hornbeck2021solving,rantanen2024competition}. In this work, we numerically and partly analytically solve the Ginzburg--Landau equations for superfluid $^3$He confined to a slab geometry in a perpendicular magnetic field as in Fig. \ref{FigSetting} to compute the ensuing phase diagram. Specifically, we assume that the superfluid is restricted in the z-direction to the interval $z\in[0,D]$, with $D$ the height of the slab, whereas the systems is idealized to be infinitely extended in the xy-plane. Pair breaking at the container surface then means that suitable boundary conditions for the order parameter have to be specified at $z=0$ and $z=D$. The magnetic field points along the z-direction with magnitude $H$. In this configuration, the confinement and magnetic field have similar tendencies in suppressing the B- over the A-phase as either $D$ is lowered or $H$ increased \cite{PhysRevB.18.4724,PhysRevB.37.5010,PhysRevB.86.094518}. Different setups, where the magnetic field is in-plane or at an arbitrary angle with the z-axis, lead to competing effects and have been studied in Refs. \cite{PhysRevB.86.094518,PhysRevLett.109.165301,He3SUSY,MizushimaJPhys,MizushimaReview}.

The setup discussed here has the following qualitative behavior. The confinement in the $z$-direction disfavors the third column of the $3\times 3$ order parameter matrix, as examined in detail in Ref. \cite{PhysRevB.108.144503}. As a result, the normal-to-superfluid transition is towards the A-phase, whose stability region in the phase diagram is enhanced. The dimensional crossover from 3D behavior to a phase structure dominated by the A-phase happens for $D \sim 100\ \text{nm}$. The magnetic field suppresses the third row of the order parameter matrix (which also enhances the stability of the A-phase, which gets replaced with the analogous A$_2$-phase in a magnetic field) and enhances equal-spin pairing of spin-down atoms over spin-up atoms (which implies that the normal-to-superfluid transition is towards a new state, the A$_1$-phase, which is a magnetized superfluid of Cooper pairs of spin-down atoms surrounded by a Fermi sea of spin-up atoms). The crossover regimes where  the two effects become relevant are $H\sim 1\ \text{kG}$ and $H\sim 50\ \text{kG} = 5\ \text{T}$, respectively. Since magnetic fields of several tesla are feasible in experimental $^3$He setups, we expect the present theoretical findings to be a useful benchmark and reference for experiment.

An intriguing aspect of superfluid $^3$He confined to slab geometries is that the presence of inhomogeneous or pair-density-wave phases \cite{agterberg2020physics} can be established theoretically for certain parameter ranges \cite{vorontsov2004domain,PhysRevLett.98.045301,wiman2016strong,levitin2019evidence,PhysRevLett.124.015301,PhysRevLett.128.015301}. By an inhomogeneous phase we refer to a superfluid order whose order parameter  under confinement does not only vary with the coordinate $z$ (as one would expect for a system with surfaces at $z=0$ and $z=D$), but also spontaneously develops an $x$- or $y$-dependence to lower the free energy. Experimentally, NMR measurements in  Ref. \cite{levitin2019evidence} for $D=1100\ \text{nm}$  are consistent with variations in the xy-plane such as the polka-dot phase or islands of domains whereas a homogeneous or stripe phase is excluded, while measurements of the superfluid density in Ref. \cite{PhysRevLett.124.015301} are consistent with a pair-density-wave phase for $D=1067, 805, 636\ \text{nm}$ that might even extend to zero temperature. In this work, we include the stripe phase as an exemplary inhomogeneous phase into the analysis at zero field for completeness and identify its range of stability. On the other hand, the suppression of both the third column and row of the order parameter in a sufficiently strong magnetic field under confinement excludes the stripe phase from the phase diagram in this experimental regime.

To disentangle confinement from magnetic-field effects and make the presentation more transparent, we first discuss the dimensional crossover for $H=0$ in Sec. \ref{SecZero} and then cover the case with $H>0$ in Sec. \ref{SecNonzero}. The magnetic fields we have in mind are $H\geq 100\ \text{G}$. In fact, our $H>0$ analysis should not be applied to small or infinitesimal $H$. The reason for this is the presence of the intrinsic dipole field of $^3$He of approximately $25\ \text{G}$. For $H\sim 25\ \text{G}$, the external magnetic field would compete with the dipolar interaction \cite{PhysRevLett.109.165301,MizushimaReview}. Since we ignore the dipolar contribution to the Ginzburg--Landau free energy in the following, we are limited to the regime $H\gg 25\ \text{G}$. Also note that NMR experiments such as in Ref. \cite{levitin2019evidence} use magnetic fields $H\sim 10\ \mu\text{T}=0.1\ \text{G}$ and thus effectively probe the $H=0$ regime of the results presented here.

\section{Ginzburg--Landau free energy}\label{SecGL}

\subsection{Free energy functional}
The order parameter of superfluid $^3$He is a $3 \times 3$ complex matrix $A=(A_{\mu i})$ without any further mathematical structure. In a confined geometry, the order parameter $A(\textbf{x})$ acquires a spatial dependence on the coordinates $\textbf{x}=(x,y,z)$. The Ginzburg--Landau free energy in an external magnetic field $\vec{H}$ and close to a second-order phase transition is given by
\begin{align}
  \nonumber F[A] &= F_{\rm kin}[A]+F_{\rm pot}[A]+F_{\rm mag}[A]\\
  \label{gl1} &= \int_\Sigma \mbox{d}^3x\ \Bigl( \text{f}_{\rm kin}+\text{f}_{\rm pot}+\text{f}_{\rm mag}\Bigr),
\end{align}
where $\Sigma$ is the region to which the superfluid is confined and the kinetic, potential, and magnetic contributions to the free energy density are given by
\begin{align}
  \nonumber \text{f}_{\rm kin} ={}&  K( \partial_i A^*_{\mu j})( \partial_iA_{\mu j}) +K(\gamma-1)( \partial_iA^*_{\mu i})( \partial_j A_{\mu j}),\\
 \nonumber \text{f}_{\rm pot} ={}& -\alpha_0\ \mbox{tr}(AA^\dagger)+\beta_1|\mbox{tr}(AA^T)|^2  \\
  \nonumber &+ \beta_2[\mbox{tr}(AA^\dagger)]^2+ \beta_3 \mbox{tr}[AA^T(AA^T)^*] \\
 \nonumber &+ \beta_4\mbox{tr}(AA^\dagger AA^\dagger)+ \beta_5\mbox{tr}[AA^\dagger(A A^\dagger)^*],\\
 \label{gl1b} \text{f}_{\rm mag} ={}&  \rmi g_{1} \vare_{\mu\nu\lambda} H_\mu (AA^\dagger)_{\nu\lambda}+ g_{2} H_\mu H_\nu (AA^\dagger)_{\mu \nu}.
\end{align}
We discuss the role of the boundary conditions below. 

In a thermodynamic regime far away from a second-order phase transition, when $A$ is not small, the integrand in Eq. (\ref{gl1}) should be replaced by a more general function of symmetry invariants along the lines of Refs. \cite{Spencer1958,Spencer1958b,ARTIN1969532,PROCESI1976306,Matteis2008,PhysRevLett.120.057002} that investigate systems other than $^3$He. Furthermore, in writing Eq. (\ref{gl1b}), we absorbed a possible third kinetic term proportional to $\partial_i A^*_{\mu j}\partial_j A_{\mu i}$ into the second term  through a partial integration \cite{volovik2003universe,vollhardt2013superfluid}, neglecting certain boundary terms that only vanish for maximally pair-breaking boundaries. In the following we will ignore both corrections, as is common, and assume that $F$ constitutes the proper free energy in all regions of the phase diagram. Computing the associated phase diagram is in itself a well-posed mathematical problem, but some deviations when comparing to experimental data might result from this assumption.

Through the parameters $K,\gamma,\alpha_0,\beta_1,\dots,\beta_5,g_1,g_2$, the free energy $F$  depends on the thermodynamic state of the system represented, for instance, by the temperature $T$ and pressure $P$ of the system. The exact functional dependencies $K=K(T,P)$, etc., are not known and their theoretical calculation constitutes a complicated quantum many-body problem. The quartic invariants $\beta_1,\dots,\beta_5$ for a homogeneous 3D system have been estimated theoretically and experimentally \cite{PhysRevB.24.183,PhysRevB.75.174503}. However, it is not known whether the bulk values also apply to confined geometries, especially when confinement is strong \cite{PhysRevB.108.144503}.

In this work, we consider a slab geometry of height $D$ along the z-direction without any confinement in the x- and y-directions. It seems natural to assume that the order parameter is independent of $x$ and $y$, but acquires a spatial dependency on $z$ due to the behavior at the container walls. Although this assumption  turns out to be false for certain boundary conditions and thermodynamic regimes (see the discussion of the stripe phase below),  let us assume for the moment that the order parameter $A(z)$ only depends on $z$. The kinetic term in the free energy then has a very characteristic form: Denoting the columns of $A(z)$ by the vectors $\vec{X}(z)$, $\vec{Y}(z)$ and $\vec{Z}(z)$, i.e. $A=(\vec{X}\vec{Y}\vec{Z})$, we have
\begin{align}
 \label{slab1} F_{\rm kin}[A] = K \int_0^D\mbox{d}z\ \Bigl( |\partial_z \vec{X}|^2 + |\partial_z \vec{Y}|^2 +\gamma |\partial_z \vec{Z}|^2\Bigr).
\end{align}
As outlined below, all relevant boundary conditions require $\vec{Z}(0)=\vec{Z}(D)=0$ at the container walls, so that $\vec{Z}(z)$ needs to either vanish or have a nontrivial $z$-dependence. For $\gamma>1$, to minimize the kinetic energy, a nonvanishing third column $\vec{Z}(z)$ comes at a higher cost than nonvanishing $\vec{X}(z)$ or $\vec{Y}(z)$. Consequently, the parameter $\gamma>1$ will qualitatively influence the phase diagram. In contrast, the superfluid order in the 3D system is homogeneous (spatially constant) and hence the value of $\gamma$ does not affect the bulk phase diagram.

Due to the external magnetic field $\vec{H}$, the free energy receives the contribution $F_{\rm mag}$ corresponding to the linear and quadratic Zeeman effect. Here, $g_{1}, g_{2}>0$ are positive phenomenological constants, whose values will be discussed below. The magnetic field explicitly breaks the spin-rotation invariance of the zero-field free energy, and the linear Zeeman term additionally breaks particle-hole-symmetry corresponding to $A\to A^*$, or $(AA^\dagger)_{\mu\nu}\to (AA^\dagger)_{\nu\mu}$. Further note that while the kinetic gradients $\partial_i$ only couple to the orbital (or second) index of $A_{\mu i}$, the magnetic field $H_\mu$ only couples to the spin (or first) index of $A_{\mu i}$.

In the following, we assume the magnetic field to point into the z-direction,
\begin{align}
 \label{mag3} \vec{H} = H \vec{e}_{\rm z}
\end{align}
with $H>0$, so that the magnetic free energy density becomes
\begin{align}
 \label{mag4} \text{f}_{\rm mag} &=\rmi g_{1} H [(AA^\dagger)_{12}-(AA^\dagger)_{21}]+ g_{2} H^2 (AA^\dagger)_{33}.
\end{align}
The main effect of the linear Zeeman term proportional to $g_1$ is to increase the critical temperature compared to the zero-field value. The main effect of the quadratic Zeeman term proportional to $g_2$ is to suppress certain matrix elements of the order parameter. Indeed, it
is proportional to $|A_{31}|^2+|A_{32}|^2+|A_{33}|^2$ and thus suppresses the third \emph{row} of the order parameter. In a slab geometry, where additionally the third \emph{column} is suppressed by $\gamma |\partial_z\vec{Z}|^2$ in Eq. (\ref{slab1}), this favors order parameters that are nonzero only on the upper-left $2\times 2$ block of the matrix $A$.

The correct order of magnitude of the coefficients $\alpha_0$ and $\beta_a$ can be inferred from the so-called weak-coupling or BCS model \cite{volovik2003universe,vollhardt2013superfluid}. There, $\alpha_0(T,P)$ close to the zero-field critical temperature $T_{\rm c0}(P)$ satisfies 
\begin{align}
 \label{weak1} \alpha_0 \simeq  \frac{N(0)}{3}\Bigl(1-\frac{T}{T_{\rm c0}}\Bigr),
\end{align}
where
\begin{align}
 \label{weak2} N(0)=\frac{m^*k_{\rm F}}{2\pi^2\hbar^2}
\end{align}
is the density of states for one spin-component. Away from criticality, the temperature-dependence is not linear, see App. \ref{AppThermo}. In the presence of an external magnetic field, it is the parameter
\begin{align}
 \label{weak2b} \alpha_H = \alpha_0 +g_1H
\end{align}
that vanishes at the critical temperature in a magnetic field, $T_{\rm cH}(P,H)$, according to
\begin{align}
 \label{weak2c} \alpha_H \simeq \frac{N(0)}{3}\Bigl(1-\frac{T}{T_{\rm cH}}\Bigr).
\end{align}
The parameter $\alpha_H$ is positive in the superfluid phase, both with and without magnetic field, and plays an important role in our choice of dimensionless units introduced below. Of course, for $H=0$ we have $\alpha_H=\alpha_0$. Since $\alpha_H\geq \alpha_0$, we have $T_{\rm cH}(P,H)\geq T_{\rm c0}(P)$, highlighting that the magnetic field enhances the tendency of pairing. Indeed, combining Eqs. (\ref{weak1})-(\ref{weak2c}) we find $T_{\rm cH}=T_{\rm c0}(1+\frac{3g_1H}{N(0)})$, highlighting the increase of the critical temperature by the linear Zeeman term, see Eq. (\ref{mbulk13}) below. The weak-coupling values of the quartic couplings are
\begin{align}
 \label{weak3} (\beta_1,\beta_2,\beta_3,\beta_4,\beta_5)_{\rm wc} = (-1,2,2,2,-2)\beta_0,
\end{align}  
where
\begin{align}
 \label{weak4} \beta_0 = N(0)\frac{7\zeta(3)}{240\pi^2(k_{\rm B}T_{\rm c0})^2}.
\end{align}
In the following, we will not assume the free energy to have the weak-coupling values, but rather keep $\alpha_0$, $\alpha_H$, and $\beta_a$ as free parameters.

In the following, we will only consider systems in the superfluid phase, where 
\begin{align}
 \label{unit1} \alpha_H>0.
\end{align}
Since the competition of distinct order parameters $A(\textbf{x})$ does not depend on the overall prefactor of $F$, we can divide the integrand of $F$ by $\alpha_H$. This, in turn, defines the correlation length $\xi=\xi(T,P,H)$ through 
\begin{align}
 \label{gl2} K=\alpha_H \xi^2,
\end{align}
and all length scales can be expressed in units of $\xi$. We emphasize once more that since the functions $\alpha_H(T,P,H)$ and  $K(T,P)$ are not known, the $T$-, $P$-, and $H$-dependence of the correlation length $\xi(T,P,H)$ is unknown as well. Several physically motivated choices are used in the literature and our analysis below, see App. \ref{AppCoherence}, but essentially they constitute elaborate guesses  and are a source of discrepancy when comparing theory and experiment.

The kinetic coefficient $\gamma$ plays an important role in the phase diagram of confined superfluid $^3$He. Weak-coupling theory predicts a value of $\gamma=3$. In a realistic setting, the true value of $\gamma$ likely deviates from this value. Strong-coupling corrections based on quasiparticle scattering predict values $\gamma\simeq 3.1$ slightly above 3 \cite{PhysRevB.17.2901,ThunebergHydro}. Close to the superfluid phase transition, fluctuations of the order parameters lead to a renormalization group flow of $\gamma$, which to two-loop order has been computed by Jones, Love, and Moore \cite{Jones:1976zz}. This analysis shows that $\gamma$ flows to a fixed-point $\gamma\to \gamma_\star=1$ in the thermodynamic limit. However, for any finite and especially tightly confined system, this fixed point is not reached and the system would have some value $\gamma>\gamma_\star$. In the following, we therefore work with the assumption $\gamma>1$, but for concreteness choose $\gamma=3$ in the plots.

The quadratic Zeeman term starts to dominate energetically over the linear one as soon as the magnetic field reaches $H \sim g_1/g_2$. The size of the individual couplings $g_1$ and $g_2$ can be determined as follows. The linear Zeeman term results in a splitting of the critical temperatures of the A$_1$- and A$_2$-phases (to be introduced below) that scales linear in $H$ according to $\Delta T = 2\frac{3g_1}{N(0)}HT_{\rm c0}\propto H$. The slope can be fitted to the experimental data, which is $\Delta T \sim 50\ \mu \text{K}$ at $H\sim 10^4\ \text{G}$. Together with $T_{\rm c0}\sim 2.5\ \text{mK}$ this yields \cite{PhysRevLett.30.81,VOLOVIK199027}
\begin{align}
 \label{mag5b} g_1H \sim \frac{N(0)}{3}\times \frac{H}{10^6\text{G}}.
\end{align}
The coefficient of the quadratic Zeeman term is $g_2 = 2 (\frac{\mu_0}{1+F_0^{\rm a}})^2\beta_0$, with $\beta_0$ from Eq. (\ref{weak4}), $\mu_0$ the magnetic moment of the $^3$He nucleus, and $F_0^{\rm a} \sim -0.7$ a temperature-independent Fermi liquid parameter \cite{vollhardt2013superfluid}. $^3$He nuclei have spin 1/2 and a gyromagnetic ratio of $\Gamma=-2\cdot 10^4\ \text{s}^{-1}\text{G}^{-1}$, hence $\mu_0 = \frac{\hbar}{2}\Gamma$. Inserting these values and $T_{\rm c0}\sim 2.5\ \text{mK}$, we obtain \cite{VOLOVIK199027}
\begin{align}
 \label{mag5d} g_2 H^2 \sim \frac{N(0)}{3} \times \Bigl(\frac{H}{10^4\text{G}}\Bigr)^2.
\end{align}
Hence the quadratic Zeeman term becomes dominant over the linear one for magnetic fields larger than $100\text{G}$. 

Note that since the gyromagnetic ratio $\Gamma$ is negative, it is favorable for the spins to anti-align with the magnetic field if the quantization axis is chosen as the z-direction, because the energy of a spin-up atom is $-\mu_0 H>0$. This means that a strong magnetic field will favor spin-down pairing and suppress spin-up pairing. One could alter this by choosing the quantization axis in the negative z-direction, but we will not make this choice here.

\subsection{Boundary conditions}

In this work, we consider a slab geometry of height $D$, where the superfluid is confined to the region $z\in[0,D]$ in the z-direction. The x- and y-directions are confined on much larger length scales $L_{\rm x},L_{\rm y}\gg D$ that are assumed infinite in the calculations. The choice of boundary conditions in the x- and y-directions is irrelevant, but it is important to specify the boundary conditions at the container walls $z=0$ and $z=D$ to study the phase diagram. In this work, we mostly focus on either maximally pair-breaking or specular boundary conditions for the order parameter $A(\vec{x})$. However, for better comparison to experiment, we also discuss diffusive boundary conditions as a third choice that interpolates between the two.

In the maximally pair-breaking case, the whole order parameter vanishes at the walls of the container according to the Dirichlet boundary conditions
\begin{align}
 \label{bc1} A(z=0) = A(z=D) =0.
\end{align}
This choice is mathematically well-defined and allows us to study the dimensional crossover from 3D to 2D in a mostly analytical fashion. The biggest computational advantage is that the stripe phase is absent for maximally pair-breaking boundaries, as is shown in App. \ref{AppStripe}. Since this also applies to the case of large nonzero magnetic field, it is a suitable starting point.

For the case of specular boundary conditions, only the components of the third column of the order parameter vanish at the container walls, whereas the remaining ones stay finite with vanishing slope. Writing again the columns of $A(\vec{x})$ as $A=(\vec{X}\vec{Y}\vec{Z})$, we have the Neumann boundary conditions
\begin{equation}\label{spec1} 
\begin{aligned}
 \partial_z\vec{X}(z=0) &= \partial_z \vec{X}(z=D) =0,\\
 \partial_z\vec{Y}(z=0) &= \partial_z \vec{Y}(z=D) =0,\\
 \vec{Z}(z=0) &= \vec{Z}(z=D) =0.
\end{aligned}
\end{equation}
Importantly, order parameters like the A-, P-, or Pol-phases (introduced below) with $\vec{Z}=0$ are unaffected by the confinement for specular boundaries, because the free energy is minimized by a homogeneous superfluid groundstate $A(\vec{x})=\text{const}$, just as in the 3D bulk case.

The specularity of the boundary can be tuned in experiments by coating the container walls with a film of $^4$He \cite{PhysRevLett.60.596,science.1233621,heikkinen2021fragility}. However, specular boundary conditions may not be realized exactly. The corresponding intermediate case of diffusive boundaries is given by the Robin boundary conditions
\begin{equation}\label{diffBC} 
\begin{aligned}
 \partial_z\vec{X}(z=0) &= \frac{1}{b_{\rm T}\xi} \vec{X}(z=0),\\
 \partial_z\vec{X}(z=D) &= -\frac{1}{b_{\rm T}\xi} \vec{X}(z=D),\\
 \partial_z\vec{Y}(z=0) &= \frac{1}{b_{\rm T}\xi} \vec{Y}(z=0),\\
 \partial_z\vec{Y}(z=D) &= -\frac{1}{b_{\rm T}\xi} \vec{Y}(z=D),\\
 \vec{Z}(z=0) &= \vec{Z}(z=D) =0,
\end{aligned}
\end{equation}
with $b_{\rm T}\geq0$ a dimensionless real parameter and $\xi$ the correlation length defined in Eq. (\ref{gl2}). For $b_{\rm T}\in(0,\infty)$, the boundary values of the components of the first and second column of $A$ are suppressed compared to their bulk values, but do no vanish identically. Instead, if the order parameter profile of these components is extrapolated linearly beyond the container walls, it would vanish at a distance $b_{\rm T}\xi$ from the boundary. Maximally pair-breaking and specular boundaries then correspond to the limits $b_{\rm T}=0$ and $b_{\rm T}=\infty$, respectively. A physically relevant choice for diffusive boundaries that is often applied in the literature is $b_{\rm T}\simeq0.5$ \cite{PhysRevA.9.2676}, though defined with the zero-temperature value of $\xi$ (see App. \ref{AppCoherence}). For concreteness, we call $b_T=0.5$ diffusive boundaries in this work, because it constitutes a representative value. Note also that our choice of $b_{\rm T}$ being dimensionless, hence in units of $\xi$, is also denoted $b_{\rm T}'$ in Ref. \cite{wiman2016strong}.

\subsection{Dimensionless units}

In this work, we measure lengths in units of the correlation length by defining
\begin{align}
 \label{real9}  \bar{z} &= z/\xi,\ \bar{D} = D/\xi.
\end{align}
We further write the order parameter as
\begin{align}
 \label{real4} A(z)=\frac{\Delta_H}{\sqrt{3}}O(\bar{z}),\ \Delta_H= \sqrt{\frac{3\alpha_H}{2(3\beta_{12}+\beta_{345})}},
\end{align}
with $O(\bar{z})$ a dimensionless $3\times 3$ complex matrix. Here, we use the common notation $\beta_{ab\dots c}=\beta_a+\beta_b+\dots+\beta_c$ and assume $3\beta_{12}+\beta_{345}>0$. Through the rescaling in Eq. (\ref{real4}), we can eliminate one of the quartic couplings from the free energy density, such that it only depends on four independent ratios $\zeta_a$ of the five quartic couplings defined by
\begin{align}
 \label{real5} \zeta_a = \frac{\beta_a}{3\beta_{12}+\beta_{345}}.
\end{align}
The mutual dependence of the $\zeta_a$ is reflected in the identity
\begin{align}
 \label{real8} 3\zeta_{12}+\zeta_{345}=1.
\end{align}
Note that we have $3\beta_{12}+\beta_{345} = \frac{3\alpha_H}{2\Delta_H^2}$ and $\beta_a =\zeta_a \frac{3\alpha_H}{2\Delta_H^2}$. Importantly though, the values of $\zeta_a$ are independent of $H$ since $\alpha_H/\Delta_H^2 = \alpha_0/\Delta_0^2$. The weak-coupling values of the $\zeta_a$ are
\begin{align}
 \label{unit2} (\zeta_1,\zeta_2,\zeta_3,\zeta_4,\zeta_5) \approx (-1,2,2,2,-2)\frac{1}{5}.
\end{align}
In the remainder of this work, we will use the notation "$\approx$" to indicate the application of the weak-coupling approximation.

Dimensionless units for the magnetic field contributions are obtained by dividing by $\alpha_H>0$. In particular, we define the dimensionless parameters
\begin{align}
 \label{unit3} h &= \frac{2g_1H}{\alpha_H},\\
 \label{unit4} r &= \frac{g_2H^2}{g_1H} \sim\frac{H}{100\text{G}}.
\end{align}
The factor of 2 in the definition of $h$ is chosen such that the critical field strengths of the A$_2$- and P$_2$-phases introduced below are at $h_{\rm c} \approx 1$ in the weak-coupling limit. The parameter $r$ measures whether the quadratic Zeeman term is small ($H\ll 100\ \text{G}$ or $r\ll 1$) or large ($H\gg 100\ \text{G}$ or $r\gg1$) compared to the linear Zeeman term.

We define the dimensionless free energy $\bar{F}[O]$ through
\begin{align}
 \label{real12b} F[A] = \frac{2L_{\rm x}L_{\rm y}\xi \alpha_H \Delta_H^2}{3}\bar{F}[O],
\end{align}
where $L_{\rm x}$ and $L_{\rm y}$ are the extensions of $\Sigma$ in the x- and y-direction. Note the overall positive prefactor of $\alpha_H\Delta_H^2$. For a given value of $H\geq 0$, it is sufficient to compare the best values of $\bar{F}$ for competing phases to determine the thermodynamically stable ground state defined as the global minimum of $\bar{F}$.

After introducing the dimensionless units as described here, the competition of superfluid orders---and hence the phase diagram---only depends on the parameters $\gamma$, four dimensionless numbers $\zeta_a$, and the parameters of the confining geometry expressed in units of $\xi$. In an external magnetic field, there is an additional dependence on $h$ and $r$. This reduced number of relevant parameters can be exploited to make concrete predictions for the phase diagram of the theory, or determine the remaining parameters by comparing to experimental data.

\subsection{Competing orders}

In the following, we list the order parameters that are relevant for the superfluid ground state in a slab geometry with and without an external magnetic field applied. For the magnetic-field case, we introduce a convenient matrix notation. We give coordinate-free expressions of the associated order parameters and a representation in terms of the spin-gap matrix in App. \ref{AppMagOrder}. The matrix order parameters listed in the following are equivalent to any matrices obtained by a global transformation that leaves the free energy invariant, i.e. $A(\textbf{x}) \to e^{\rmi \Phi}RA(\textbf{x})S$ with $e^{\rmi \Phi}\in\text{U}(1)$, $R\in\text{SO}(2)_{\rm L}\subset\text{SO}(3)_{\rm L}$, and $S\in\text{SO}(3)_{\rm S}$ \cite{vollhardt2013superfluid}.

In the absence of a magnetic field, the competing superfluid orders are the A-phase, B-phase, planar phase (P), polar phase (Pol), and planar-distorted B-phase (pdB). Their order parameters read
\begin{align}
\label{gl3} A_{\rm A} &=\begin{pmatrix} B & \rmi B & 0 \\ && \\ & &  \end{pmatrix},\  A_{\rm B} = \begin{pmatrix} B & & \\ & B & \\ & & B \end{pmatrix},\\
 \label{gl5} A_{\rm P} &= \begin{pmatrix} B & & \\ & B & \\ & & 0 \end{pmatrix},\ A_{\rm Pol} = \begin{pmatrix} B & & \\ & 0 & \\ & & 0 \end{pmatrix},\\
 \label{gl6}  A_{\rm pdB} &= \begin{pmatrix} B & & \\ & B & \\ & & C \end{pmatrix}.
\end{align}
Herein, $B$ and $C$ are real functions that depend on $z$ and can be chosen positive without loss of generality. Note that the B- and P-phases are special cases of the pdB-phase with $C=B$ and $C=0$, respectively. The A-phase, due to the presence of both real and imaginary terms in $A$, spontaneously breaks the time-reversal symmetry of $F$. The remaining orders listed here preserve time-reversal symmetry, with order parameters that are real and diagonal.

The stripe phase (S) spontaneously breaks translation symmetry in the y-direction by forming a stripe-patterned modulation of the order parameter \cite{vorontsov2004domain,PhysRevLett.98.045301,wiman2016strong}. The order parameter reads
\begin{align}
\label{gl16b} A_{\rm S} =\begin{pmatrix} A_{11} & & \\ & A_{22} & A_{23} \\ & A_{32} & A_{33} \end{pmatrix}.
\end{align}
Herein, the coefficients are real functions of $z$ and $y$. This superfluid order preserves translation symmetry in the x-direction and is only thermodynamically stable for certain parameters regimes to be discussed in Sec. \ref{SecStripe}. Of course, the choice of translation invariance in the x-direction is an arbitrary choice made here for concreteness. In experiment, any direction in the xy-plane will be chosen spontaneously.

In the presence of an external magnetic field in the z-direction, imaginary parts $\pm\rmi M$ appear in the order parameter $A$ to lower the free energy due to the linear Zeeman term. The sign of $\rmi M$ equals the sign of $g_1 H$. We choose $g_1 H >0$ in the following, hence $M>0$, but the effect of the reversed sign of $g_1 H$ is obtained by replacing $A\to A^*$. To obtain a better understanding of the following order parameters, we denote the pairing amplitudes in the $|\up\up\rangle$, $|\down\down\rangle$, and $\frac{1}{\sqrt{2}}|\up\down+\down\up\rangle$ triplet states by the real numbers $\Delta_{\up\up}$, $\Delta_{\down\down}$, and $\Delta_{\up\down}$. We then have
\begin{align}
 \label{mag6b} B= \frac{\Delta_{\down\down}+\Delta_{\up\up}}{2},\ M=\frac{\Delta_{\down\down}-\Delta_{\up\up}}{2},\ C=\Delta_{\up\down}.
\end{align}
These equations also apply to the case of vanishing magnetic field, where $M=0$ and thus $\Delta_{\down\down}=\Delta_{\up\up}$.

For nonvanishing field, the A-phase gets replaced by the A$_2$-phase with order parameter
\begin{align}
 \label{mag12b} A_{\rm A_2} &= \begin{pmatrix} B & \rmi B & \\ -\rmi M & M & \\  &  & 0 \end{pmatrix},
\end{align}
where $B,M$ are real functions of $z$ as in Eq. (\ref{mag6b}) that satisfy $B> M> 0$. Because of the magnetic field, $M>0$ (or $\Delta_{\down\down}>\Delta_{\up\up}$), reflecting the fact that spin-down pairing is favored. For $M=0$ (or $\Delta_{\up\up}=\Delta_{\down\down}$) we obtain the representation of the A-state from Eq. (\ref{gl3}). The limit $B=M$ (or $\Delta_{\up\up}=0$) constitutes  the A$_1$-phase with order parameter
\begin{align}
 \label{mag12} A_{\rm A_1} &= \begin{pmatrix} B &\rmi B &  \\ -\rmi B & B &  \\  &  & 0 \end{pmatrix}.
\end{align}
It is realized in strong magnetic fields and describes a magnetized superfluid where only atoms with down-spins pair, while atoms with up-spins remain in the normal phase forming a Fermi liquid.

The real B- and pdB-phases get replaced by the complex B$_2$-phase in a magnetic field. The associated order parameter is given by
\begin{align}
 \label{mag21} A_{\rm B_2} &= \begin{pmatrix} B & \rmi M & \\ -\rmi M & B & \\ & & C \end{pmatrix}.
\end{align}
Here $B,M,C$ are real functions of $z$ with $B> M> 0$, $C>0$, that relate to the triplet states as in Eq. (\ref{mag6b}). In the limit $C=0$, we obtain a phase that replaces the P-phase in a magnetic field, which we call the P$_2$-phase hereafter. Its order parameter reads
\begin{align}
 \label{mag23} A_{\rm P_2} &= \begin{pmatrix} B & \rmi M & \\ -\rmi M & B & \\ & & 0 \end{pmatrix}.
\end{align}
While the P$_2$-phase is likely not realized in experiment, as is explained in Sec. \ref{SecThinH}, it is conceptually important. For $B=M$, the P$_2$-state is equivalent to the A$_1$-state. Hence there is no P$_1$-state. Since the polar state is not relevant in a slab geometry with physical parameters for the free energy coefficients, we do not consider its modifications in a magnetic field in this work.

\section{Zero-field phase diagram}\label{SecZero}

\subsection{Order parameter profiles as elliptic functions}\label{SecEllip}

The equilibrium order parameter $O(\bar{z})$ in a slab geometry can be found as the solution to the equations of motion. Order parameters of the form 
\begin{align}
 \label{ell1} O(\bar{z})=f(\bar{z})M,
\end{align}
with $M$ a suitably normalized constant $3\times 3$-matrix whose entries vanish on the third column, result in identical equations of motion for the function $f(\bar{z})$. The ensuing solution $f(\bar{z})$ can be determined analytically. Since many of the most competitive orders are of the above form---in particular the A-phase, P-phase, Pol-phase, A$_1$-phase---this yields many analytical statements about the phase diagram of the confined system, despite the complications due to the boundary conditions. Throughout this work, $f(\bar{z})$ will always refer to this specific function, with the appropriate boundary conditions depending on the context.

To put the function $f(\bar{z})$ into a concrete context, we consider the case of the B-phase solution for $\gamma=1$. The order parameter reads
\begin{align}
 \label{real13a} O_{\rm B}(\bar{z}) = f(\bar{z}) \begin{pmatrix} 1 && \\ & 1 & \\ && 1\end{pmatrix}.
\end{align}
Although this is not of the form (\ref{ell1}), the same function $f(\bar{z})$ applies. The equations of motion for $f(\bar{z})$, which follow from Eq. (\ref{real11}) below with $f_1=f_2=f_3=f$ and $\gamma=1$, are given by
\begin{align}
 \label{real13} -f'' -f +f^3 = 0,
\end{align}
where a prime denotes a derivative with respect to $\bar{z}$. Note that the equation is independent of the quartic couplings because of $3\zeta_{12}+\zeta_{345}=1$.  In the following, we only consider positive solutions without nodes in the interval $0< \bar{z}<\bar{D}$, as these have the lowest free energies. For specular boundaries, we have $f'(0)=f'(\bar{D})=0$, and the solution is simply
\begin{align}
 \label{ell1b}  f(\bar{z}) = 1\ \text{(specular)}.
\end{align}
For maximally pair-breaking boundary conditions, however, we have
\begin{align}
 \label{real13b} f(0)=f(\bar{D})=0.
\end{align} 
The corresponding solution $f(\bar{z})$ has a nontrivial functional form and is an elliptic function in the mathematical sense (a meromorphic function with two complex periods). In the following, we examine this function.

Some properties of the solution to Eqs. (\ref{real13}) and (\ref{real13b}) can be inferred directly. First, $f=0$ is a solution, corresponding to a system without superfluid order. Further, if the boundary condition was not there, $f=1$ would also be a solution (as in the specular case). Due to the confinement, the solution deviates from $f=1$ in a region of size $z \sim \xi$ close to the boundaries. In the limit $\bar{D}\to \infty$, where the second boundary is removed, the solution is given by
\begin{align}
 \label{real14} f(\bar{z}) = \tanh\Bigl(\frac{\bar{z}}{\sqrt{2}}\Bigr),
\end{align}
which can be verified through direct calculation. Note that this solution satisfies $f'(0)=1/\sqrt{2}$.

\begin{figure}[t!]
\centering
\includegraphics[width=8.6cm]{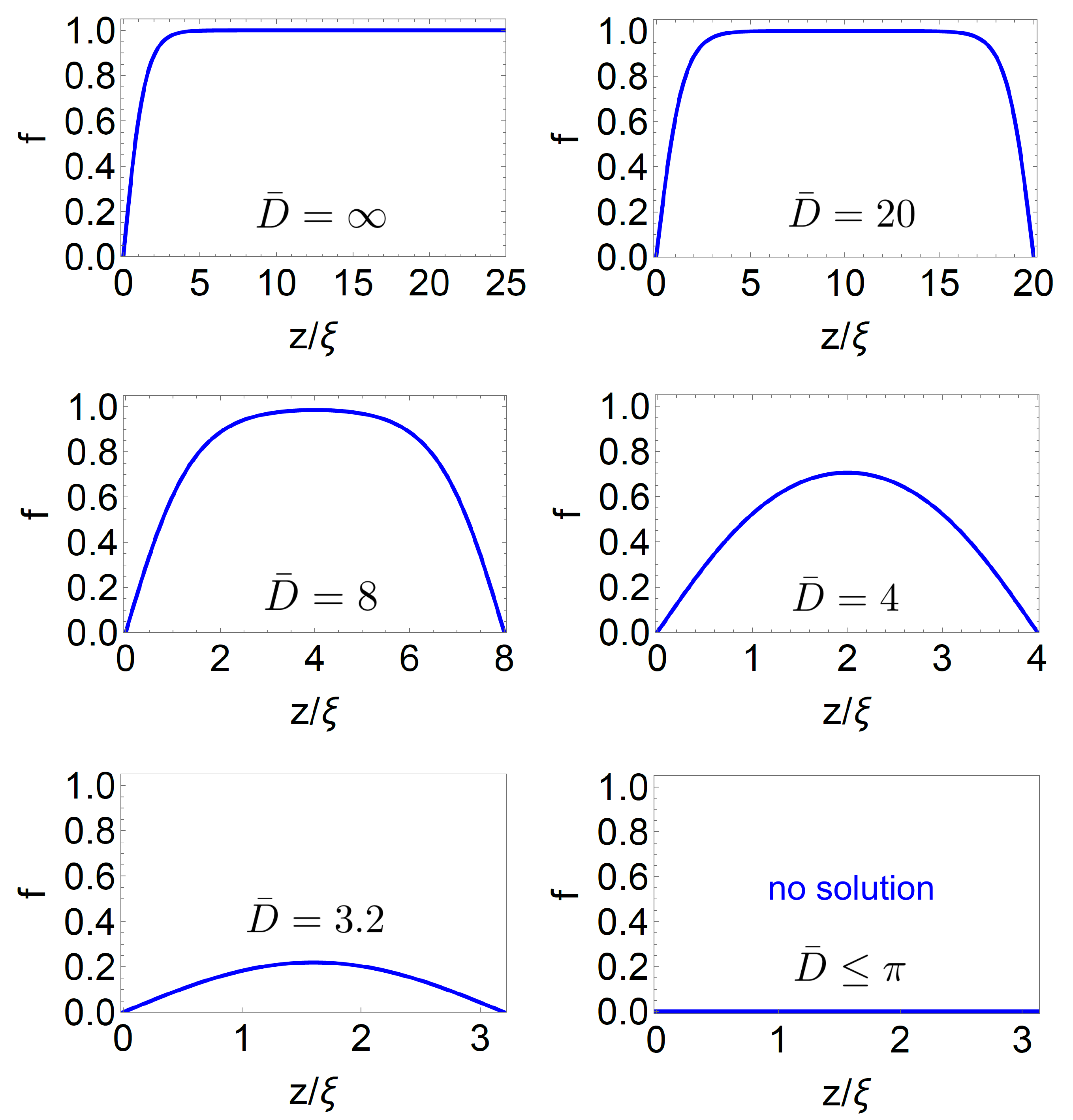}
\caption{Order parameter profiles in a slab geometry of height $D$ with maximally pair-breaking boundaries, for various ratios $\bar{D}=D/\xi$. The functions $f(\bar{z})$ are solutions to Eqs. (\ref{real13}), (\ref{real13b}) and parametrize a variety of superfluid orders. They are elliptic functions in the mathematical sense with real period $2\bar{D}$ and follow the analytical expression (\ref{real16}). Solutions only exist for $\bar{D}>\pi$. Note that the functions approach the bulk value $f= 1$ in regions sufficiently far from the boundaries.}
\label{FigBplots}
\end{figure}

Let us discuss two views on Eqs. (\ref{real13}) and (\ref{real13b}) that point towards its full solution. For one, the equation can be understood as a \emph{boundary value problem}. It may be solved numerically by discretizing the interval $z\in[0,\bar{D}]$ and starting with an initial guess of, say, $f=1$. On the other hand, we can understand the equation as an \emph{initial value problem} for the evolution of the function $f(\bar{z})$ with the time-coordinate $\bar{z}$. The initial conditions are $f(0)=0$ and $f'(0)=\sigma$, where the parameter $\sigma$ has to be fine-tuned such that $f(\bar{D})=0$ (shooting method). Evaluating $f(\bar{z})$ for values $\bar{z}>\bar{D}$, we obtain a periodic function with period $2\bar{D}$. Such periodic solutions exist for any $0<\sigma< 1/\sqrt{2}$, where the limit $\sigma=1/\sqrt{2}$ corresponds to the solution in Eq. (\ref{real14}) with $\bar{D}=\infty$, whereas $\sigma \to 0$ corresponds to $\bar{D}\to \pi^+$ and $f(\bar{z})\propto \sin(\bar{z})$. Any intermediate value $0<\sigma<1/\sqrt{2}$ smoothly corresponds to some confinement $\pi <\bar{D}<\infty$, whereas no solution exists for $\bar{D}\leq \pi$. The situation is summarized in Fig. \ref{FigBplots}.

The solution to Eq. (\ref{real13}) is given by
\begin{align}
 \label{real16} f(\bar{z}) = a\ \text{sn}\Bigl(\frac{\sigma \bar{z}}{a},k^2\Bigr),
\end{align}
where $\text{sn}(x,k^2)$ is the Jacobi elliptic function,
\begin{align}
 \label{real17} a=\sqrt{1-\sqrt{1-2\sigma^2}},\ k^2 = \frac{a^4}{2\sigma^2},
\end{align}
and the parameter $\sigma$ is determined such that
\begin{align}
 \label{real18} \frac{\sigma \bar{D}}{a} = 2K(k^2).
\end{align}
Here, $K(k^2)$ is the complete elliptic integral of the first kind defined by
\begin{align}
 \label{real19} K(k^2) =\int_0^1 \frac{\mbox{d}t}{\sqrt{(1-t^2)(1-k^2t^2)}}.
\end{align}
All parameters of $f(\bar{z})$ are determined by the value of $\bar{D}$. The numbers $\sigma$ and $a$ have the following interpretation: $\sigma=f'(0)$ is the slope at $\bar{z}=0$, and $a=f_{\rm max}=f(\bar{D}/2)$ is the maximal value or oscillation amplitude. The special functions $\text{sn}(x,k^2)$ and $K(k^2)$ are implemented in Mathematica as $\textsc{JacobiSN}[x,k^2]$ and $\textsc{EllipticK}[k^2]$, as well as in other software packages.

The proof of Eq. (\ref{real16}) is presented in App. \ref{AppProofTh} for a more general case. It relies on the observation that Eq. (\ref{real13}) describes a softening-spring Duffing oscillator with time coordinate $\bar{z}$ and period $2\bar{D}$. For such a system, the quantity
\begin{align}
\label{real19b} e(\bar{z}) = \frac{1}{2}f'(\bar{z})^2+\frac{1}{2}f(\bar{z})^2-\frac{1}{4}f(\bar{z})^4
\end{align}
is independent of $\bar{z}$ when evaluated on the solution, reflecting the fact that the energy of a Duffing oscillator is conserved.

\begin{figure}[t!]
\centering
\includegraphics[width=7cm]{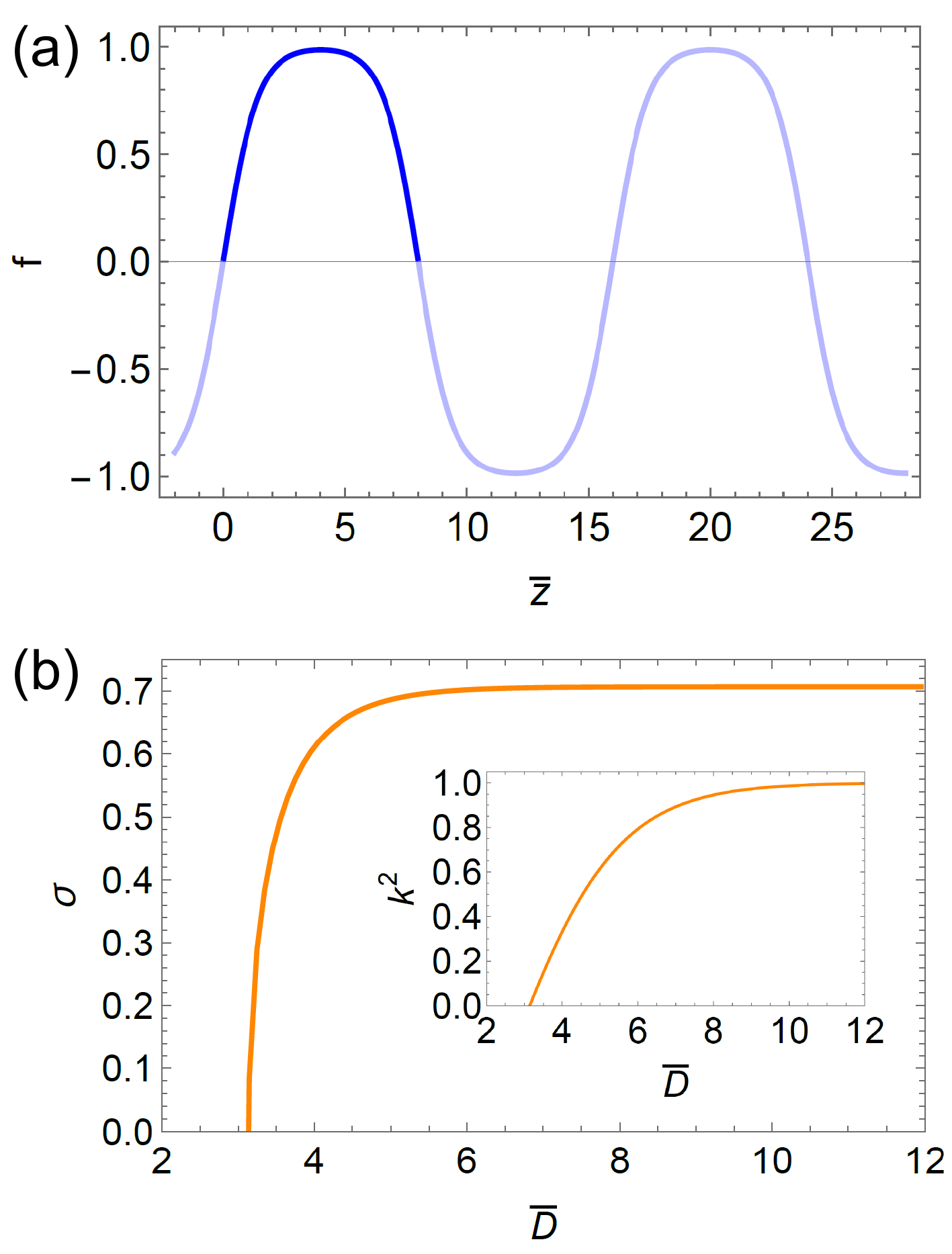}
\caption{Properties of the elliptic function $f(\bar{z})$. \emph{Panel (a).} The function $f(\bar{z})$ with $\bar{z}\in[0,\bar{D}]$ can be extended to a periodic function on $\bar{z}\in\mathbb{R}$. It then corresponds to a Duffing oscillator with softening spring and period $2\bar{D}$. \emph{Panel (b).} Solutions to Eq. (\ref{real13}) are Jacobi elliptic functions parametrized by $f'(0)=\sigma$ or, equivalently, the modular parameter $k^2$. The condition $0\leq k^2< 1$ for proper oscillations translates to $0\leq \sigma<1/\sqrt{2}$ and $\pi \leq \bar{D}<\infty$, although solutions with $\bar{D}=\pi$ have vanishing amplitude.}
\label{FigBsigmak2}
\end{figure}

The functions $\sigma(\bar{D})$ and $k^2(\bar{D})$ are shown in Fig. \ref{FigBsigmak2}. The reason why suitable solutions only exist for $\sigma\leq 1/\sqrt{2}$ can be understood as follows. The potential energy
\begin{align}
 \label{real27} e_{\rm pot}(\bar{z}) = \frac{1}{2}f(\bar{z})^2-\frac{1}{4}f(\bar{z})^4
\end{align}
represents an inverted double-well potential with maximal value $e=1/4$. The kinetic energy $e_{\rm kin}=\frac{1}{2}f'^2$ is always non-negative. Periodic solutions $f(\bar{z})$ with $f(\bar{z}+2\bar{D})=f(\bar{z})$ exist when $e < 1/4$, which means the evolution does not leave the potential energy trough around the origin. The limit $e=1/4$ corresponds to the solution for $\bar{D}=\infty$ in Eq. (\ref{real14}). Indeed, Eq. (\ref{real19}) then yields $k^2=1$, $\sigma=1/\sqrt{2}$, and $a=1$, and we can utilize the special property of the elliptic function that $\text{sn}(x,1)=\tanh(x)$, i.e.,
\begin{align}
\label{real28} f(\bar{z}) =  \text{sn}\Bigl(\frac{\bar{z}}{\sqrt{2}},1\Bigr) = \tanh\Bigl(\frac{\bar{z}}{\sqrt{2}}\Bigr).
\end{align}
Solutions with $\pi<\bar{D}<\infty$ corresponds to $0<e<1/4$ and $0<k^2 <1$. For $\bar{D}\to \pi^+$ we note that the solution to Eq. (\ref{real19}) yields $k^2\to 0^+$. In this limit, $a\simeq \sigma$, but both vanish: $a,\sigma\to 0$. However, the ratio 
\begin{align}
 \label{real29} \frac{f(\bar{z})}{a} \to \text{sn}(\bar{z},0) = \sin(\bar{z})
\end{align}
remains finite. In conclusion, Eq. (\ref{real16}) provides a full analytic understanding of the results shown in Fig. \ref{FigBplots}.

The free energy of the B-phase solution for $\gamma=1$ is given by
\begin{align}
 \label{real29b} \bar{F}_{\rm B,\gamma=1} = 3 \bar{F}_0(\bar{D}).
\end{align}
Here we define the reference free energy
\begin{align}
 \label{real29c} \bar{F}_0(\bar{D}) = \int_0^{\bar{D}}\mbox{d}\bar{z}\ \Bigl(\frac{1}{2}f'^2-\frac{1}{2}f^2+\frac{1}{4}f^4\Bigr)<0,
\end{align}
where $f(\bar{z})$ is the solution given in Eq. (\ref{real16}), which is fully determined by the value of $\bar{D}$.  The relevance of the quantity $\bar{F}_0$ is that all order parameters of the form (\ref{ell1}) lead to a free energy that is proportional to $\bar{F}_0(\bar{D})$, and the comparison of their prefactors gives the energetically favored configuration among them, even in a confined geometry. An analytic expression for $\bar{F}_0(\bar{D})$ is derived in App. \ref{AppClosedEnergy}.

\subsection{Phase structure of real order parameters}\label{SecRealEOM}

Consider superfluid orders with real order parameter matrix $O(\bar{z})$ and maximally pair-breaking boundaries. The equations of motion are then given by the matrix differential equation
\begin{equation}\label{real7}
\begin{aligned}
 0 ={}&-O_{\mu i}''-(\gamma-1)\delta_{i3}O_{\mu 3}''-O_{\mu i}\\
&+\zeta_{12}\mbox{tr}(OO^T)O_{\mu i}+\zeta_{345}(OO^TO)_{\mu i},
\end{aligned}
\end{equation}
with $O(0)=O(\bar{D})=0$. Crucially, $\zeta_{12}$ and $\zeta_{345}=1-3\zeta_{12}$ are not independent due to Eq. (\ref{real5}). Hence the phase diagram of real orders only depends on three dimensionless parameters: $\zeta_{12}$, $\gamma$, and $\bar{D}$. We further assume that the order parameter is diagonal and write
\begin{align}
 \label{real10} O(\bar{z}) = \begin{pmatrix} f_1(\bar{z}) & & \\ & f_2(\bar{z}) & \\ & & f_3(\bar{z}) \end{pmatrix}.
\end{align}
For $\gamma=1$, any real order parameter $A(z)$ that minimizes the free energy in a slab geometry can be assumed to be diagonal, because matrices with non-diagonal entries that minimize $F$ are related to diagonal matrices by an $\text{SO}(3)_{\rm S}\times\text{SO}(3)_{\rm L}$ transformation. For $\gamma>1$, since the orbital symmetry is reduced to $\text{SO}(2)_{\rm L}$, we were not able to show whether this property is generally true, and so Eq. (\ref{real10}) may constitute a truncation. Note, however, that  it simultaneously captures the B-, P-, Pol-, and pdB-phases.

The dimensionless free energy for real diagonal order is given by
\begin{equation}\label{real12c}
 \begin{aligned}
 \bar{F} ={}& \int_0^{\bar{D}}\mbox{d}\bar{z}\ \Biggl( \frac{1}{2}(f_1'^2+f_2'^2+\gamma f_3'^2)-\frac{1}{2}(f_1^2+f_2^2+f_3^2)\\
&+\frac{\zeta_{12}}{4}(f_1^2+f_2^2+f_3^2)^2+\frac{\zeta_{345}}{4}(f_1^4+f_2^4+f_3^4)\Biggr).
\end{aligned}
\end{equation}
The equations of motion read
\begin{equation}\label{real11}
\begin{aligned}
 &0 =- \begin{pmatrix} f_1'' & & \\ & f_2'' & \\ & & \gamma f_3''\end{pmatrix}-\begin{pmatrix} f_1 & & \\ & f_2 & \\ & & f_3\end{pmatrix}\\
 &+\zeta_{12}(f_1^2+f_2^2+f_3^2)\begin{pmatrix} f_1 & & \\ & f_2 & \\ & & f_3\end{pmatrix}+\zeta_{345}\begin{pmatrix} f_1^3 & & \\ & f_2^3 & \\ & & f_3^3\end{pmatrix}.
\end{aligned}
\end{equation}
The boundary conditions are given by $f_i(0) = f_i(\bar{D})=0$ for $i=1,2,3$.

Several properties of the solution can be inferred from the form of these equations. We first observe that the equation for each $f_i$ can be solved by setting $f_i=0$. Furthermore, all equations are invariant under $f_i\to -f_i$ for the components individually, and we fix $f_i\geq 0$ for concreteness. The equations for $f_1$ and $f_2$ are independent of $\gamma$, and so the P- and Pol-phase are independent of $\gamma$. The equations for $f_1$ and $f_2$ are identical, hence either $f_1=f_2\neq 0$ are identical, or at least one of them has to vanish. For $\gamma=1$, the equation for $f_3$ is also identical to the others, thus either $f_1=f_2=f_3=f$, which is the B-phase, or some of the components vanish. For $\gamma>1$, the equation for $f_3$ is manifestly different from the other two and suppresses $f_3$ because of the higher kinetic-energy costs, thus $f_1=f_2\neq f_3$ if all components are nonzero. Importantly, for $\gamma>1$ the B-phase cannot be a solution to the equations of motion in a slab of finite height, although the difference between $f_1$ and $f_3$ would be immeasurably small for  large heights.

\begin{figure}[t!]
\centering
\includegraphics[width=8.6cm]{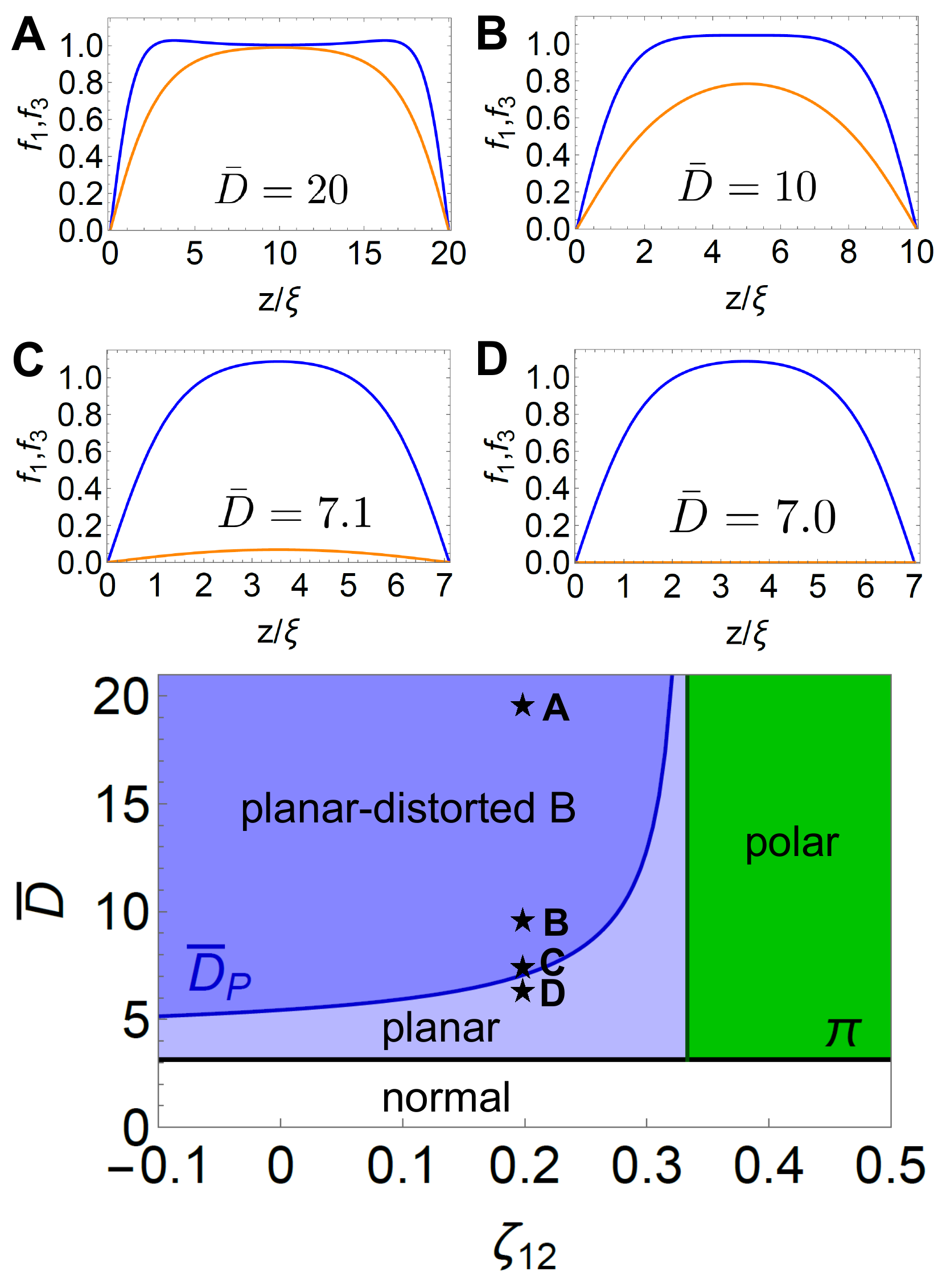}
\caption{Phase diagram for real order parameters with maximally pair-breaking boundaries. For sufficiently large $\bar{D}=D/\xi$ and $\zeta_{12}<\frac{1}{3}$, the system is in the pdB-phase with $f_1=f_2>f_3$. The phase boundary $\bar{D}_{\rm P}(\gamma,\zeta_{12})$ is determined analytically by Eq. (\ref{real41}). For tight confinement, the system is in the P- or Pol-phase, depending on $\zeta_{12}$, whereas no superfluid order is possible for $\bar{D}\leq \pi$. Representative order parameter profiles are shown in panels A--D along the weak-coupling line $\zeta_{12}=\frac{1}{5}$, with $f_1$ ($f_3$) shown in blue (orange), corresponding to the stars in the phase diagram. The phase diagram is shown here for $\gamma=3$, but has the same shape for all $\gamma>1$. For $\gamma\to1$, $\bar{D}_{\rm P}\to \pi$, the P-phase vanishes, and the pdB-phase becomes the B-phase.} 
\label{FigRealPD}
\end{figure}

The phase diagram for real orders determined by Eqs. (\ref{real11}) for $\gamma>1$ is shown in Fig. \ref{FigRealPD}, together with representative order parameter profiles. In the following, we summarize its key features, while the derivation of these statements is left to the remainder of this section. Stable superfluid orders are found for $\zeta_{12}<\frac{1}{2}$ and $\bar{D}>\bar{D}_{\rm c}$, with
\begin{align}
 \label{real32b} \bar{D}_{\rm c}=\pi,
\end{align}
independently of $\gamma$ and $\zeta_{12}$. For $\frac{1}{3}<\zeta_{12}<\frac{1}{2}$, the Pol-phase is realized for all $\bar{D}>\pi$. For $\zeta_{12}<\frac{1}{3}$, the phase structure has a more intriguing behavior. For large enough $\bar{D}>D_{\rm P}$, with $\bar{D}_{\rm P}=\bar{D}_{\rm P}(\gamma,\zeta_{12})$ given in Eq. (\ref{real41}), the system is in the pdB-phase. The function $\bar{D}_{\rm P}(\gamma,\zeta_{12})$ satisfies
\begin{align}
 \label{slabDPextra} \bar{D}_{\rm P}(\gamma,0) &=\sqrt{\gamma} \pi,\ \bar{D}_{\rm P}\Bigl(\gamma,\frac{1}{3}\Bigr)=\infty.
\end{align}
For $\bar{D} \in(\pi, \bar{D}_{\rm P})$, the system is in the P-phase. For even tighter confinements, $\bar{D}<\pi$, the order parameter vanishes and the system is in the normal phase. (For $\gamma=1$, the phase structure is considerably simpler: 
the P- and pdB-phase regions are replaced by the B-phase, whereas the Pol-phase region persists.)

We now derive the statements from the previous paragraph. For the P- and Pol-phase, it is convenient to parametrize the order parameters as
\begin{align}
 \label{slab3} O_{\rm P}(\bar{z}) &= \frac{f(\bar{z})}{\sqrt{1-\zeta_{12}}} \begin{pmatrix} 1 && \\ & 1& \\ && 0 \end{pmatrix},\\
\label{slab4} O_{\rm Pol}(\bar{z}) &= \frac{f(\bar{z})}{\sqrt{1-2\zeta_{12}}} \begin{pmatrix} 1 && \\ & 0& \\ && 0 \end{pmatrix},
\end{align}
respectively, because the equations of motion for $f(\bar{z})$ are then given by Eqs. (\ref{real13}), (\ref{real13b}) with the corresponding elliptic function solution from Eq. (\ref{real16}). This implies that the dimensionless free energies are given by
\begin{align}
 \label{slab5} \bar{F}_{\rm P} &= \frac{2}{1-\zeta_{12}}\bar{F}_0(\bar{D}),\\
  \label{slab6} \bar{F}_{\rm Pol}&=\frac{1}{1-2\zeta_{12}}\bar{F}_0(\bar{D}).
\end{align}
Hence we have $|\bar{F}_{\rm Pol}|=\frac{1-\zeta_{12}}{2(1-2\zeta_{12})}|\bar{F}_{\rm P}|$ and, since both free energies are negative, the Pol-phase is energetically favored over the P-phase for $\zeta_{12} >\frac{1}{3}$. The transition between both phases is  of first order.

To analyze the second-order transition between the P- and pdB-phase, we set $f_1=f_2$ in Eq. (\ref{real10}) and obtain the coupled pdB-phase equations
\begin{equation}\label{real30}
\begin{aligned}
 0 &=-f_1''-f_1 +(1-\zeta_{12}) f_1^3 +\zeta_{12} f_1f_3^2,\\
 0 &=-\gamma f_3'' -f_3+(1-2\zeta_{12})f_3^3 + 2\zeta_{12} f_1^2f_3.
\end{aligned}
\end{equation}
The boundary conditions are
\begin{equation}\label{real31}
 \begin{aligned}
 0 &=f_1(0) = f_1(\bar{D}),\\
 0&=f_3(0) = f_3(\bar{D}).
\end{aligned}
\end{equation}
This system of coupled Duffing oscillators  does not seem to admit an analytic solution, except in special cases, and hence needs to be solved numerically as a boundary value problem.
Both $f_1(\bar{z})$ and $f_3(\bar{z})$ approach $1$ in regions far away from the boundary if $\bar{D}$ is large enough, mimicking the bulk B-phase. Solutions with $f_3>0$ exist for $\bar{D}>\bar{D}_{\rm P}$, whereas $f_3=0$ for all $\bar{D}<\bar{D}_{\rm P}$, corresponding to the pdB- and P-phases. The transition between both phases is of second order since $f_3\to 0$ continuously at $\bar{D}_{\rm P}$. The dimensionless free energy in the pdB-phase is given by
\begin{align}
 \nonumber \bar{F}_{\rm pdB} ={}& \int_0^{\bar{D}}\mbox{d}\bar{z}\ \Biggl( \frac{1}{2}(2f_1'^2+\gamma f_3'^2)-\frac{1}{2}(2f_1^2+f_3^2)\\
 \label{real12d}&+\frac{1-\zeta_{12}}{2}f_1^4+\frac{1-2\zeta_{12}}{4}f_3^4+\zeta_{12}f_1^2f_3^2\Biggr).
\end{align}
When evaluated for the solution to the equations of motion for either maximally pair-breaking or specular boundaries, we have
\begin{align}
 \bar{F}_{\rm pdB} ={}& -\int_0^{\bar{D}}\mbox{d}\bar{z} \Biggl( \frac{1-\zeta_{12}}{2}f_1^4+\frac{1-2\zeta_{12}}{4}f_3^4+\zeta_{12}f_1^2f_3^2\Biggr).
\end{align}
This expression is not needed for the analysis here, but determines the competition between the A- and pdB-phases below.

Although the solutions to Eqs. (\ref{real30}) need to be determined numerically, the resulting phase boundary $\bar{D}_{\rm P}$ can be determined analytically by mapping the system of equations (\ref{real30}) close to the phase boundary to the Lam\'{e} equation. This yields
\begin{align}
 \label{real41} \bar{D}_{\rm P}(\gamma,\zeta_{12}) = 2\sqrt{k_{1\rm c}^2+1}\ K(k_{1\rm c}^2).
\end{align}
Herein, $k_{1\rm c}^2$ is the critical value of $k_1^2$ such that a nonzero solution $f_3$ to Eq. (\ref{real30}) exists in the presence of $f_1\neq 0$. Its value is determined by 
\begin{align}
 \label{real39} \frac{k_{1\rm c}^2+1}{\gamma} = a_\nu^{(1)}(k_{1\rm c}^2),
\end{align}
where $a_\nu^{(2m+1)}(k^2)$ is the Lam\'{e} eigenvalue and
\begin{align}
 \label{real40} \nu = \frac{1}{2}\Biggl(-1+\sqrt{1+\frac{16\zeta_{12}}{\gamma(1-\zeta_{12})}}\Biggr).
\end{align}
To solve Eq. (\ref{real39}) with Mathematica, use $a_\nu^{(2m+1)}(k^2)=\textsc{LameEigenvalueA}[\nu,2m+1,k^2]$, and similarly in other software packages. Some properties of the Lam\'{e} equation are reviewed in Appendix \ref{AppReal}, together with the derivation of Eq. (\ref{real41}). For the weak-coupling values $\gamma=3$ and $\zeta_{12}=\frac{1}{5}$, we have $\bar{D}_{\rm P}(3,\frac{1}{5}) = 7.08675$.

\subsection{Competition with the A-phase}\label{SecA}

We now consider the A-phase for maximally pair-breaking boundary conditions. We parametrize the order parameter by
\begin{align}
 \label{A2} O_{\rm A}(\bar{z}) = \frac{f(\bar{z})}{\sqrt{2\zeta_{245}}}\begin{pmatrix} 1& \rmi & 0\\ && \\ & &  \end{pmatrix},
\end{align} 
which ensures that the dimensionless real function  $f(\bar{z})$ satisfies the equations of motion (\ref{real13}), (\ref{real13b}). The corresponding dimensionless free energy is
\begin{align}
 \label{A3} \bar{F}_{\rm A} = \frac{1}{\zeta_{245}}\bar{F}_0(\bar{D}).
\end{align}
Note that this expression is independent of $\gamma$, since the order parameter does not have a third column and so does not contribute to the second kinetic energy  term in Eq. (\ref{gl1}). It is irrelevant whether the pair $(1,\rmi)$ appears on the first, second, or third row of Eq. (\ref{A2}), or in other symmetry-related permutations. In this work, we choose the first row for concreteness.

\begin{figure}[t!]
\centering
\includegraphics[width=8.6cm]{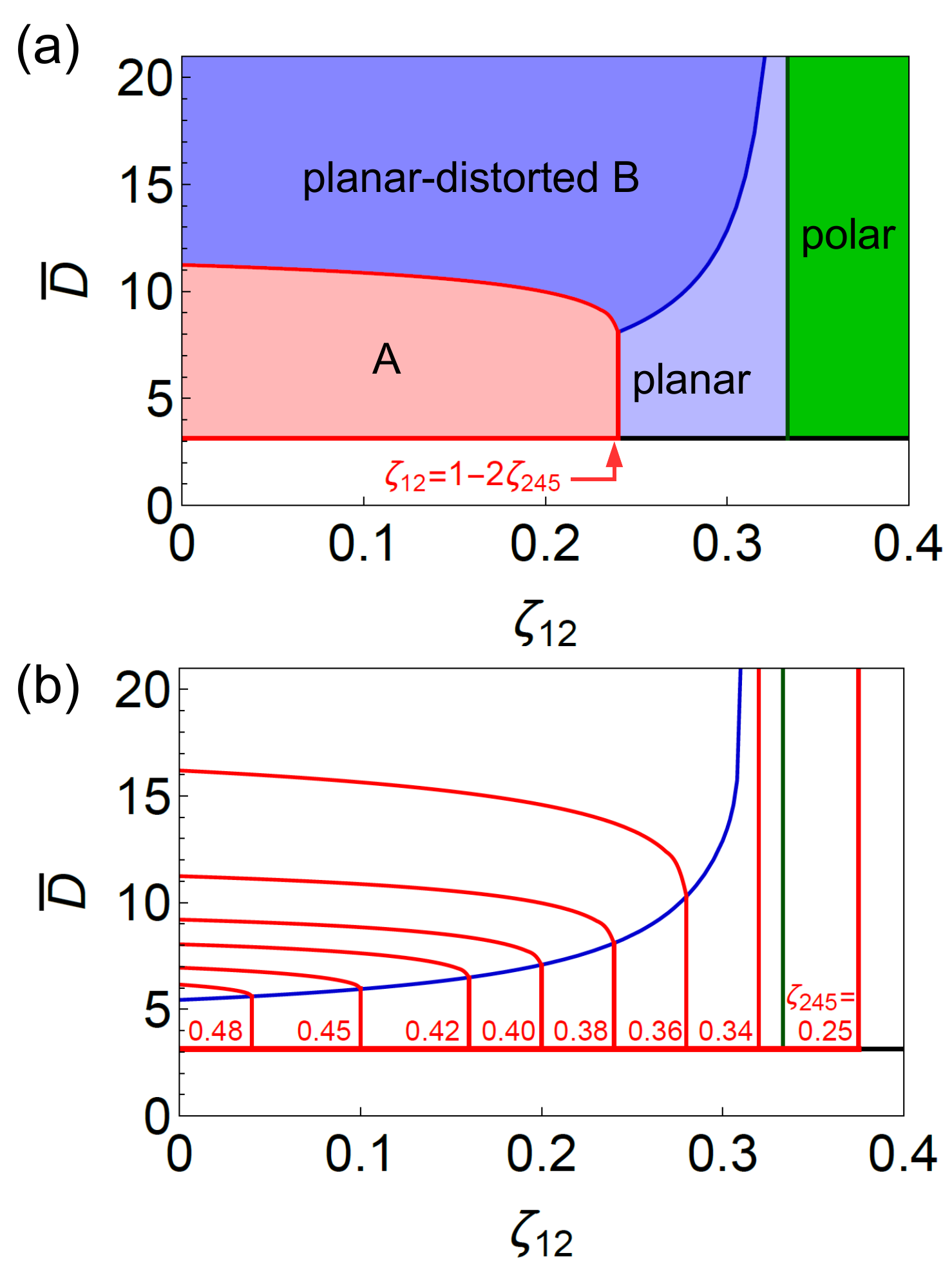}
\caption{Phase diagram for maximally pair-breaking boundary conditions. \emph{Panel (a).} The time-reversal symmetry breaking A-phase competes with the real orders of Fig. \ref{FigRealPD}, resulting in first-order phase transitions, shown here for $\zeta_{245}=0.38$ and $\gamma=3$. For $\zeta_{245}>\frac{1}{3}$, the A-phase is energetically favorable over the P-phase in the region $\zeta_{12}<1-2\zeta_{245}$ (red arrow). \emph{Panel (b).}  The competition between the A- and pdB-phase has to be determined numerically as function of $\zeta_{12},\zeta_{245},\gamma$, and $\bar{D}$. We show the phase boundary between the A-phase and the real orders for several values of $\zeta_{245}$. For $\zeta_{245}<\frac{1}{3}$, the phase diagram consist of only two phases, the A- and Pol-phase, separated by a vertical line at $\zeta_{12}=\frac{1}{2}(1-\zeta_{245})$. The phase diagrams in the 3D and quasi-2D limits, shown in Fig. \ref{FigBulkPD}, can be read off from the behavior at $\bar{D}=\infty$ and $\bar{D}=\pi$, respectively.}
\label{FigAPD}
\end{figure}

While the phase diagram of real orders in Fig. \ref{FigRealPD} only depends on the parameters $\gamma$ and $\zeta_{12}$, the competition with the A-phase introduces $\zeta_{245}$ as another relevant control parameter. In the following, we determine the phase structure that arises from the free energy competition between the A-, P-, Pol-, and pdB-phases for $\gamma>1$ and maximally pair-breaking boundaries. The resulting phase diagram is shown in Fig. \ref{FigAPD}.

The transition from the normal into the superfluid phase occurs at $\bar{D}_{\rm c}=\pi$, where the system enters either the A-, P-, or Pol-phase. Since in each of these phases the free energy is proportional to $\bar{F}_0$, the relative size of the prefactors determines the phase boundaries analytically and independently of $\bar{D}$. The A-phase competes with the P-phase or Pol-phase if $\zeta_{245}>\frac{1}{3}$ or $\zeta_{245}<\frac{1}{3}$, respectively. If $\zeta_{12}<\frac{1}{3}$, then the A-phase wins over the P-phase for $\zeta_{12}<1-2\zeta_{245}$. If $\zeta_{12}>\frac{1}{3}$, then the A-phase wins over the Pol-phase for $2\zeta_{12}<1-\zeta_{245}$. All of these transitions are of first order. For the weak-coupling values $\zeta_{12}=\frac{1}{5}$ and $\zeta_{245}=\frac{2}{5}$, A- and P-phase are energetically degenerate.

To determine the phase boundary between the A- and pdB-phase, we consider the region of the phase diagram with $\zeta_{12}<\frac{1}{3}$ and $\bar{D}>\bar{D}_{\rm P}$. The ensuing superfluid order is then determined by solving the equations of motion (\ref{real30})  and comparing the free energies $\bar{F}_{\rm pdB}$ and $\bar{F}_{\rm A}$ from Eqs. (\ref{real12d}) and (\ref{A3}). We find that the A-phase is favorable in a region that increases as $\zeta_{245}$ is decreased. Since the boundary between the A- and pdB-phase always touches the curve $\bar{D}_{\rm P}(\zeta_{12})$ (at the value $\zeta_{12}=1-2\zeta_{245}$), the pdB- and P-phases vanish from the phase diagram when $\zeta_{245}<\frac{1}{3}$, because $\bar{D}_{\rm P}\to \infty$ as $\zeta_{12}\to \frac{1}{3}$ (see Eq. (\ref{slabDPextra})). In this case, the phase diagram only consists of the A- and Pol-phase, realized for all $\bar{D}>\pi$ and separated by a vertical line in the phase diagram. Various phase boundaries as a function of $\zeta_{245}$ are shown in Fig. \ref{FigAPD}.

\subsection{Specular boundaries and the stripe phase}\label{SecStripe}

Specular boundary conditions imply that the order parameters $A(z)$ for the A-, P-, and Pol-phases are constant. Replacing $\bar{F}_0(\bar{D})\to -\frac{1}{4}\bar{D}$ with its bulk value, see App. \ref{AppClosedEnergy}, we obtain
\begin{align}
 \label{spec2} \bar{F}_{\rm A} &= -\frac{1}{4\zeta_{245}}\bar{D},\\
 \label{spec3} \bar{F}_{\rm P} &= -\frac{1}{2(1-\zeta_{12})}\bar{D},\\
 \label{spec4} \bar{F}_{\rm Pol} &= -\frac{1}{4(1-2\zeta_{12})}\bar{D}.
\end{align}
There is no obstruction to superfluidity for tight confinement and ordering in either one of the three phases exists for all $\bar{D}>\bar{D}_{\rm c}=0$. 

\begin{figure}[t!]
\centering
\includegraphics[width=8.6cm]{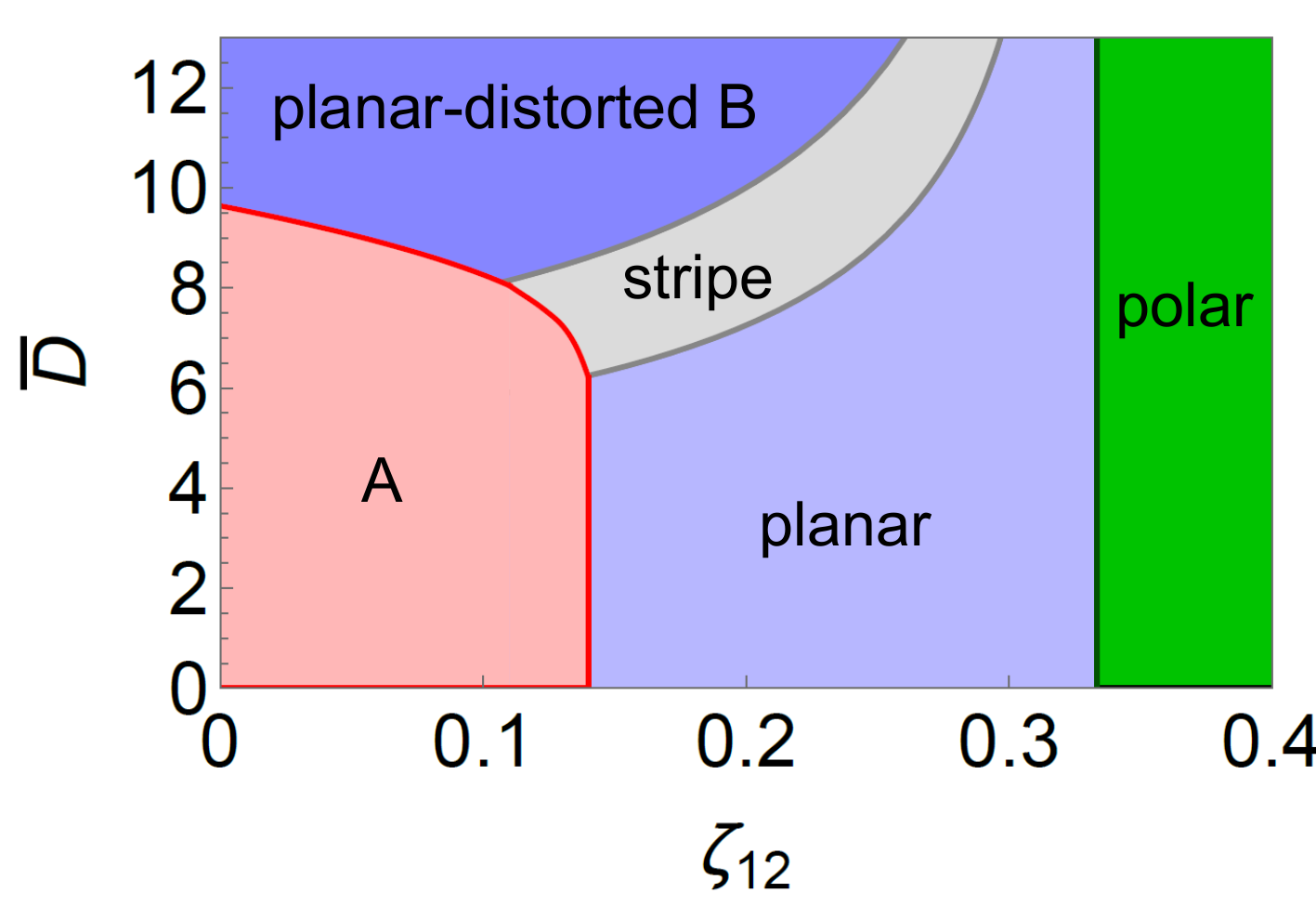}
\caption{Phase diagram for specular boundary conditions, shown here for $\gamma=3$ and $\zeta_{245}=0.43$. The phase structure is similar to the one with maximally pair-breaking conditions in Fig. \ref{FigAPD}, but with two key differences: (i) Superfluidity persists for all confinements $\bar{D}>\bar{D}_{\rm c} = 0$, in contrast to $\bar{D}_{\rm c}=\pi$ in the other case. (ii) The P-pdB transition is preempted by a time-reversal symmetry preserving stripe phase where the system spontaneously breaks translation symmetry in one of the unconfined directions. For $\zeta_{245}<\frac{1}{3}$ (not shown), the system has only two phases, the A- and Pol-phase, separated by a vertical line in the plot.}
\label{FigSpecPD}
\end{figure}

The order parameter profiles in the pdB-phase are nontrivial and need to be determined numerically from Eq. (\ref{real30}) with the boundary conditions
\begin{align}
 \label{spec5} 0 &= f_1'(0)=f_1'(\bar{D}),\\
 \label{spec6} 0 &= f_3(0)=f_3(\bar{D}).
\end{align}
The transition from the P- to the pdB-phase is preempted by the stripe phase discussed below. However, restricting to order parameters that only depend on the $z$-coordinate, $A(z)$, the boundary $\bar{D}_{\rm P}$ of the putative P-pdB-transition can be computed analytically.
For this, we consider again the limit where we neglect terms of order $f_3^2$ and $f_3^3$ in Eqs. (\ref{real30}). Then $f_1=\frac{1}{\sqrt{1-\zeta_{12}}}$ for the first component. For $f_3$ close to the transition, instead of a Lam\'{e} equation, we obtain a simple harmonic oscillator
\begin{align}
 \label{spec7} 0 &\approx  f_3'' +\frac{1-3\zeta_{12}}{\gamma(1-\zeta_{12})}f_3.
\end{align}
The critical confinement is the half-period of the oscillator, hence
\begin{align}
 \label{spec8} \bar{D}_{\rm P}(\gamma,\zeta_{12}) = \pi \sqrt{\frac{\gamma(1-\zeta_{12})}{1-3\zeta_{12}}}.
\end{align}
As in the maximally pair-breaking case, we have
\begin{align}
 \label{spec9} \bar{D}_{\rm P}(\gamma,0) =\sqrt{\gamma} \pi,\  \bar{D}_{\rm P}\Bigl(\gamma,\frac{1}{3}\Bigr)=\infty,
\end{align}
and for the weak-coupling point $\bar{D}_{\rm P}(3,\frac{1}{5})=\sqrt{6}\pi=7.6953$. In fact, the transition lines $\bar{D}_{\rm P}$ in the maximally pair-breaking and specular cases look quite similar in the phase diagram. The full phase diagram including the S- and A-phases is presented in Fig. \ref{FigSpecPD}. The phase diagram of real orders including the line $\bar{D}_{\rm P}$ for specular boundaries (neglecting the S-phase) is shown in the appendix in Fig. \ref{FigAppRealSpecular}.

The S-phase constitutes a superfluid phase where translation symmetry in the unconfined x- and y-directions is spontaneously broken. Assuming for concreteness that translation symmetry in the y-direction is broken, the order parameter develops stripes of width $\sim L_{\rm S}$ in the y-direction, which have alternating sign in the diagonal component $O_{33}(y,z)$ and are separated by domain walls. Why would this lower the free energy, if the formation of domain walls costs kinetic energy? The answer lies in the fact that the domain walls in $O_{33}(y,z)$ are only part of the story: the system also develops a modulated off-diagonal component $O_{32}(y,z)$ that is maximal when $O_{33}(y,z)$ is zero, and vice versa. The interplay of both components in the \emph{second} kinetic term proportional to $\gamma-1$ in the free energy leads to a situation where a stripe modulation in y-direction can be energetically favored over a homogeneous profile in y-direction. Importantly, this effect can only be effective for (at least partly) specular boundaries, since otherwise both $O_{32}$ and $O_{33}$ would have to vanish at the container walls, making the mechanism ineffective. For this reason the S-phase is not favorable for maximally pair-breaking boundary conditions.

To make these statements explicit, we use the symmetry properties of the stripe arrangement discussed above, together with time-reversal symmetry, to parametrize the order parameter in the S-phase as
\begin{align}
 \label{spec10} O_{\rm S}(\bar{y},\bar{z}) = \begin{pmatrix} O_1(\bar{y},\bar{z}) & 0 & 0 \\ 0 & O_2(\bar{y},\bar{z}) & O_{23}(\bar{y},\bar{z}) \\ 0 & O_{32}(\bar{y},\bar{z}) & O_3(\bar{y},\bar{z})\end{pmatrix},
\end{align}
with real matrix elements \cite{wiman2016strong}. Since the order parameter is real, as outlined in Sec. \ref{SecRealEOM}, the order parameter profiles and free energy of the S-phase only depend on the values of $\zeta_{12}$, $\gamma$, and $\bar{D}$. They need to be found numerically by solving the partial differential equations of motion in the yz-plane, for instance by means of the finite-element method applied to the associated boundary value problem. We present this numerical approach  in App. \ref{AppStripe}. Here we concentrate on the mechanism due to the kinetic energy term, which reads
\begin{equation}\label{spec11}
\begin{aligned}
 &\bar{F}_{\rm kin}[O_{\rm S}(\bar{y},\bar{z})] \propto \int\mbox{d}\bar{y} \int_0^{\bar{D}}\mbox{d}\bar{z} \Biggl( \frac{1}{2}\Bigl[(\partial_{\bar{y}} O_1)^2+(\partial_{\bar{y}} O_{23})^2\\
 &+(\partial_{\bar{y}} O_3)^2+(\partial_{\bar{z}} O_1)^2+(\partial_{\bar{z}} O_2)^2+(\partial_{\bar{z}} O_{32})^2\Bigr]\\
&+\frac{\gamma}{2}\Bigl[(\partial_{\bar{y}}O_2)^2+(\partial_{\bar{y}}O_{32})^2+(\partial_{\bar{z}}O_{23})^2+(\partial_{\bar{z}}O_3)^2\Bigr]\\
&+(\gamma-1)\Bigl[(\partial_{\bar{y}} O_2)(\partial_{\bar{z}}O_{23})+(\partial_{\bar{y}} O_{32})(\partial_{\bar{z}}O_3)\Bigr]\Biggr).
\end{aligned}
\end{equation}
The first three lines are obviously positive semi-definite and cannot be minimized by modulations. The last line, however, can be negative. For example, close to the transition from the P- to the S-phase, the matrix components of $O_{\rm S}$ that solve the equation of motions behave as
\begin{equation}\label{spec12}
\begin{aligned}
 O_1(\bar{y},\bar{z})&\simeq O_2(\bar{y},\bar{z}) \simeq \text{const},\\
 O_{23}(\bar{y},\bar{z}) &\simeq 0\\
 O_{32}(\bar{y},\bar{z}) &\simeq \mathcal{A}_{32}\cos\Bigl(\frac{\pi \bar{y}}{\bar{L}_{\rm S}}\Bigr)\cos\Bigl(\frac{\pi \bar{z}}{\bar{D}}\Bigr),\\
 O_3(\bar{y},\bar{z}) &\simeq \mathcal{A}_3\sin\Bigl(\frac{\pi \bar{y}}{\bar{L}_{\rm S}}\Bigr)\sin\Bigl(\frac{\pi \bar{z}}{\bar{D}}\Bigr),
\end{aligned}
\end{equation}
with amplitudes $\mathcal{A}_{32},\mathcal{A}_3>0$. With this ansatz, the last line in Eq. (\ref{spec11}) becomes
\begin{align}
 \label{spec13} -(\gamma-1) \mathcal{A}_{32}\mathcal{A}_3 \Bigl[\sin\Bigl(\frac{\pi \bar{y}}{\bar{L}_{\rm S}}\Bigr)\cos\Bigl(\frac{\pi \bar{z}}{\bar{D}}\Bigr)\Bigr]^2,
\end{align}
which is manifestly negative and can lower the free energy because of the nonvanishing modulation.

Of course, eventually the free energy of the stripe solution to the equations of motion has to be computed numerically and compared to the phases described by order parameters of the form $A(z)$. One finds that the S-phase is energetically favorable for $\gamma>2$ in a range of confinements 
\begin{align}
 \label{spec14} \bar{D}_{\rm S1} < \bar{D} < \bar{D}_{\rm S2}.
\end{align}
Importantly, $\gamma$ needs to be sufficiently large to give rise to the S-phase, because lowering of the free energy according to Eq. (\ref{spec13}) is insufficient for $\gamma\leq 2$. Furthermore, the free energy of the S-phase compared to the ones without $y$-dependence is better by less than a percent at its peak. This may imply that the S-phase regime might be smaller or even vanish when corrections to the free energy functional beyond Ginzburg--Landau theory or order-parameter fluctuations are included. It is known from other paired-fermion superfluids such as ultracold atoms or relativistic condensates that mean-field theories often overestimate the regime of phases with spontaneously broken translation symmetry, see for instance  Refs. \cite{PhysRevA.91.053611,BoettcherSarma,PhysRevD.91.116006,PhysRevA.95.063626,Kinnunen_2018}.

Following Wiman and Sauls \cite{wiman2016strong}, a few analytical insights into the S-phase can be gained by treating the parametrization in Eq. (\ref{spec12}) as a variational ansatz close to the lower transition at $\bar{D}_{\rm S1}$. We present the full calculation in App. \ref{AppDS1}. One finds that, for $\gamma>2$, the optimal solution close to $\bar{D}_{\rm S1}$ satisfies
\begin{align}
 \label{spec15} \frac{\mathcal{A}_{32}}{\mathcal{A}_3} = \sqrt{\frac{\gamma-2}{\gamma}},\ \bar{L}_{\rm S} = \sqrt{\frac{\gamma}{\gamma-2}}\bar{D},
\end{align}
and we have
\begin{align}
 \label{spec16} \bar{D}_{\rm S1} = \frac{2\sqrt{\gamma-1}}{\gamma} \bar{D}_{\rm P}.
\end{align}
In particular, $\bar{D}_{\rm S1}<\bar{D}_{\rm P}$ for $\gamma>2$. For $\gamma=3$ we obtain $\bar{L}_{\rm S}=\sqrt{3}\bar{D}$ and $\bar{D}_{\rm S1}=0.94\bar{D}_{\rm P}$. The upper transition $\bar{D}_{\rm S2}$ cannot be determined analytically. We find that the numerical solution to the equations of motion is rather well captured by the parametrization in Eq. (\ref{spec12}), but with sine and cosine in the y-direction replaced by elliptic Jacobi-sn and -cn functions. The transition into the pdB-phase for $\gamma=3$ occurs at $\bar{D}_{\rm S2}\simeq 1.3 \bar{D}_{\rm P}$.

For the case of diffusive boundary conditions, we find that even for small values of $b_{\rm T}$ such as $b_{\rm T}=0.5$, 
the phase diagram features a stripe phase, although in a narrower strip of $\bar{D}$-values. We discuss the interplay between $b_{\rm T}$ and the S-phase in App. \ref{AppStripe}.

\subsection{Dimensional crossover from 3D to quasi-2D}

\begin{figure}[t!]
\centering
\includegraphics[width=8.6cm]{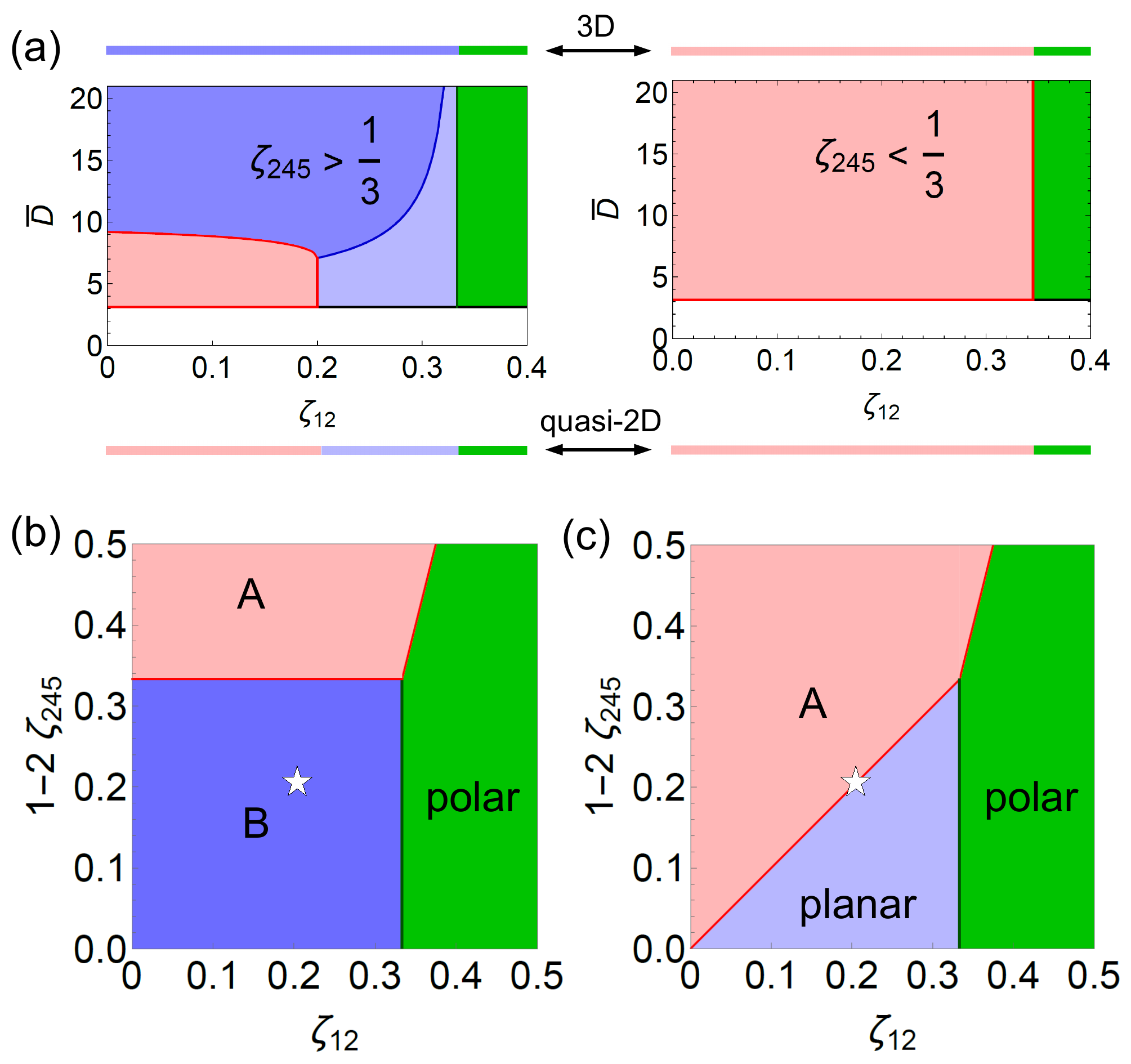}
\caption{\emph{Panel (a).} The 3D bulk and quasi-2D phase diagrams are obtained from the phase diagram under confinement in Figs. \ref{FigAPD} and \ref{FigSpecPD} for $\bar{D}\to \infty$ and $\bar{D}\to \bar{D}_{\rm c}$, respectively. Here, $\bar{D}_{\rm c}=\pi$ for maximally pair-breaking and $\bar{D}_{\rm c}=0$ for specular boundary conditions. Both 3D and quasi-2D limits are independent of the boundary conditions. The topology of the phase structure is different for $\zeta_{245}>\frac{1}{3}$ and $\zeta_{245}<\frac{1}{3}$, which is shown here schematically for maximally pair-breaking boundaries and fixed values of $\zeta_{245}$. \emph{Panel (b).} 3D bulk phase diagram: The P- and S-phases are absent in the bulk. The star indicates the weak-coupling point $\zeta_{12}=\zeta_{245}=0.2$. \emph{Panel (c).} Quasi-2D phase diagram for tight confinement: The star indicates again the weak-coupling point, where P and A are energetically degenerate. Beyond the quasi-2D limit, the phases displayed here are realized at the normal-to-superfluid transition under confinement for a given set of critical values $\zeta_{12,\rm c}$ and $\zeta_{245,\rm c}$ at $T_{\rm c}(P)$. Note that all stable quasi-2D orders have order parameters that do not possess a third column in accordance with Ref. \cite{PhysRevB.108.144503}.}
\label{FigBulkPD}
\end{figure}

In the 3D bulk system, the superfluid orders are homogeneous and the value of $\gamma$ is irrelevant. The corresponding orders that connect to our analysis are the A-, B-, P-, and Pol-phases, each described by the function $f(\bar{z})=1$. To determine the associated bulk phase diagram, we use the expressions (\ref{A3}), (\ref{real29b}), (\ref{slab5}),\ (\ref{slab6}) for the dimensionless free energies of the A-, B-, P-, and Pol-phases and set $f=1$. This yields $\bar{F}_0(\bar{D})=-\frac{1}{4}\bar{D}$, see App. \ref{AppClosedEnergy}, and the bulk values
\begin{align}
 \label{bulk1} \bar{F}_{\rm A} &= -\frac{1}{4\zeta_{245}}\bar{D},\\
 \label{bulk2} \bar{F}_{\rm pdB} &= \bar{F}_{\rm B} = -\frac{3}{4}\bar{D},\\
 \label{bulk3} \bar{F}_{\rm P} &= -\frac{1}{2(1-\zeta_{12})}\bar{D},\\
 \label{bulk4} \bar{F}_{\rm Pol} &= -\frac{1}{4(1-2\zeta_{12})}\bar{D}.
\end{align}
Restricting to the stability region with $\zeta_{12}<\frac{1}{2}$, we find that the B-phase wins over the Pol-phase for $\zeta_{12}<\frac{1}{3}$, the A-phase wins over the B-phase for $\zeta_{245}<\frac{1}{3}$ (equivalently $1-2\zeta_{245}>\frac{1}{3}$), and the A-phase wins over the Pol-phase for $\zeta_{245}<1-2\zeta_{12}$ (equivalently $1-2\zeta_{245}>4\zeta_{12}-1$). The P-phase is never energetically favorable in the bulk. The 3D bulk phase diagram is shown in Fig. \ref{FigBulkPD}. The weak-coupling point $\zeta_{12}=1-2\zeta_{245}=\frac{1}{5}$ is located in the B-phase region. This is related to the well-known fact that weak-coupling theory predicts the normal-to-superfluid transition to always be into the B-phase, not reproducing the experimentally observed A-phase at high pressure.

We recover the bulk phase diagram in the slab geometry for $\bar{D}\to \infty$ for the following reason. Although sending $\bar{D}\to \infty$ produces $f(\bar{z}) = \tanh(\bar{z}/\sqrt{2})$ (see Eq. (\ref{real14}), which is not constant, the difference from $f=1$ is limited to a small region of size $\bar{z}\sim 1$ close to the origin, which is negligible compared to $\bar{D}\to \infty$. Consequently, the free energies for finite $\bar{D}<\infty$ approach their bulk limits as $\bar{D}\to \infty$.

More concretely, consider the case of maximally pair-breaking boundaries (the specular one is analogous). The topology of the  phase diagram is then as depicted in Fig. \ref{FigAPD}. The bulk phase can be read off by extrapolating the phase under confinement to $\bar{D}\to \infty$, i.e. the top edge of the plot. Assume first that $\zeta_{245}>\frac{1}{3}$, so that  the A-phase is only realized for finite $\bar{D}<\infty$ and does not survive in the limit $\bar{D}\to \infty$. For $\zeta_{12}<\frac{1}{3}$, we obtain the pdB-phase with $f_1=f_2\simeq f_3$, i.e. the B-phase, while for $\zeta_{12}>\frac{1}{3}$ we end up in the Pol-phase. The P-phase is reduced to a single line on the B-Pol-transition, hence is effectively absent from the bulk phase diagram. On the other hand, if $\zeta_{245}>\frac{1}{3}$, the confined phase diagram in the lower panel of Fig. \ref{FigAPD} only consists of the A- and the Pol-phase, separated by the vertical line $\zeta_{12}=\frac{1}{2}(1-\zeta_{245})$. The $\bar{D}\to \infty$ limit is then trivial, yielding the A- or Pol-phase in their associated region. Hence we exactly reproduce the bulk phase diagram in Fig. \ref{FigBulkPD}. 

The crucial point for the A-B-transition in the bulk to be reproduced in the $\bar{D}\to \infty$ limit is that  the A-pdB-transition under confinement touches the P-pdB-line $\bar{D}_{\rm P}$. For $\zeta_{245}>\frac{1}{3}$, $\bar{D}_{\rm P}<\infty$, and thus the A-phase is not seen in the limit $\bar{D}\to \infty$. However, as $\zeta_{245}\to \frac{1}{3}$ is decreased from above, the touching point $\bar{D}_{\rm P}\to \infty$, and the A-phase becomes relevant in the bulk. Since the divergence of $\bar{D}_{\rm P}$ at $\zeta_{12}=\frac{1}{3}$ has been established for both maximally pair-breaking and specular boundaries (Eqs. (\ref{real41}) and (\ref{spec8})) independently of $\gamma$, this shows that the correct limit is obtained independent of the boundary conditions or the value of $\gamma$.

For specular boundaries and $\gamma=3$ we numerically established that the S-phase transition lines satisfy $\bar{D}_{\rm S1}\simeq 0.94 \bar{D}_{\rm P}$ and $\bar{D}_{\rm S2}\simeq 1.3 \bar{D}_{\rm P}$ for values $\bar{D}<12$, see App. \ref{AppStripe}. Assuming this trend to continue as $\bar{D}\to \infty$, the S-phase is reduced to a single line along the B-Pol-transition and hence effectively absent from the bulk phase diagram, just as the P-phase. In our simulations we also found no indication that the proportionality $\bar{D}_{\rm S1,2} \propto \bar{D}_{\rm P}$ depends on $\gamma$, except for the prefactor, which we have shown at least for $\bar{D}_{\rm S1}$ in Eq. (\ref{spec16}). We conclude that the S-phase disappears in the limit $\bar{D}\to \infty$.

The quasi-2D limit of tight confinement is obtained by sending $\bar{D}\to \bar{D}_{\rm c}$ from above, with $\bar{D}_{\rm c}=\pi$ for maximally pair-breaking and $\bar{D}_{\rm c}=0$ for specular boundary conditions. The resulting quasi-2D phase diagram is not affected by the boundary conditions. For $\zeta_{245}>\frac{1}{3}$, the A- and P-phase compete; the A-phase wins over the P-phase for $\zeta_{12}<1-2\zeta_{245}$. For $\zeta_{245}<\frac{1}{3}$, the A- and Pol-phase compete; the A-phase wins over the Pol-phase for $\zeta_{245}<1-2\zeta_{12}$ (or $1-2\zeta_{245}>-1+4\zeta_{12}$), which is the bulk 3D line. Finally, the transition between the P- and Pol-phase is at $\zeta_{12}=\frac{1}{3}$. Note that the quasi-2D phase boundaries are not affected by the value of $\gamma>1$.

\subsection{$P$-$T$-$D$ phase diagram}\label{SecPTD}

To compute the phase diagram in terms of the thermodynamic variables pressure and temperature, we have to specify the pressure- and temperature-dependence of the free energy coefficients $\zeta_a(T,P)$, $\gamma(T,P)$, and $\xi(T,P)$. Since these functions are not known, a suitable parametrization or estimate has to be chosen and the resulting $P$-$T$ phase diagrams have a residual dependence on this choice. In principle, all functions $\zeta_a$, $\gamma$, $\xi$ would also depend on $D$ (especially if $D$ is small), which we ignore here.

\begin{figure}[t!]
\centering
\includegraphics[width=8.6cm]{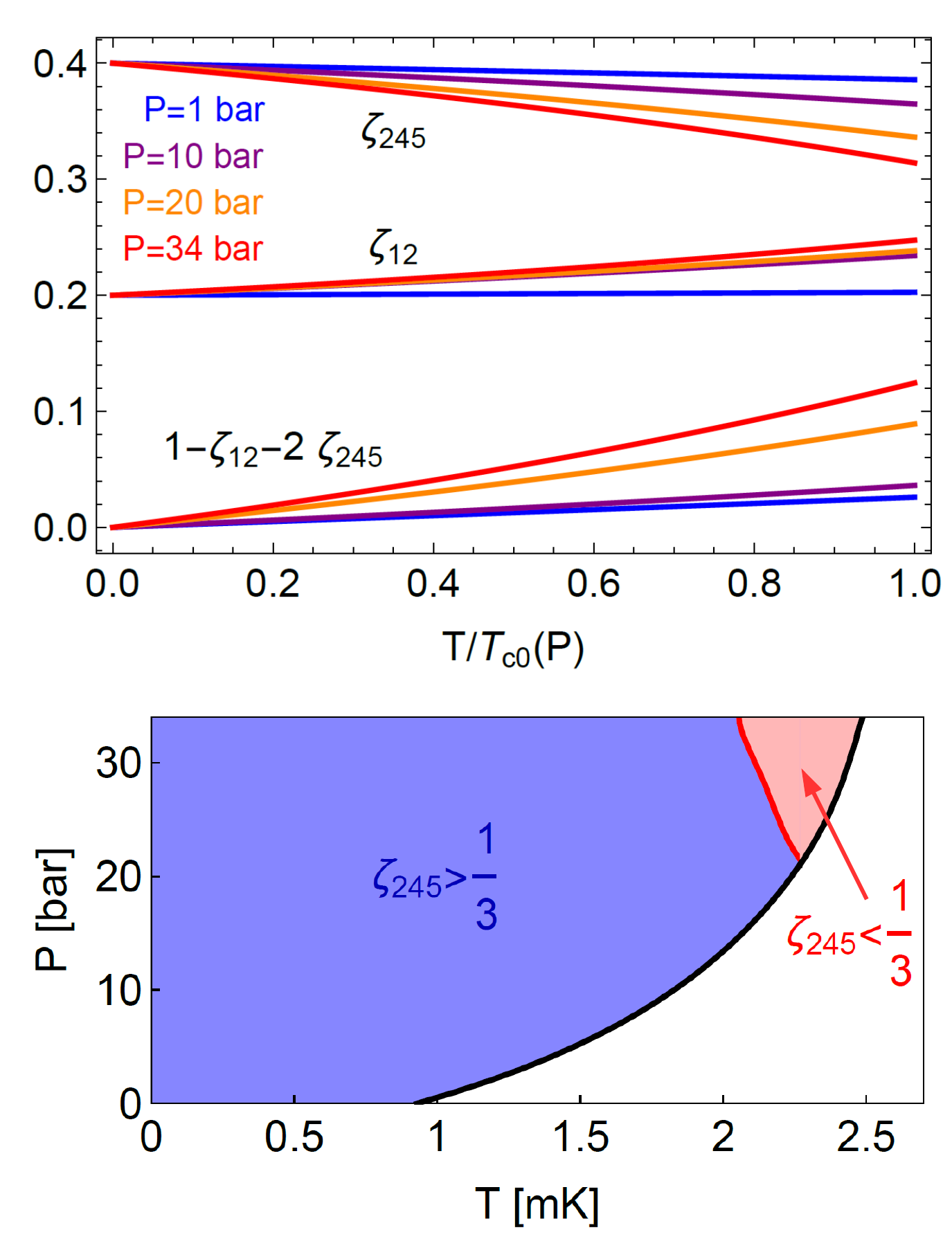}
\caption{\emph{Upper panel.} We show the strong-coupling values of the quartic couplings $\zeta_{12}$ and $\zeta_{245}$ defined by Eq. (\ref{pt9}). In the plot, $T_{\rm c0}$ is the critical temperature of the 3D system without magnetic field. In contrast to weak-coupling theory, where $\zeta_{12,\rm wc}=\frac{1}{5}$ and $\zeta_{245}=\frac{2}{5}$, there is a variation in $P$ and $T$. The fact that $\zeta_{245}$ is below $\frac{1}{3}$ at high pressures and temperatures allows for the A-phase to be stable in the 3D bulk system. Furthermore, we observe that $1-\zeta_{12}-2\zeta_{245}>0$ for all $P$ and $T$ so that the A-phase is always energetically favored over the P-phase in a slab geometry. \emph{Lower panel.} Bulk phase diagram computed from the strong-coupling values. The transition between the A (red) and B phase (blue) is determined by $\zeta_{245}=\frac{1}{3}$, while $\zeta_{12}<\frac{1}{3}$ always disfavors the polar phase.}
\label{FigStrong}
\end{figure}

Some qualitative statements can be inferred from the well-established phase structure of the bulk phase diagram together with Figs. (\ref{FigAPD}) and (\ref{FigSpecPD}). In the 3D bulk system, the superfluid transition appears in the pressure range $0\leq P \lesssim 34\ \text{bar}$ at a temperature $T_{\rm c0}(P)$ in the mK-regime, see Fig. \ref{FigAppThermo}a. Most of the $P$-$T$ phase diagram is occupied by the B-phase, whereas the A-phase is confined to a region at high pressures and close to $T_{\rm c0}$ defined by $\zeta_{245}(P,T)<\frac{1}{3}$, see Fig. \ref{FigStrong}. The Pol-phase is absent in 3D, meaning the interplay of $\zeta_{12}(P,T)$ and $\zeta_{245}(P,T)$ disfavors the Pol-phase for all $P$ and $T$. Since the $\zeta_{12}$- and $\zeta_{245}$-dependence of the phase boundary between the Pol-phase and the A- or P-phase is identical for all $\bar{D}\leq\infty$, this implies that the Pol-phase cannot be stabilized in a slab geometry.

The bulk normal-to-superfluid phase transition is of second order and thus the correlation length for fixed $P$ behaves as
\begin{align}
 \label{pt1} \xi(T,P) &\to \xi_0(P)\ \text{for}\ T\to 0,\\
 \label{pt2} \xi(T,P) &\to \infty\ \text{for}\ T\to  T_{\rm c0}(P).
\end{align}
Here $\xi_0(P)$ is the correlation length for the low-temperature system, which is around 70 nm at low pressures and 20 nm at high pressures, see Fig. \ref{FigAppThermo}c. This implies that, for fixed $P$ and $D$, the ratio $\bar{D}=D/\xi$ behaves as
\begin{align}
 \label{pt3} \bar{D} &\to D/\xi_0\ \text{for}\ T\to 0,\\
 \label{pt4} \bar{D} &\to 0\ \text{for}\ T\to  T_{\rm c0}(P).
\end{align}
Consequently, for a given choice of $P$ and $D$, the system at low temperatures is placed at a value $\bar{D}\simeq D/\xi_0$ in the $\bar{D}$ vs. $\zeta_{12}$ phase diagram. If $D$ is large enough, this value will be in the pdB-phase. Then, as $T$ is increased towards $T_{\rm c0}$, the value $\bar{D}$ is lowered and the system enters the A- or P-phase for maximally pair-breaking boundaries, or might pass through the S-phase before entering A or P for specular and diffusive boundaries. Whether the A- or P-phase is reached for large $T$ depends on the sign of $1-\zeta_{12}-2\zeta_{245}$, see below. Furthermore, if $D$ is small, the ratio $D/\xi_0$ might not be large enough for the system to stabilize the pdB-phase even at low temperatures, in which case the A-, P-, or S-phase is realized at low temperatures. Eventually, as $T$ is further increased and thus $\bar{D}$ lowered, the system enters the normal phase when $\bar{D} = \bar{D}_{\rm c}$ at a temperature $T_{\rm c}<T_{\rm c0}$ for maximally pair-breaking and diffusive boundaries, whereas $\bar{D}_{\rm c}=0$ for specular boundaries, and thus $T_{\rm c}=T_{\rm c0}$ in the specular case.

The critical temperature under confinement, $T_{\rm c}=T_{\rm c}(P,D,b_{\rm T})$, depends both on the height of the slab and the boundary conditions. It is defined from the condition $\bar{D}(T=T_{\rm c})=\bar{D}_{\rm c}$, where $\bar{D}_{\rm c}$ is the lowest value of $\bar{D}$ for Eq. (\ref{real13}) to have a nonzero solution under the according boundary conditions. For maximally pair-breaking boundaries ($b_{\rm T}=0$) we have $\bar{D}_{\rm c}=\pi$, see Fig. \ref{FigBplots}, whereas for specular boundaries ($b_{\rm T}=\infty$) we have $\bar{D}_{\rm c}=0$. The latter is due to the fact that $f=1$ is always a solution for specular boundaries. If $\bar{D}_{\rm c}>0$, then the system enters the normal phase at a temperature $T_{\rm c}<T_{\rm c0}$ where $\xi$ is still finite. The value of $\bar{D}_{\rm c}$ can be found numerically for any fixed value of $b_{\rm T}$. For $b_{\rm T}=0.5$ we have $\bar{D}_{\rm c}=2.2143$, while the whole function is well approximated by the empirical expression
\begin{align}
 \label{pt5} \bar{D}_{\rm c}(b_{\rm T}) \simeq \frac{\pi}{(1+1.75 b_{\rm T}^{1/0.7})^{0.7}},
\end{align}
see Fig. \ref{FigAppbT}. The associated shift in critical temperature is found from solving $\bar{D}_{\rm c}(b_{\rm T})=\frac{D}{\xi_0}\Phi(T/T_{\rm c})$ with $\Phi(t)$ from Eq. (\ref{pt8}), which yields
\begin{align}
 \label{pt6} \frac{T_{\rm c}(P,D,b_{\rm T})}{T_{\rm c0}(P)} \simeq 1- \Bigl(\frac{\bar{D}_{\rm c}(b_{\rm T})}{1.7}\frac{\xi_0(P)}{D}\Bigr)^2.
\end{align}

We parametrize the correlation length by 
\begin{align}
 \label{pt7} \xi(T,P) = \frac{\xi_0(P)}{\Phi(\frac{T}{T_{\rm c0}(P)})},
\end{align}
with  the phenomenological function 
\begin{align}
  \label{pt8} \Phi(t) = \begin{cases} 1 & t<0.3\\ \sqrt{1-t}(0.9663+0.7733t) & 0.3\leq t\leq 1\end{cases},
\end{align}
which satisfies the constraints in Eqs. (\ref{pt1}) and (\ref{pt2}). The choice of $\Phi(t)$ is motivated by the phenomenological temperature dependence of the gap amplitude $\Delta(T)$ and $\xi \sim \Delta(T)^{-1}$ \cite{Carless,PhysRevB.104.094520}. A plot of $\Phi(t)$ is given in Fig. \ref{FigAppThermo}d. For alternative choices of defining $\xi$ that are used in the literature, see App. \ref{AppCoherence}. The freedom in parameterizing $\xi$ adds to the many uncertainties in the coefficients of the free energy functional, but does not change the overall picture of the phase diagram in the dimensional crossover.

The quartic coefficients $\beta_a(T,P)$ are estimated using the strong-coupling corrections from Choi et al \cite{PhysRevB.75.174503}. The associated $\beta_a$ interpolate linearly between the weak-coupling values, assumed to hold at $T=0$, and measured 3D critical values at $T_{\rm c0}$ according to
\begin{align}
 \label{pt9} \beta_a(P,T) = \beta_{a,\rm wc} + \frac{T}{T_{\rm c0}(P)}\Delta \beta_a(P).
\end{align}
Here $\beta_{a,\rm wc}$ are the weak-coupling values from Eq. (\ref{weak3}) and the functions $\Delta \beta_a(P)$ are described in App. \ref{AppThermo} and Fig. \ref{FigAppThermo}b. We refer to the coefficients from Eq. (\ref{pt9}) as the strong-coupling values. They reproduce the experimental 3D bulk phase diagram and should give a plausible estimate for the phase structure under confinement. Furthermore, since they are close to the measured values at criticality, they should give a rather reliable estimate of the region around the phase transition (ignoring critical phenomena relevant very close to $T_{\rm c}$). Of course, the assumption that weak-coupling theory describes the quantum many-body system at $T=0$ is likely insufficient, which should be kept in mind when interpreting the phase structure at low temperatures. 
 We further choose $\gamma(T,P)=3$ at its weak-coupling value, mostly because there is no better guess from experimental constraints.

The strong-coupling values of the functions $\zeta_a(T,P)$ are not too far off from the weak-coupling values in Eq. (\ref{unit2}). Importantly, the sign of the strong-coupling coefficients always agrees with the weak-coupling ones. However, two important aspects of the $^3$He phase structure are supported by Eq. (\ref{pt9}), which are not visible in weak-coupling theory. First, since weak-coupling theory predicts $\zeta_{245,\rm wc} = \frac{2}{5} = 0.4$, the 3D bulk system would always be in the B-phase since the A-phase criterion $\zeta_{245}<\frac{1}{3}$ is never satisfied, in contradiction to the experimentally observed phase diagram. Second, the A- and P-phase are energetically degenerate in weak-coupling theory, both in 3D and under confinement. Indeed, the A-phase is favored (disfavored) over the P-phase if $1-\zeta_{12}-2\zeta_{245}>0$ ($1-\zeta_{12}-2\zeta_{245}<0$), because $\bar{F}_{\rm A} = \frac{1}{\zeta_{245}}\bar{F}_0(\bar{D})$ and $\bar{F}_{\rm P} = \frac{2}{1-\zeta_{12}}\bar{F}_0(\bar{D})$ independent of the boundary conditions. The weak-coupling value  $\zeta_{12,\rm wc} = \frac{1}{5} =0.2$ then implies $\bar{F}_{\rm A}=\bar{F}_{\rm P}$. In contrast, using the strong-coupling corrections, we find
\begin{align}
 \label{pt10} 1-\zeta_{12}(T,P)-2\zeta_{245}(T,P)>0
\end{align}
for all $T$ and $P$. Consequently, the A-phase is always energetically superior to the P-phase. We show the behavior of the functions $\zeta_{245}$ and $\zeta_{12}$ as pressure and temperature are varied in Fig. \ref{FigStrong}.

\begin{figure*}[t!]
\centering
\includegraphics[width=15.6cm]{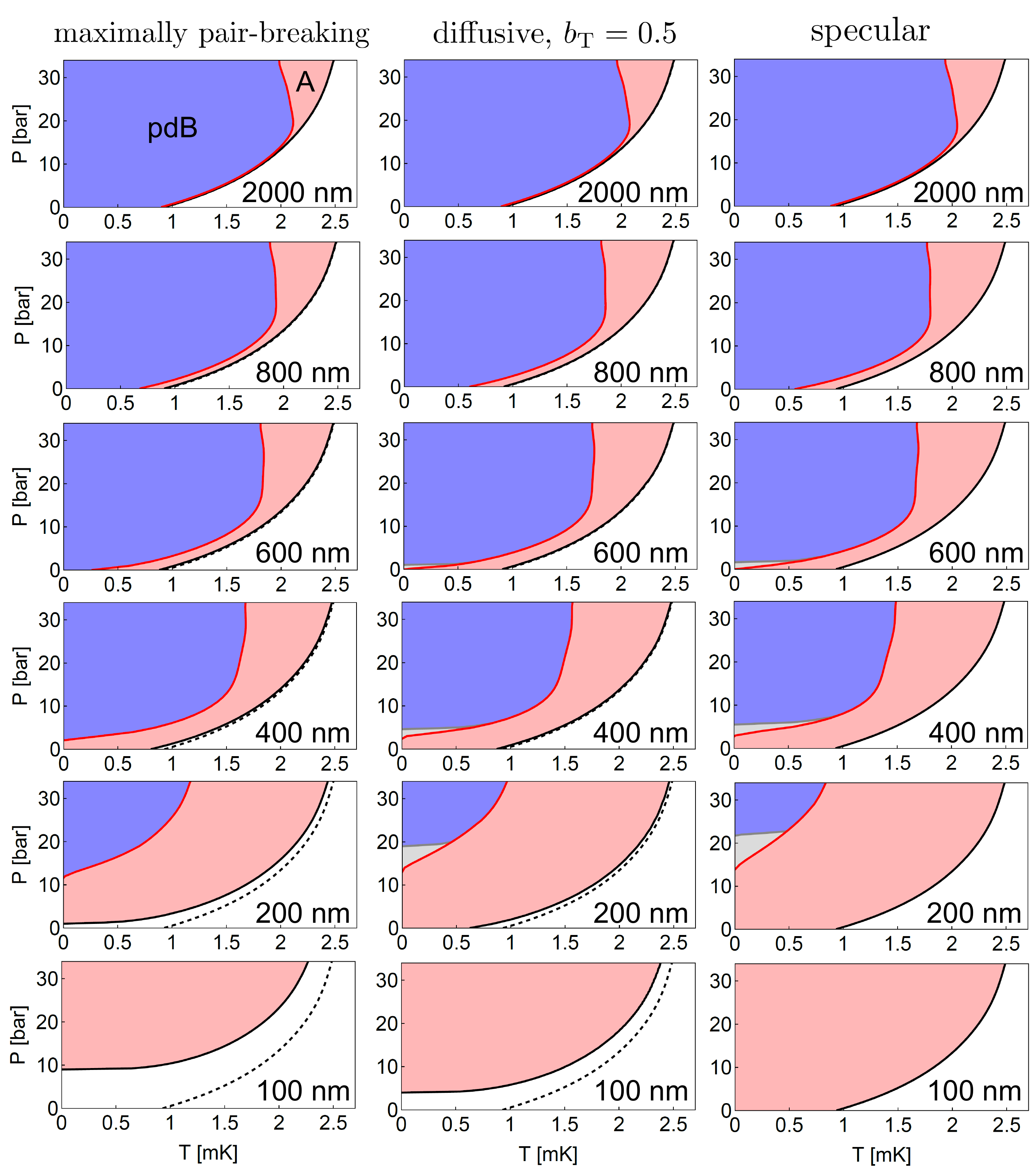}
\caption{$P$-$T$ phase diagrams as a function of slab height $D$ ranging from 2000 nm to 100 nm. The pdB- and A-phases are shown in blue and red, the S-phase in gray, and the normal phase in white. The superfluid phase transition under confinement, indicated by the continuous black line, is of second order. All other transitions are of first order. The dashed black line shows the 3D bulk transition for reference. The dimensional crossover from 3D to quasi-2D happens when $D$ is a few hundred nanometres. We show the crossover for three choices of boundary conditions: maximally pair-breaking, where the critical temperature is reduced due to confinement and the S-phase is absent; specular, where the critical temperature agrees with the 3D bulk value and the S-phase appears at low temperatures for a range of $D$-values; and diffusive with the experimentally relevant value $b_{\rm T}=0.5$, which interpolates between the other two cases and shows features of both.}
\label{FigPDCrossover}
\end{figure*}

With the parameters of the free energy described as functions of pressure and temperature, the $P$-$T$ phase diagram can be computed in the crossover from large to small values of $D$. We show the result for $D$ ranging from 2000 nm to 100 nm for all three boundary conditions in Fig. \ref{FigPDCrossover}. For the diffusive case, we choose $b_{\rm T}=0.5$. To determine the normal-to-A phase boundary, we use the criterion $\bar{D}=\bar{D}_{\rm c}$ to find $T_{\rm c}$. The pdB-to-S boundary is determined by $\bar{D}=\bar{D}_{\rm S2}$, where we approximate $\bar{D}_{\rm S2}\simeq 1.3\bar{D}_{\rm P}$ and $\bar{D}_{\rm S2}\simeq 1.2\bar{D}_{\rm P}$ for specular and diffusive boundaries, respectively, see Fig. \ref{FigAppRealSpecular}. There is no S-phase for maximally pair-breaking boundaries. The pdB-to-A boundary, since it is a first-order transition,  has to be obtained from comparing the free energies $\bar{F}_{\rm A}$ to $\bar{F}_{\rm pdB}$. Similarly, to find the S-to-A boundary, we compare $\bar{F}_{\rm A}$ and $\bar{F}_{\rm S}$. For computational ease, we approximate $\bar{F}_{\rm S}\simeq \bar{F}_{\rm pdB}$ even for $\bar{D}<\bar{D}_{\rm S2}$, which is accurate to better than a percent.

Some common trends are discernible from the crossover phase diagrams in Fig. \ref{FigPDCrossover}. First, the phase structure is independent of the boundary conditions and closely resembles the 3D bulk phase diagram for $D\gtrsim2000\ \text{nm}$. However, the transition from the normal to the superfluid phase is towards the A-phase, never towards the pdB-phase. This is in contrast to the 3D phase diagram, where the transition is towards the B-phase at low pressures. These two findings are not conflicting as, for large $D$, the A-phase is restricted to a small sliver around $T_{\rm c}$ for low pressures so that the 3D bulk phase diagram is recovered for $D\to \infty$, where the sliver becomes infinitely thin. Second, the transition from the normal into the superfluid phase is always towards the A-phase, not the P-phase, because of the constraint in Eq. (\ref{pt10}). This is not obvious from the phase diagrams in Figs. \ref{FigAPD} and \ref{FigSpecPD} alone, which would allow for a normal-to-P transition. Third, the critical temperature is reduced for both maximally pair-breaking and diffusive boundaries, whereas $T_{\rm c}=T_{\rm c0}$ for specular boundaries, in agreement with the discussion above and experiment \cite{science.1233621,levitin2019evidence,PhysRevLett.124.015301,PhysRevLett.134.136001}. Fourth, the S-phase appears in a small region of the phase diagram for diffusive and specular boundary conditions for $D\lesssim 600\ \text{nm}$, whereas there is no S-phase for maximally pair-breaking boundaries. At last, for small enough $D\lesssim 600\ \text{nm}$, other phases than the pdB-phase are stabilized at $T\simeq 0$ and the pdB-phase is eventually expelled from the phase diagram for thin slabs with $D\lesssim 100\ \text{nm}$.

\begin{figure}[t!]
\centering
\includegraphics[width=8.6cm]{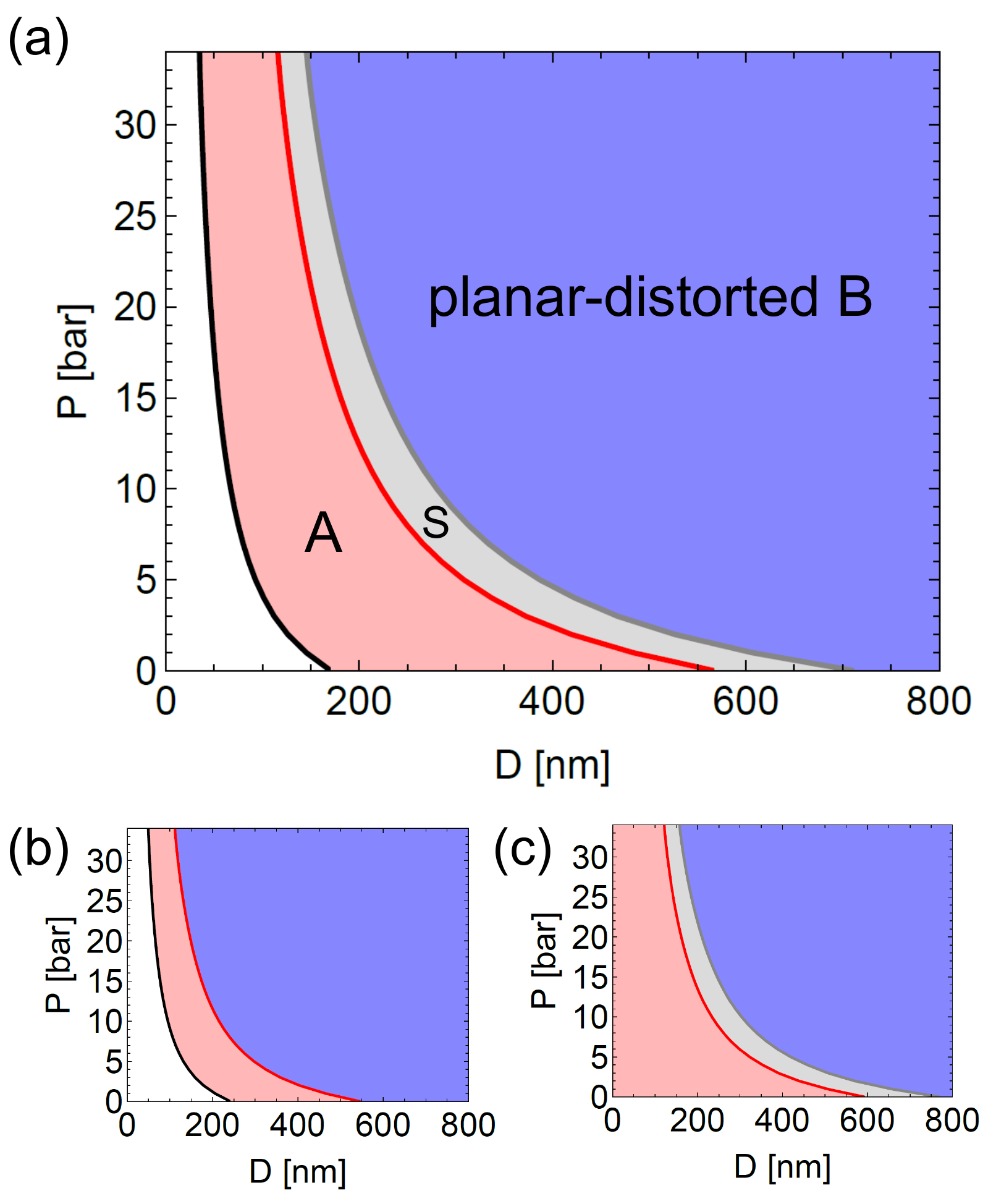}
\caption{Weak-coupling quantum phase diagram obtained from the phase structure shown in Fig. \ref{FigPDCrossover} extrapolated to $T=0$. Note that the strong-coupling coefficients in Eq. (\ref{pt9}) are parametrized such that they yield the weak-coupling values at $T=0$, hence the phase diagram shown here is the weak-coupling prediction. We show the case of diffusive boundaries with $b_{\rm T}=0.5$ in (a), whereas maximally pair-breaking and specular boundaries are shown in (b) and (c), respectively. The A-phase dominates the quantum phase structure for small values of $D$. For (a) and (b), tight confinement can destroy the superfluid coherence at zero temperature, resulting in a normal fluid (white). For (a) and (c), the S-phase occurs in a small window of pressure and confinement values.}
\label{FigQPD}
\end{figure}

In Fig. \ref{FigQPD}, we present a quantum (or low-temperature) phase diagram of the phases that are realized at $T\simeq 0$ in dependence of $P$ and $D$. One should keep in mind though that the strong-coupling ansatz in Eq. (\ref{pt9}) reproduces weak-coupling theory at $T=0$, and so Fig. \ref{FigQPD} is really a weak-coupling prediction. Nonetheless, some features of the $P$-$D$ phase diagrams are conveniently summarized in this manner. We observe that the A- and S-phases can be stabilized at low temperatures, in accordance with experiments \cite{PhysRevLett.124.015301}. Furthermore, for non-specular boundaries, the system is in the normal state for sufficiently small $D\lesssim 200\ \text{nm}$. The confinement is felt stronger at low pressures, because the correlation lengths are longer, see Fig. \ref{FigAppThermo}c.

\section{Phase diagram in a magnetic field}\label{SecNonzero}

\subsection{Bulk phase diagram}

In the presence of any nonzero magnetic field $H>0$, the phases A, P, and B/pdB get replaced by A$_2$, P$_2$, and B$_2$, respectively, as these have a lower free energy due to the linear Zeeman term. Additionally, the A$_1$-phase is stabilized close to the superfluid transition. Within the strong-coupling approximation for the quartic couplings $\beta_a$, no other phases are relevant. However, even within this limited space of relevant orders, all four independent coefficients $\zeta_a$ are needed to compute the phase structure. To discuss the phase diagram in general terms, we thus make the following assumptions on the quartic coefficients, which are satisfied by both the weak-coupling and all strong-coupling values: We assume $\beta_1,\beta_5<0$ and $\beta_2,\beta_3,\beta_4>0$ (which implies $\beta_{24}>0$), and $ \beta_{245},\ \beta_{345}>0$. Furthermore, we assume
\begin{equation}\label{mbulk0}
\begin{aligned}
 \mathcal{C}_1 &= 2\beta_{12} + \beta_{345} -2\beta_{245} \\
 &=(3\beta_{12}+\beta_{345})(1-\zeta_{12}-2\zeta_{245}) >0,
\end{aligned}
\end{equation}
while
\begin{equation}\label{mbulk0b}
\begin{aligned}
 \mathcal{C}_2 &= 2\beta_{12} + \beta_{345} -2\beta_{24}\\
 &=(3\beta_{12}+\beta_{345})(1-\zeta_{12}-2\zeta_{24}) <0.
\end{aligned}
\end{equation}
(Note the slight difference in the last term.) The relation $\mathcal{C}_1>0$ is true for the strong-coupling values, and leads to the A$_2$-phase being preferred over P$_2$, whereas weak-coupling theory would predict $\mathcal{C}_1=0$. The relation $\mathcal{C}_2<0$ is valid for both weak- and strong-coupling coefficients.

We determine the 3D bulk phase diagram in a magnetic field in terms of the bare parameters of the free energy, $\alpha_0$, $g_1H$, $g_2H^2$, and $\beta_a$. This may be more transparent than using the dimensionless units as done in the subsequent sections. Of course, both formulations are equivalent. The free energy densities of the A$_1$-, A$_2$-, and P$_2$-phases are given by
\begin{align}
 \label{mbulk3} \text{f}_{\rm A_1} ={}& -\frac{(\alpha_0+g_1 H)^2}{4\beta_{24}}\ \text{for }\alpha_0>-g_1 H,\\
 \label{mbulk4} \text{f}_{\rm A_2} ={}& -\frac{\alpha_0^2}{4\beta_{245}} +\frac{(g_1 H)^2}{4\beta_5}\ \text{for }\alpha_0>\alpha_{\rm cA},\\
 \nonumber \text{f}_{\rm P_2} ={}& -\frac{\alpha_0^2}{2(2\beta_{12}+\beta_{345})}\\
 \label{mbulk5} &+\frac{(g_1 H)^2}{2(2\beta_{12}+\beta_{345}-2\beta_{24})}\ \text{for } \alpha_0>\alpha_{\rm cP},
\end{align}
see App. \ref{AppMag}. Importantly, the A$_1$-phase can exists for slightly negative $\alpha_0$ as long as $\alpha_H=\alpha_0+g_1H>0$, whereas the A$_2$- and P$_2$-phases require a sufficiently large, positive $\alpha_0$ with critical values
\begin{align}
 \label{mbulk6b} \alpha_{\rm cA} &= \frac{\beta_{245}}{|\beta_5|} g_1H >0,\\
 \label{mbulk6c} \alpha_{\rm cP} &= \frac{2\beta_{12}+\beta_{345}}{|2\beta_{12}+\beta_{345}-2\beta_{24}|}g_1H>0.
\end{align}
In the weak-coupling limit, we have 
\begin{align}
 \label{mbulk6d} \alpha_{\rm cA}\approx \alpha_{\rm cP} \approx g_1H.
\end{align}
However, using the strong-coupling coefficients, we find
\begin{align}
 \label{mbulk6e}  \alpha_{\rm cA} < \alpha_{\rm cP}
\end{align}
for all $T$ and $P$, see Eq. (\ref{fs17c}). The free energy for the B$_2$-phase needs to be determined numerically by minimizing $\text{f}_{\rm B_2}$ from Eq. (\ref{fs24}) with respect to $B,M,C$, see App. \ref{AppMag}. Importantly, the B$_2$-phase with nonzero $C$ can only be stabilized by having a sufficiently large, positive $\alpha_0>\alpha_{\rm c B} \simeq g_2 H^2$.

In the following, we consider sufficiently large magnetic fields with $H\gg 100\ \text{G}$. In the region of negative $\alpha_0\in(-g_1 H,\alpha_{\rm cA})$, only the A$_1$-phase is possible, which means that in a region close to the transition temperature in magnetic field, $T_{\rm cH}$, the A$_1$-phase is realized. As soon as $\alpha_0>\alpha_{\rm cA}$ for lower temperatures, however, the A$_2$-state is superior to the A$_1$-state, because
\begin{align}
 \label{mbulk7} \text{f}_{\rm A_2} -\text{f}_{\rm A_1} = \frac{(\beta_5\alpha_0+\beta_{245}g_1 H)^2}{4\beta_{24}\beta_5\beta_{245}} <0.
\end{align}
Note that for $\alpha_0=\alpha_{\rm cA}$ we have $\text{f}_{A_1}=\text{f}_{A_2}$, and the A$_1$- and A$_2$-phases are separated by a second-order transition. Regarding the P$_2$-state, we  find for $\alpha_0>\alpha_{\rm cP}$ that
\begin{align}
 \nonumber \text{f}_{\rm P_2}-\text{f}_{\rm A_1} &= \frac{[(2\beta_1+\beta_{35}-\beta_4)\alpha_0+(2\beta_{12}+\beta_{345})g_1H]^2}{4\beta_{24}(2\beta_{12}+\beta_{345})(2\beta_1+\beta_{35}-\beta_{24})}\\
 \label{mbulk8} &<0.
\end{align}
Similar to the case of A$_2$ versus A$_1$, this means that as soon as $\alpha_0$ is large enough to support the P$_2$-state, it is energetically favored over the A$_1$-state.

To understand the competition between the A$_2$- and P$_2$-phases, we note that for $\alpha_0>\alpha_{\rm cP}$ we have
\begin{align}
 \nonumber \text{f}_{\rm A_2}-\text{f}_{\rm P_2} ={}& \frac{\mathcal{C}_1}{4} \Bigl(\frac{-\alpha_0^2}{\beta_{245}(2\beta_{12}+\beta_{345})}\\
 \label{mbulk10} &+\frac{(g_1 H)^2}{\beta_5(2\beta_{12}+\beta_{345}-2\beta_{24})}\Bigr).
\end{align} 
In the weak-coupling limit, $\mathcal{C}_1=0$ and A$_2$ and P$_2$ are once more energetically degenerate, just as A and P for zero field. Since the A-phase wins over the P-phase in the bulk, we have $\mathcal{C}_1>0$, which is supported by the strong-coupling values. For small $\alpha_0$, the term in round brackets is positive, which suggests that the P$_2$-state could be energetically favorable in a certain parameter range. However, the right-hand side is only negative for $\alpha_0<\alpha_{\rm AP}$ with a critical value given by
\begin{align}
 \alpha_{\rm AP} = \sqrt{\alpha_{\rm cA}\alpha_{\rm cP}}.
\end{align}
Since we already established that $\alpha_{\rm cA}<\alpha_{\rm cP}$, we conclude that $\alpha_{\rm AP}<\alpha_{\rm cP}$. Consequently, in the regime $\alpha_0>\alpha_{\rm cP}$ where the P$_2$-phase is possible, the right-hand side is positive, and hence the P$_2$ is always energetically inferior to the A$_2$. As a result, the P$_2$-phase does not appear in the bulk phase diagram in a magnetic field.

The qualitative 3D phase structure in a magnetic field $H\gg 100\ \text{G}$ is thus as follows. Close to the zero-field critical temperature $T_{\rm c0}$ we have 
\begin{align}
 \label{mbulk12} \alpha_0(T,P) \simeq \frac{N(0)}{3}\Bigl(1-\frac{T}{T_{\rm c 0}(P)}\Bigr).
\end{align}
In a small temperature interval around $T_{\rm c0}$, the A$_1$-phase is stable, namely for $\alpha_0\in(-g_1 H,\alpha_{\rm cA})$ or $T\in(T_{\rm A_1A_2},T_{\rm cH})$, where the normal-to-superfluid critical temperature $T_{\rm cH}$ is determined by $\alpha_0(T_{\rm cH},P)=-g_1H$. Thus we have
\begin{align}
 \label{mbulk13} T_{\rm cH} \sim T_{\rm c0}\Bigl(1+\frac{H}{10^6\text{G}}\Bigr).
\end{align}
The critical temperature of the A$_1$-A$_2$-transition, $T_{\rm A_1A_2}$, satisfies $\alpha_0(T_{\rm A_1A_2},P)=\alpha_{\rm cA}\approx g_1 H$, which yields the linear-in-$H$ relation
\begin{equation}
\begin{aligned}
 \Delta T &= T_{\rm cH} - T_{\rm A_2A_2} = \frac{3T_{\rm c0}}{N(0)} (\alpha_{\rm cA}+g_1H) \approx 2 \frac{3 g_1 H}{N(0)}T_{\rm c0}
\end{aligned}
\end{equation}
that was used in Eq. (\ref{mag5b}) to determine the coefficient $g_1H$. We have
\begin{align}
 \label{mbulk13b} T_{\rm A_1A_2} \sim T_{\rm c0}\Bigl(1-\frac{H}{10^6\text{G}}\Bigr).
\end{align}
The A$_2$-phase is stable in the region with $\alpha_0\in(\alpha_{\rm cA},\alpha_{\rm cB})$ or $T\in(T_{\rm A_2B_2},T_{\rm A_1A_2})$, where $\alpha_{\rm cB}\simeq g_2H^2$ is the critical value to allow for the B$_2$-phase. When the B$_2$-phase is available, it is typically energetically favored over A$_2$, because the free energy of A$_2$ is close to that of P$_2$. The B$_2$-phase is stable for $\alpha_0>\alpha_{\rm cB}$ and $T<T_{\rm A_2B_2}$. The A$_2$- and B$_2$-phases are separated by a first-order transition at
\begin{align}
 \label{mbulk15} T_{\rm A_2B_2} \sim T_{\rm c0}\Bigl[1-\Bigl(\frac{H}{10^4\text{G}}\Bigr)^2\Bigr].
\end{align}
For low temperatures away from $T_{\rm cH}$, the above statements are qualitatively still true, but the expression for $\alpha_0$ in Eq. (\ref{mbulk12}) has to be replaced with a more realistic function that is approximately constant for small $T$, see App. \ref{AppThermo}. Furthermore, for large $H \gtrsim 10^4\ \text{G}=1\ \text{T}$ the stability region of the B$_2$-phase at low temperatures shrinks to zero.

We briefly comment on the regime $H\sim 100\ \text{G}$ that we will not explore further below. In this regime, there is a competition between the linear and quadratic Zeeman term, hence $\alpha_{\rm cA} \sim \alpha_{\rm cB}$, and the stability regions of the A$_2$- and B$_2$-phases are not as simple as in the picture described above. Depending on the pressure, either $\alpha_0=\alpha_{\rm cA}$ or $\alpha_0=\alpha_{\rm cB}$ may be satisfied at higher temperatures. In the second case, a first-order transition from the A$_1$- into the B$_2$-phase is possible. In this field range, the phase diagram then has the following qualitative structure: a small sliver of A$_1$-phase appears at the normal-to-superfluid boundary. The tricritical point of the zero-field phase diagram (located at the normal-A-B intersection) is replaced by a tricritical point located at the A$_1$-A$_2$-B$_2$ intersection. As $H\gg 100\ \text{G}$ is increased, the tricritical point merges with the $P=0$ axis and the A$_2$-phase is eventually stabilized close to $T_{\rm c0}$. In this range, the effects of confinement and magnetic field towards stabilizing the A- and A$_2$-phases are comparable.

\subsection{Thin-slab geometries or large magnetic fields: A$_1$-, A$_2$-, and P$_2$-phase}\label{SecThinH}

In a slab geometry, the confining walls suppress the third column of the order parameter, and therefore $C=\Delta_{\up\down}$. For $H>0$, this suppression is enhanced by the quadratic Zeeman term. To stabilize the component $C>0$, a sufficiently large $\alpha_H$ is needed to increase the condensation energy. Consequently, for either thin slabs, near the critical temperature, or in high magnetic fields, the B$_2$-phase is unstable and the phase diagram is comprised only of A$_1$, A$_2$, and P$_2$. In the following, we discuss their competition and the associated phase structure, before including the B$_2$-phase in the next section.

We write the order parameter of the A$_1$-phases as
\begin{align}
 \label{thin1} O_{\rm A_1}(\bar{z}) = \frac{f(\bar{z})}{\sqrt{\zeta_{24}}}\begin{pmatrix} 1 & \rmi & \\ -\rmi & 1 & \\ & & 0 \end{pmatrix}
\end{align}
and find that $f(\bar{z})$ solves the usual equation of motion (\ref{real13}) with the appropriate boundary conditions. For specular boundaries, the solution exists for all $\bar{D}>0$, whereas for maximally pair-breaking boundaries, it only exists for $\bar{D}>\pi$. The dimensionless free energy is given by
\begin{align}
 \label{thin2} \bar{F}_{\rm A_1} = \frac{1}{\zeta_{24}}\bar{F}_0(\bar{D}),
\end{align}
with $\bar{F}_0(\bar{D})$ from App. \ref{AppClosedEnergy}. In particular, for specular boundaries, we have $\bar{F}_{\rm A_1} = -\frac{1}{4\zeta_{24}}\bar{D}$. The A$_1$-solution exists for any $\alpha_H>0$ or, equivalently,  any $h<\infty$.

The order parameters of the A$_2$- and P$_2$-phase are parametrized as
\begin{align}
 \label{thin3} O_{\rm A_2}(\bar{z}) &= \frac{1}{\sqrt{\zeta_{24}}}\begin{pmatrix} b(\bar{z}) & \rmi b(\bar{z}) & \\ - \rmi m(\bar{z}) & m(\bar{z}) & \\ & & 0 \end{pmatrix},\\
 \label{thin4} O_{\rm P_2}(\bar{z}) &= \frac{1}{\sqrt{\zeta_{24}}}\begin{pmatrix} b(\bar{z}) & \rmi m(\bar{z}) & \\ - \rmi m(\bar{z}) & b(\bar{z}) & \\ & & 0 \end{pmatrix},
\end{align}
with
\begin{align}
 \label{thin5} b(\bar{z}) &= \frac{f_\down(\bar{z}) +f_\up(\bar{z})}{2},\\
 \label{thin6} m(\bar{z}) &= \frac{f_\down(\bar{z}) -f_\up(\bar{z})}{2}.
\end{align}
The equations of motion for the functions $f_{\down,\up}(\bar{z})$ have an identical form for the A$_2$- and P$_2$-states. They read
\begin{equation}\label{thin7} 
\begin{aligned}
 0 &= -f_\down ''-f_\down + f_\down^3 +(1-h_{\rm c})f_\up^2f_\down,\\
  0 &= -f_\up'' -(1-h)f_\up +f_\up^3 +(1-h_{\rm c}) f_\down^2f_\up,
\end{aligned}
\end{equation}
where for A$_2$ and P$_2$ we have to replace
\begin{align}
  \label{thin9} h_{\rm c} \to h_{\rm cA} &= \frac{-2\zeta_5}{\zeta_{24}}>0,\\
  \label{thin10} h_{\rm c} \to h_{\rm cP} &= \frac{-(1-\zeta_{12}-2\zeta_{24})}{\zeta_{24}}>0,
\end{align}
respectively. In the weak-coupling limit, we have
\begin{align}
 \label{thin11}  h_{\rm cA} \approx h_{\rm cP} \approx 1,
\end{align}
but with strong-coupling corrections we find that 
\begin{align}
 \label{thin12} h_{\rm cA} > h_{\rm cP},
\end{align}
which is equivalent to Eq. (\ref{mbulk6e}).

Solutions to the equations of motion (\ref{thin7}) only exists for 
\begin{align}
 h<h_{\rm c},
\end{align}
which, for a fixed value of $H$, requires a sufficiently large value of $\alpha_H$. The free energies in the A$_2$- and P$_2$-phases for $h<h_{\rm c}$ are given by
\begin{align}
 \nonumber \bar{F}_{\rm A_2/P_2} = \frac{1}{\zeta_{24}}\int_0^{\bar{D}}\mbox{d}\bar{z}\ &\Bigl[\frac{1}{2}(f_\down'^2+f_\up'^2)-\frac{1}{2}f_\down^2-\frac{1-h}{2}f_\up^2 \\
 \label{thin13} &+ \frac{1}{4}(f_\down^4+f_\up^4)+\frac{1}{2}(1-h_{\rm c})f_\down^2f_\up^2\Bigr],
\end{align}
with the same replacement of $h_{\rm c}$ as before. For either maximally pair-breaking or specular boundaries, when evaluated for the solution to the equations of motion, we have
\begin{align}
 \label{thin14} \bar{F}_{\rm A_2/P_2} = -\frac{1}{\zeta_{24}}\int_0^{\bar{D}}\mbox{d}\bar{z}\ \Bigl[\frac{1}{4}(f_\down^4+f_\up^4)+\frac{1}{2}(1-h_{\rm c})f_\down^2f_\up^2\Bigr].
\end{align}
For specular boundaries, the order parameter profiles are constant functions given by
\begin{align}
  \label{thin15}  \begin{pmatrix} f_\down(\bar{z})^2 \\ f_\up(\bar{z})^2 \end{pmatrix} &= \begin{pmatrix} f_{\down 0}^2 \\ f_{\up0}^2 \end{pmatrix} = \frac{1}{h_{\rm c}(2-h_{\rm c})}\begin{pmatrix} h+h_{\rm c}-hh_{\rm c} \\ h_{\rm c}-h\end{pmatrix},
\end{align}
which yields
\begin{align}
 \label{thin16}  \bar{F}_{\rm A_2/P_2} &= -\frac{1}{\zeta_{24}} \frac{2h_{\rm c}-2hh_{\rm c}+h^2}{4h_{\rm c}(2-h_{\rm c})}\bar{D}
\end{align}
for $h<h_{\rm c}$.

We have seen that the A$_2$- and P$_2$-phases are only stable for $h<h_{\rm cA}$ and $h<h_{\rm cP}$, respectively, whereas the A$_1$-phases is stable for all $h<\infty$. This implies that the transition at $T_{\rm cH}$ is always from the normal into the A$_1$-phase. Due to $h_{\rm cA}>h_{\rm cP}$, once $h$ is lowered to the critical value $h_{\rm cA}$ (or, equivalently, $\alpha_0$ is increased to $\alpha_{\rm cA}$), there is a second-order transition from the A$_1$- into the A$_2$-phase. If $h$ is decreased below $h_{\rm cP}$, the P$_2$-phase is technically possible. However, because of $h_{\rm cA}>h_{\rm cP}$, it is never energetically favored over the A$_2$-phase, which can be inferred from comparing $\bar{F}_{\rm A_2}$ and $\bar{F}_{\rm P_2}$.  As a result, the phase diagram for thin slabs only features the A$_1$- and A$_2$-phases. If the critical values $h_{\rm cA}$ and $h_{\rm cP}$ were such that $h_{\rm cP}>h_{\rm cA}$, then the identical shape of the equations of motion would imply that the thin-slab phase structure only consisted of the A$_1$- and P$_2$-phase, connected by a second-order transition.

\begin{figure}[t!]
\centering
\includegraphics[width=8.6cm]{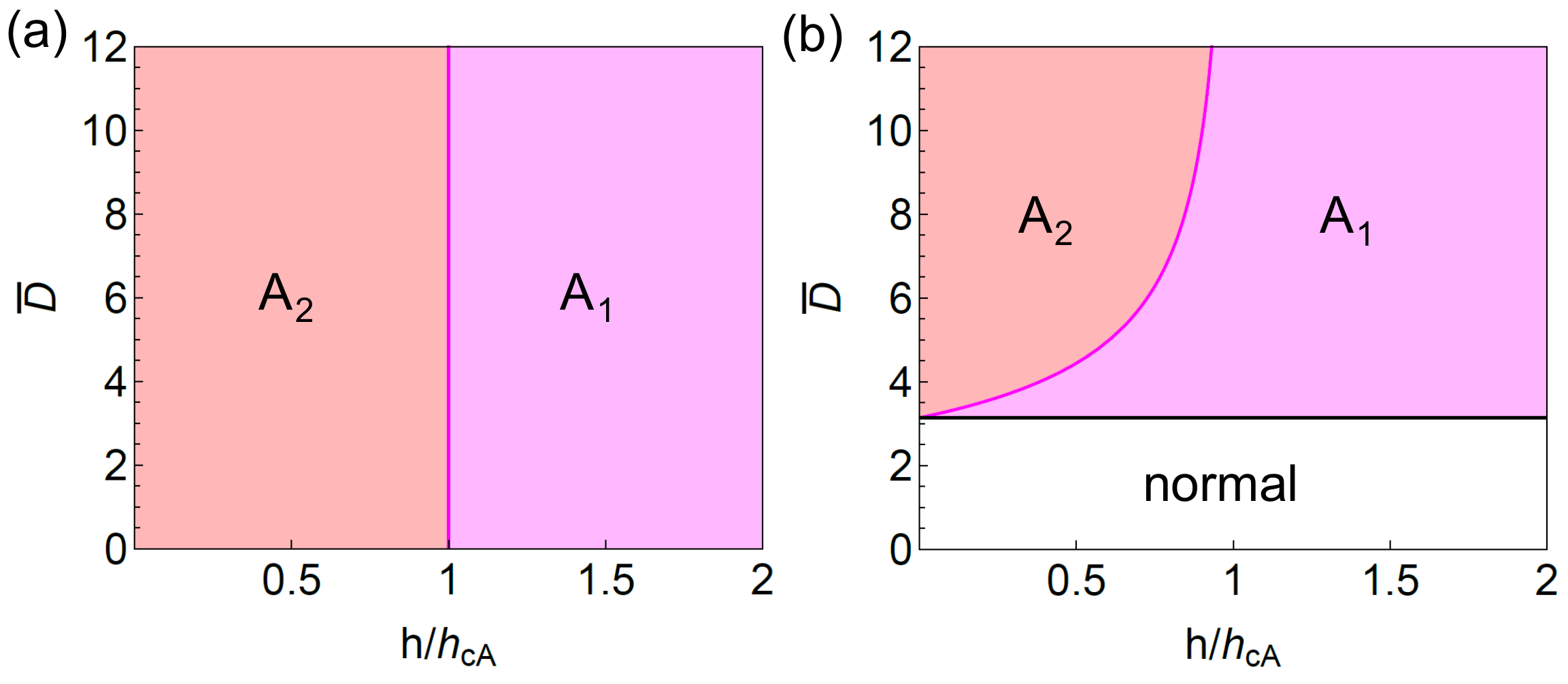}
\caption{Phase structure for thin-slab geometries or large magnetic fields. We show the phase diagram for specular boundary conditions in (a) and maximally pair-breaking boundary conditions in (b). Only the A$_1$- and A$_2$-phases are realized. Close to the critical temperature, $T_{\rm cH}$, we have $\alpha_H\simeq 0$ and thus $h\simeq \infty$. As a result, the superfluid transition is always towards the A$_1$-phase. For maximally pair-breaking boundaries, the system is in the normal phase for $\bar{D}<\bar{D}_{\rm c}=\pi$. At the left edge of the plot, for $h=0$, the A$_2$-phase becomes the A-phase and the phase diagram connects to the low-$\bar{D}$ region of Figs. \ref{FigAPD} and \ref{FigSpecPD} for zero field.}
\label{FigThinSlab}
\end{figure}

Let us now determine the transition line between the A$_1$- and A$_2$-phase in the phase diagram of $\bar{D}$ versus $h$. For specular boundaries, we use  Eqs. (\ref{thin2}) and (\ref{thin16}) to write
\begin{align}
 \label{thin17} \bar{F}_{\rm A_2}-\bar{F}_{\rm A_1} = -\frac{1}{\zeta_{24}} \frac{(h_{\rm cA}-h)^2}{4h_{\rm cA}(2-h_{\rm cA})}\bar{D},
\end{align}
which shows that the A$_2$-phase is energetically favorable as soon as it is available, $h\leq h_{\rm cA}$, and thus the transition is at $h=h_{\rm cA}$. Another way to understand this is as follows: Since both order parameters are continuously connected through $f_\up$ being zero or nonzero, the mere existence of a nontrivial solution to the equations of motion with $f_\up>0$ indicates that the free energy is lowered by a nonzero component $f_\up>0$, hence favoring the A$_2$- over the A$_1$-phase. This situation is fully analogous to the zero-field transition from the P- to the pdB-phase at $\bar{D}_{\rm P}$ as soon as $f_3>0$ in Eq. (\ref{real30}). In the case of maximally pair-breaking boundaries, a nonvanishing component $f_\up >0$ requires a sufficiently wide slab as in Fig. \ref{FigBplots}. Therefore, the A$_1$-phase is stable for small values of $\bar{D}$ even if $h<h_{\rm cA}$ (right panel in Fig. \ref{FigThinSlab}), because a solution with $f_\up>0$ does not exist in this parameter range: the $f_\up$-component is disfavored over the $f_\down$-component because of the additional linear-Zeeman factor $1-h$ in the quadratic part of Eq. (\ref{thin13}).

To determine the A$_1$-A$_2$-transition line for maximally pair-breaking boundaries, we thus apply the strategy that helped us to identify $\bar{D}_{\rm P}$ in the zero-field setting. Assuming $f_\up$ is small, the equations of motion (\ref{thin7}) can be approximated by
\begin{align}
 \label{thin17a} 0 &\simeq  -f_\down ''-f_\down + f_\down^3 ,\\
 \label{thin17b} 0 &\simeq  -f_\up'' -\Bigl[ (1-h) +(1-h_{\rm cA}) f_\down^2\Bigr] f_\up.
\end{align}
The first equation is solved by $f_\down(\bar{z})=f(\bar{z})$ from Eq. (\ref{real16}). This turns the second equation into the Lam\'{e} equation and an analysis analogous to the one outlined in App. \ref{AppReal} yields the critical value $\bar{D}_{\rm A_1A_2}(h)$ as
\begin{align}
 \label{thin8b} \bar{D}_{\rm A_1A_2}(h) = 2 \sqrt{k_{\rm c\up}^2+1}K(k_{\rm c\up}^2),
\end{align}
where $k_{\rm c\up}^2$ solves
\begin{align}
 \label{tinh8c} &(1-h)(k_{\rm c\up}^2+1) = a_{\nu_\up}^{(1)}(k_{\rm c\up}^2),\\
 \label{thin8d} &\nu_\up = \frac{1}{2}\Bigl(-1+\sqrt{9-8h_{\rm cA}}\Bigr).
\end{align}
Here $a_\nu^{(1)}(k^2)$ is the first Lam\'{e} eigenvalue as in Eq. (\ref{real39}). Note that $\bar{D}_{\rm A_1A_2}$ depends on $h$. For $h\to 0$, when the stability regime of the A$_1$-phase shrinks to zero and the A$_2$-phase turns into the A-phase, we recover the critical value $\bar{D}_{\rm c}$ for the transition from the normal to the A-phase with maximally pair-breaking boundaries, namely $k_{\rm c\up}^2\to0$ and 
\begin{align}
 \label{thin8e} \bar{D}_{\rm A_1A_2}(h\to 0) = \bar{D}_{\rm c}=\pi.
\end{align}
The phase structure for thin slabs and large magnetic fields is summarized in Fig. \ref{FigThinSlab}.

\subsection{Competition with the B$_2$-phase}\label{SecB2}

We parametrize the B$_2$-phase order parameter under confinement as
\begin{align}
 \label{b1} O_{\rm B_2}(\bar{z}) &= \frac{1}{\sqrt{\zeta_{24}}}\begin{pmatrix} b(\bar{z}) & \rmi m(\bar{z}) & \\ - \rmi m(\bar{z}) & b(\bar{z}) & \\ & & f_3(\bar{z}) \end{pmatrix},
\end{align}
with $b(\bar{z}),m(\bar{z})$ expressed in terms of $f_{\down}(\bar{z}), f_\up(\bar{z})$ as in Eqs. (\ref{thin5}) and (\ref{thin6}). The equations of motion read
\begin{align}
 \nonumber  0 ={}& -f_\down''-f_\down+f_\down^3\\
  \label{b2} &+\frac{1-\zeta_{12}-\zeta_{24}}{\zeta_{24}}f_\down f_\up^2+\frac{\zeta_2}{\zeta_{24}}  f_\down f_3^2 +\frac{\zeta_1}{\zeta_{24}} f_\up f_3^2,\\
 \nonumber 0 ={}& -f_\up''-(1-h)f_\up + f_\up^3 \\
 \label{b3} &+\frac{1-\zeta_{12}-\zeta_{24}}{\zeta_{24}}f_\down^2 f_\up+\frac{\zeta_2}{\zeta_{24}}  f_\up f_3^2+\frac{\zeta_1}{\zeta_{24}} f_\down f_3^2,\\
 \nonumber 0 ={}&-\gamma f_3''-\Bigl(1-\frac{1+r}{2}h\Bigr)f_3+\frac{1-2\zeta_{12}}{\zeta_{24}}f_3^3\\
 \label{b4} &+\frac{\zeta_2}{\zeta_{24}} (f_\down^2+f_\up^2)f_3+2\frac{\zeta_1}{\zeta_{24}} f_\down f_\up f_3.
\end{align}
Note that for $f_3=0$ we recover the equations (\ref{thin7}) for the P$_2$-state, while for $h=0$ we recover the pdB-equations (\ref{real30}) for $f_\down=f_\up\to\sqrt{\zeta_{24}}f_1$ and $f_3 \to \sqrt{\zeta_{24}}f_3$. 

The dimensionless free energy of the B$_2$-state reads
\begin{align}
 \nonumber \bar{F}_{\rm B_2} ={}& \frac{1}{\zeta_{24}}\int_0^{\bar{D}}\mbox{d}\bar{z}\ \Biggl[ \frac{1}{2}(f_\down'^2+f_\up'^2+\gamma f_3'^2)-\frac{1}{2}f_\down^2\\
 \nonumber &-\frac{1}{2}(1-h)f_\up^2-\frac{1}{2}\Bigl(1-\frac{1+r}{2}h\Bigr)f_3^2+\frac{1}{4}(f_\down^4+f_\up^4)\\
 \nonumber &+\frac{1-\zeta_{12}+\zeta_{24}}{2\zeta_{24}}f_\down^2f_\up^2+\frac{1-2\zeta_{12}}{4\zeta_{24}}f_3^4\\
 \label{b5}&+\frac{\zeta_2}{\zeta_{24}}(f_\down^2+f_\up^2)f_3^2+\frac{\zeta_1}{\zeta_{24}} f_\down f_\up f_3^2\Biggr].
\end{align}
When evaluated for the solution to the equations of motion for maximally pair-breaking or specular boundaries, the free energy becomes
\begin{align}
 \nonumber \bar{F}_{\rm B_2} ={}& -\frac{1}{\zeta_{24}}\int_0^{\bar{D}}\mbox{d}\bar{z}\ \Biggl[\frac{1}{4}(f_\down^4+f_\up^4)+\frac{1-\zeta_{12}+\zeta_{24}}{2\zeta_{24}}f_\down^2f_\up^2\\
 \label{b6} &+\frac{1-2\zeta_{12}}{4\zeta_{24}}f_3^4+\frac{\zeta_2}{\zeta_{24}}(f_\down^2+f_\up^2)f_3^2+\frac{\zeta_1}{\zeta_{24}} f_\down f_\up f_3^2\Biggr].
\end{align}
In Eq. (\ref{b5}), note that a large negative quadratic coefficient enhances superfluidity through the condensation energy. Having $h>0$ suppresses the components $f_\up$ and $f_3$ in comparison to $f_\down$. In addition, $f_3$ is also suppressed by the quadratic Zeeman term through $r>0$.

\begin{figure}[t!]
\centering
\includegraphics[width=7.6cm]{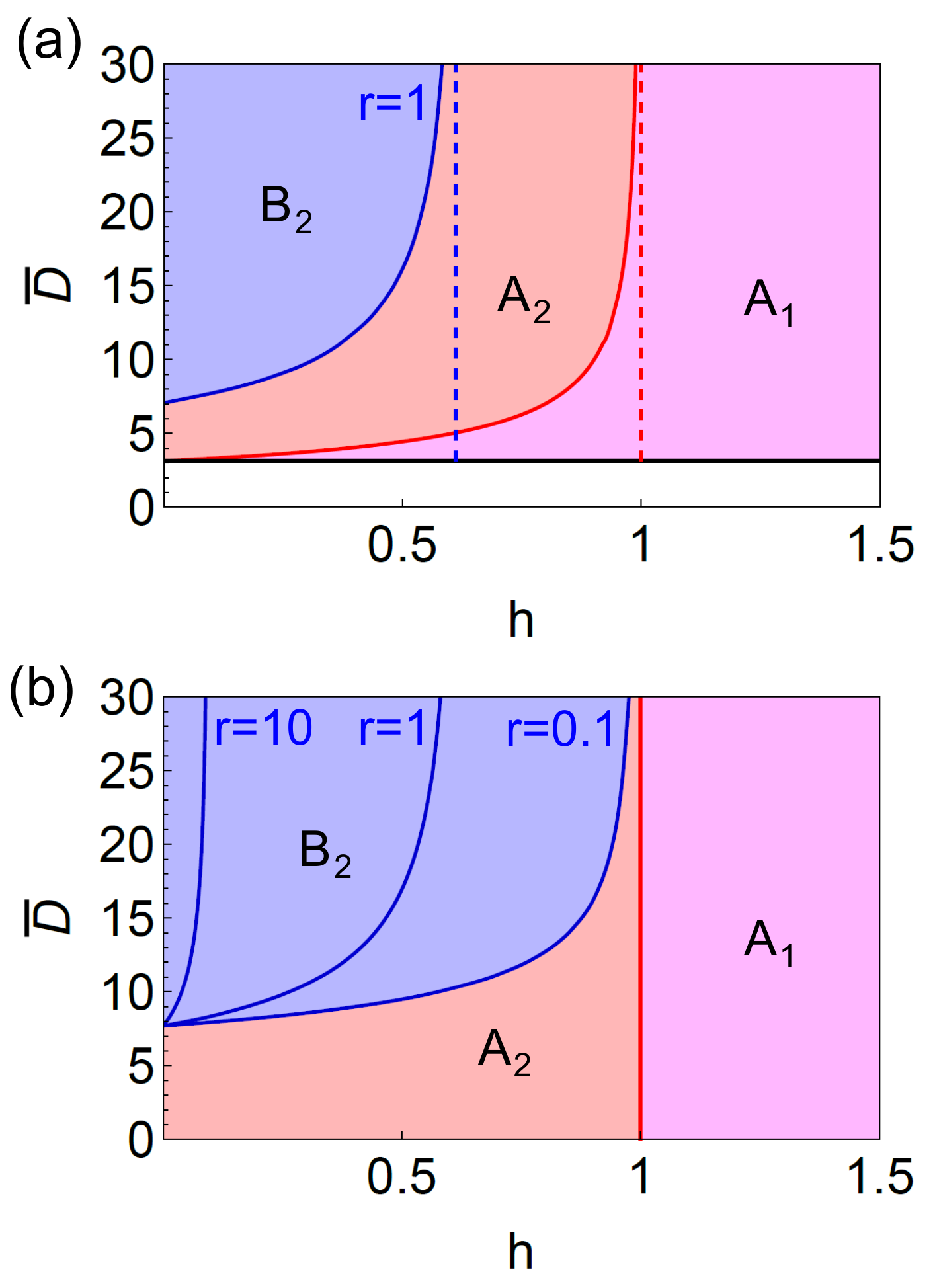}
\caption{Phase diagram in a magnetic field for maximally pair-breaking and specular boundaries in (a) and (b), respectively.  We observe that the B$_2$-phase is restricted to a region at sufficiently large $\bar{D}$ and sufficiently small $h$, whereas the thin-slab phase structure from Fig. \ref{FigThinSlab} is reproduced otherwise. The size of the B$_2$-region depends on the parameter $r=H/100\text{G}$, which resembles the importance of the quadratic Zeeman term. For concreteness, we use the weak-coupling values for the coefficients $\zeta_a$ and $\gamma=3$ in the plot, but ignore the P$_2$-state, which is disfavored over the A$_2$-state when strong-coupling corrections are incorporated. The critical value of the A$_2$-A$_1$-transition is at $h_{\rm cA}\approx 1$. For specular boundaries, the transition line $\bar{D}_{\rm B_2A_2}(h,r)$  between the B$_2$- and A$_2$-phase (continuous blue line) follows Eqs. (\ref{b9}) and (\ref{b11}). For $\bar{D}\to \infty$, the B$_2$- and A$_2$-phases only exists for $h<h_{\rm cB}$ and $h<h_{\rm cA}\approx 1$ (dashed blue and red lines). For the specular case, we have ignored the S-phase that will only be stable for small $h$ and $r$.}
\label{FigPhaseH}
\end{figure}

The equations of motion (\ref{b2})-(\ref{b4}) need to be solved numerically, and the free energy compared to the competing A$_2$-state. In Fig. \ref{FigPhaseH} we show the resulting phase diagram for both maximally pair-breaking and specular boundaries. We use weak-coupling values for the $\zeta_a$ and $\gamma$ in the plot, but assume $h_{\rm cA}>h_{\rm cP}$ so that the degeneracy between the A$_2$- and P$_2$-states is lifted. We observe that the B$_2$-phase is only stabilized for sufficiently large $\bar{D}>\bar{D}_{\rm B_2A_2}$, with the critical value depending on $h$ and $r$. For large $\bar{D}\gg1$, it is restricted to low magnetic fields according to $h<h_{\rm cB}(r)$. For small $\bar{D}$ or large $h$, the phase structure agrees with the one of thin slabs and large magnetic fields discussed in Sec. \ref{SecThinH}.

The competition between the B$_2$- and A$_2$-phase is similar for both maximally pair-breaking and specular boundary conditions, as is clear from Fig. \ref{FigPhaseH}. To study the critical line $\bar{D}_{\rm B_2A_2}$, we therefore focus on specular boundaries. Recall that, in the zero-field case, the expressions for $\bar{D}_{\rm P}$ for either boundary conditions are barely distinguishable in the plots. Furthermore, we assume that the $\zeta_a$ are close to the weak-coupling values, so that $h_{\rm cA}\approx h_{\rm cP}\approx 1$ and the free energies of the A$_2$ and P$_2$-phases are identical. This trick allows us to estimate $\bar{D}_{\rm B_2A_2}$ from the competition between the B$_2$- and P$_2$-phase, although with strong-coupling corrections this would underestimate $\bar{D}_{\rm B_2A_2}$, since then $\bar{F}_{\rm A_2}<\bar{F}_{\rm P_2}$.

In Eqs. (\ref{b2})-(\ref{b4}), we identify the onset of the B$_2$-phase from the P$_2$-phase as the point where $f_3>0$ starts to solve the equations of motion, and hence leads to a lower free energy. At this transition point, $f_\down,f_\up$ are approximately constant for specular boundaries and given by Eq. (\ref{thin15}). Inserting these values into the equation for $f_3$ and ignoring nonlinear terms in $f_3\approx 0$, we arrive at the oscillator equation
\begin{align}
 \label{b7} f_3'' + \omega_{\rm H}^2 f_3 \approx 0
\end{align} 
with
\begin{align}
 \label{b8} \omega_{\rm H}^2 = \frac{1-\frac{1+r}{2}h-\frac{1}{\zeta_{24}}[\zeta_2(f_{\down 0}^2+f_{\up 0}^2)+2\zeta_1 f_{\down 0}f_{\up 0}]}{\gamma}.
\end{align} 
The critical value of $\bar{D}$ is the half-period of the oscillator and thus
\begin{align}
 \label{b9} \bar{D}_{\rm B_2A_2}(h,r) \approx \frac{\pi}{\omega_{\rm H}(h,r)}.
\end{align}
Note that for $h\to 0$ and $h_{\rm c}\to h_{\rm cP}$ we recover the specular value of the pdB-P-transition from Eq. (\ref{spec8}),
\begin{align}
 \label{b10} \bar{D}_{\rm B_2A_2}(0,r) \approx \bar{D}_{\rm P}= \sqrt{\frac{\gamma(1-\zeta_{12})}{1-3\zeta_{12}}}.
\end{align}
Inserting the weak-coupling values for the $\zeta_a$, we arrive at
\begin{align}
 \label{b11} \omega_{\rm H}^2 \approx \frac{1}{2\gamma}\Bigl(-rh +\sqrt{1-h}\Bigr).
\end{align}
Furthermore, a necessary condition for the existence of a nontrivial solution to Eq. (\ref{b7}) is $\omega_{\rm H}^2>0$, which yields the critical field
\begin{align}
 \label{b12} h_{\rm cB}(r) \approx \frac{-1+\sqrt{1+4r^2}}{2r^2}.
\end{align}
This is strictly smaller than $h_{\rm cA}\approx 1$. For instance,  for $r=1$ we obtain the inverse golden ratio $h_{\rm cB}\approx \frac{\sqrt{5}-1}{2} = 0.618$.

\subsection{$P$-$T$-$H$-$D$ phase diagram}\label{PTHD}

To estimate the crossover as observed in experiment, we compute the $P$-$T$ phase diagrams in dependence of $H$ and $D$ by solving the equations of motions for the dimensionless order parameters and comparing $\bar{F}_{\rm B_2}$, $\bar{F}_{\rm A_2}$, and $\bar{F}_{\rm A_1}$. We use the strong-coupling coefficients $\beta_a(P,T)$ from Eq. (\ref{pt9}), $\gamma=3$, and assume that $g_1$ and $g_2$ are given by Eqs. (\ref{mag5b}) and (\ref{mag5d}). To normalize with $\alpha_H$, we write
\begin{align}
 \label{b13} K & = \frac{N(0)}{3}\xi_0^2,\\
 \alpha_0 &= \frac{N(0)}{3} \Phi\Bigl(\frac{T}{T_{\rm c0}(P)}\Bigr)^2,
\end{align}
where $\xi_0$ is the zero-temperature correlation length and
\begin{align}
  \label{b14} \Phi(t)^2 = \begin{cases} 1 & t<0.3\\ (1-t)(0.9663+0.7733t)^2 & t \geq 0.3\end{cases}
\end{align}
is the square of $\Phi(t)$ in Eq. (\ref{pt8}). Importantly, the expression $\Phi(t)^2$ is well-defined for $t>1$, where $\alpha_0$ is negative, see Fig. \ref{FigAppThermo}d. Hence
\begin{align}
 \xi(T,P,H)= \sqrt{\frac{K}{\alpha_H}} = \frac{\xi_0(P)}{\sqrt{\Phi(T/T_{\rm c0}(P))^2+H/10^6\text{G}}},
\end{align}
which is used to compute $\bar{D}=D/\xi$. We further have
\begin{align}
 h & = \frac{2H/10^6\text{G}}{\Phi(T/T_{\rm c0})^2+H/10^6\text{G}}
\end{align}
and $r=H/100\text{G}$.

\begin{figure*}[t!]
\centering
\includegraphics[width=16.6cm]{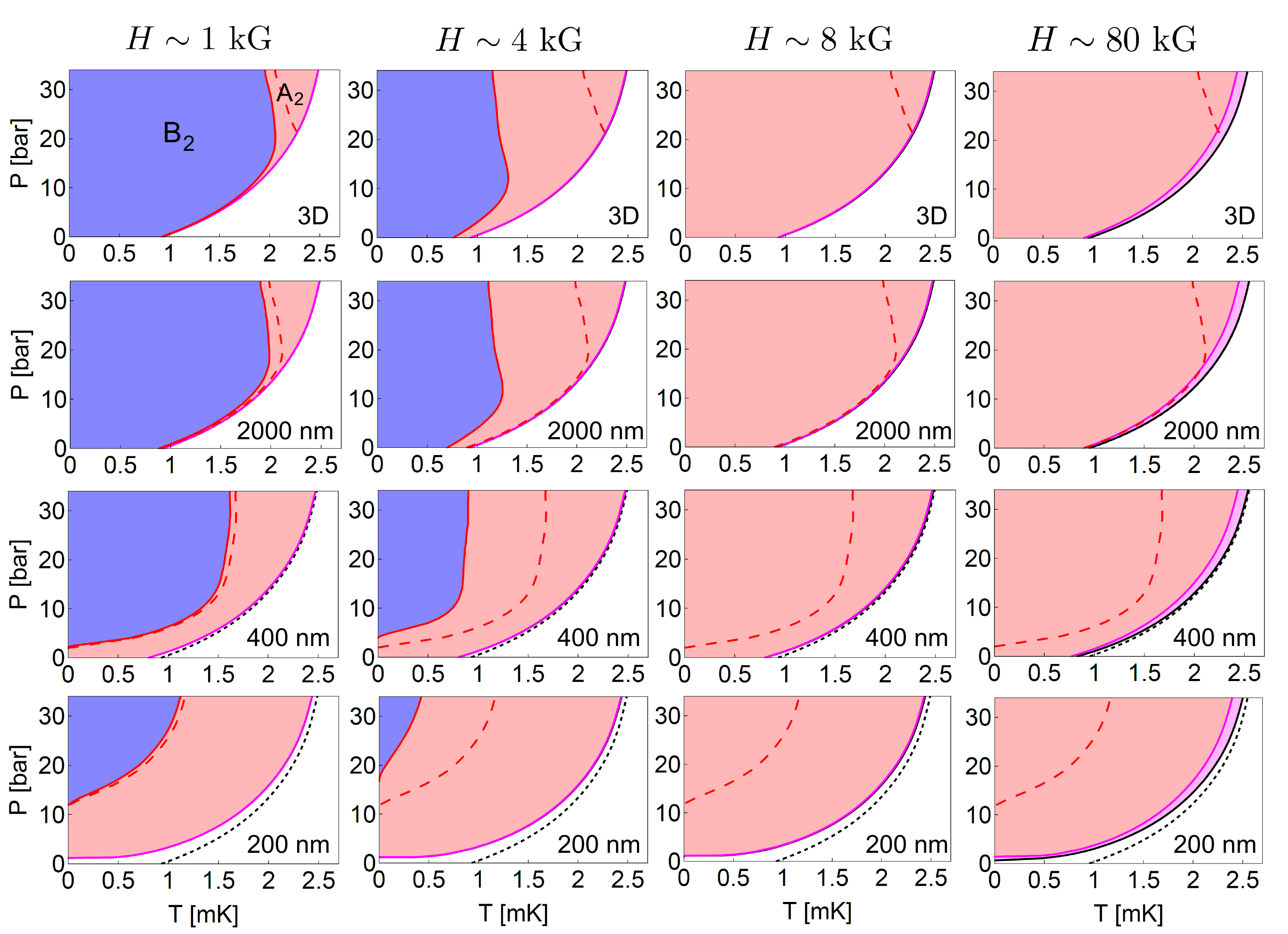}
\caption{Dimensional crossover of $P$-$T$ phase diagrams in an external magnetic field $H$ for maximally pair-breaking boundary conditions. The columns correspond to the magnetic fields as indicated, whereas the rows correspond to various slab heights $D$, with the first row being the 3D bulk limit.  The colors follow Fig. \ref{FigPhaseH}, with phases B$_2$ shown in blue, A$_2$ in red, and A$_1$ in magenta. The A$_1$-phase is limited to a narrow region close to the critical temperature, which becomes sizable once $H\sim 50\ \text{kG}=5\ \text{T}$. The A$_1$-A$_2$-transition is of second order, the A$_2$-B$_2$-transition is of first order. For each $D$, the dashed red line indicates the zero-field A-pdB transition to highlight the effect of the magnetic field. In contrast, the black dashed line is the bulk critical temperature $T_{\rm cH}$ in a magnetic field to highlight the effect of the spatial confinement.}
\label{FigPhasePTH}
\end{figure*}

The resulting dimensional crossover of $P$-$T$ phase diagrams in a magnetic field is shown in Fig. \ref{FigPhasePTH}. We find that the phase diagram in a magnetic field starts to deviate from the zero-field case for $H\sim 100\ \text{G}$ ($r\sim 1$), but the difference is small and not shown here. The strongest effects are visible in the regime $H\sim 1\ \text{kG}$ ($r\sim 10$), which is the magnetic field where the quadratic Zeeman term becomes effective and suppresses the B$_2$-phase. The B$_2$-phase disappears entirely from the phase diagram for $H\sim 7\ \text{kG}$ for our choice of parameters. The phase structures of the 3D bulk system and for $D=2000\ \text{nm}$ closely resemble each other. The phase diagram for tight confinement and large magnetic fields is comprised of the A$_2$- and A$_1$-phases. The superfluid transition at $T_{\rm cH}$ is always towards the A$_1$-phase. The A$_1$-phase is restricted to a narrow region around the critical temperature, which only becomes substantial for $H\sim 50 \text{kG} = 5\ \text{T}$, which is the region where the linear Zeeman term becomes comparable to $\alpha_0$.

While Fig. \ref{FigPhasePTH} is shown for maximally pair-breaking boundaries, a similar crossover is observed for specular and diffusive boundaries, with the zero-field limit shown in Fig. \ref{FigPDCrossover}. Although we have not considered the effect of the magnetic field on the S-phase, we expect that the S-phase is confined to a small region between the B$_2$- and A$_2$-phases for specular and diffusive boundaries. This expectation results from extrapolating the zero-field phase diagram in Fig. \ref{FigSpecPD}. In fact, the S-phase is additionally suppressed by the quadratic Zeeman term once $r\gtrsim 1\ \text{kG}$. Recall that the S-phase is energetically favored over the pdB-phase in the zero-field case by having a nonzero component $A_{32}$ of the order parameter. This mechanism is preempted by the quadratic Zeeman term that suppresses the third row of the order parameter.

\section{Conclusion and outlook}\label{SecConcl}

In this work, we have studied the phase structure of superfluid $^3$He confined to a slab geometry in a perpendicular magnetic field. Our analysis was built on the Ginzburg--Landau free energy functional, which led to a rich interplay between the complex $3\times 3$ matrix order parameter and various boundary conditions at the container walls. Although the solution to the underlying equations of motion had to be carried out numerically in most cases, many analytical statements for certain order parameter profiles and, importantly, the location of phase transition lines could be made by relying on the theory of Jacobi and Lam\'{e} elliptic functions. In particular, by introducing a unifying function $f(\bar{z})$ and a reference free energy $\bar{F}_0(\bar{D})$, many exact statements could be made about the phase structure of the confined system, especially close to the critical temperature. 

We found that the combined effect of confinement and magnetic field is to suppress the third column and third row of the order parameter, respectively, which favors the A-, A$_2$-, and A$_1$-phases in thin slabs and strong magnetic fields. We also observed that diffusive boundary conditions with $b_{\rm T}=0.5$ are an interesting situation interpolating between maximally pair-breaking and specular boundaries, showing features of both, such as a suppression of $T_{\rm c}$ or the existence of the stripe phase for zero field. While the planar and related P$_2$-phase in a magnetic field are energetically degenerate with the corresponding A-phases in weak-coupling theory, they are generically disfavored by physical strong-coupling corrections, and hence unlikely to be observed in experiment. This, however, is under the assumption of a negligible dipole effect that may stabilize the planar phase for field values $H\lesssim 25\ \text{G}$ \cite{Jones:1976zz,VOLOVIK199027} close to criticality.

In the following, we point out two directions of future research that build on our findings. First, the tightly confined superfluid in high magnetic fields constitutes an intriguing quantum system in itself. It has been known for a long time that the excitations of 2D p-wave superfluids feature exotic Abelian and non-Abelian exchange statistics \cite{PhysRevB.61.10267,volovik2003universe}, which may be used to build a topological quantum computation platform \cite{10.1063/1.2337825}. Viewed as a quantum many-body system, the superfluid is described by a complex $2\times 2$ matrix order parameter
\begin{align}
 \label{con1} A = \begin{pmatrix} A_{11} & A_{12} \\ A_{21} & A_{22} \end{pmatrix},
\end{align} 
with the free energy functional close to the critical point given by Eq. (\ref{gl1}). While the 9 complex components of the usual $^3$He order parameter make a full treatment of the phase structure even in mean-field theory intractable, the reduced number of 4 complex component may simplify the computation. For instance, for the cases of Refs. \cite{PhysRevB.104.L180507,PhysRevB.108.144503}, where either the third column or third row of the order parameter is neglected, the weak-coupling free energy can be minimized exactly. Furthermore, going beyond the polynomial expansion of the free energy, the full functional within mean-field theory has been minimized in Ref. \cite{PhysRevB.100.104503} for the case of d-wave pairing with 5 complex components. Studies of the field theory implied by Eq. (\ref{con1}) could, therefore, shed new light onto the physics of broken symmetries with complex matrix order parameters.

A second application of our work consists in using the theoretical predictions about the phase structure in dependence of $D$ and $H$ to experimentally constrain the parameters of the free energy. For instance, the parameter $\gamma$ is currently only poorly constrained and a closer analysis of the A-pdB-transition under confinement could determine its value, because it enters the transition condition $\bar{F}_{\rm A}=\bar{F}_{\rm pdB}(\gamma)$. Or the specularity of surfaces coated with $^4$He, i.e. the value of $b_{\rm T}$, can be estimated from the suppression of the critical temperature together with Eqs. (\ref{pt5}) and (\ref{pt6}). Measuring the phase structure at low temperatures as in Fig. \ref{FigQPD} would allow to determine the coefficients $\beta_a(T,P)$ at $T=0$. Recall that a drawback of the commonly employed strong-coupling corrections in Eq. (\ref{pt9}) is the assumption that they reduce to weak-coupling theory for $T=0$. Importantly, assuming that the free energy coefficients only weakly depend on $D$, one can utilize the measurements for various values of $D$ to infer the temperature- and pressure-dependence of the bulk parameters.

\begin{acknowledgements}
The authors thank Frank Marsiglio and Canon Sun for inspiring discussions. They acknowledge funding from the Natural Sciences and Engineering Research Council of Canada (NSERC) through grants RGPIN-2021-02534, DGECR2021-00043, RGPIN-2022-03078, and ALLRP 602120-2024, the University of Alberta startup grant UOFAB Startup Boettcher, and additional support provided by Alberta Innovates (AI) grant \#242506367. AM acknowledges funding through the IBET Fellowship of the UofA ELITE program.

\end{acknowledgements}

\begin{appendix}

\section{Derivation of Eq. (\ref{real16})}\label{AppProofTh}

To compute the solution $f(\bar{z})$ of Eqs. (\ref{real13}), (\ref{real13b}) for maximally pair-breaking boundaries given in Eq. (\ref{real16}), we consider the more general problem of finding the function $f(\bar{z})\geq 0$ that solves
\begin{equation}\label{proof1}
 \begin{aligned}
 0&= -\gamma f'' -f +\kappa f^3,\\
 0&= f(0)=f(\bar{D})
 \end{aligned}
\end{equation}
for $\bar{z}\in[0,\bar{D}]$, with $\gamma,\kappa>0$. Note, however, that through a proper rescaling of $\bar{z}$ and $f$, Eq. (\ref{proof1}) can always be reduced to Eq. (\ref{real13}). The solution to Eq. (\ref{proof1}) is given by
\begin{align}
 \label{proof2} f(\bar{z}) = a\ \text{sn}\Bigl(\frac{\sigma \bar{z}}{a},k^2\Bigr),
\end{align}
with $a, k^2, \sigma$ determined such that
\begin{align}
 \label{proof3} a&=\sqrt{\frac{1-\sqrt{1-2\gamma\kappa \sigma^2}}{\kappa}},\\
 k^2 &= \frac{\kappa a^4}{2\gamma \sigma^2},\\
 \label{proof4} \frac{\sigma \bar{D}}{a} &= 2K(k^2).
\end{align}
Note that the definition of $k^2$ implies 
\begin{align}
k^2+1 = \frac{a^2}{\gamma \sigma^2}
\end{align}
and Eq. (\ref{proof4}) can be recast into the form 
\begin{align}
\sqrt{\gamma(k^2+1)}K(k^2)=\bar{D}/2.
\end{align}

To prove Eq. (\ref{proof2}), first note that Eq. (\ref{proof1}) constitutes the time evolution of a softening-spring Duffing oscillator, with time-coordinate $\bar{z}$ and period $2\bar{D}$. As for any conservative system, the energy of the Duffing oscillator is conserved. In our case, the function
\begin{align}
 \label{real20} e(\bar{z}) = \frac{\gamma}{2}f'(\bar{z})^2+\frac{1}{2}f(\bar{z})^2-\frac{\kappa}{4}f(\bar{z})^4
\end{align}
is constant when evaluated for the solution of Eq. (\ref{proof1}). Indeed,
\begin{align}
 \label{real21} e'(\bar{z}) = f'(\bar{z})\Bigl(-\gamma f''(\bar{z})-f(\bar{z})+\kappa f(\bar{z})^3\Bigr)=0.
\end{align}
Since $e$ is constant, it coincides with the values at $\bar{z}=0$ and $\bar{z}=\bar{D}/2$, where the contribution is purely kinetic or potential, respectively. Hence
\begin{align}
 \label{real22} e = \frac{\gamma}{2}\sigma^2 = \frac{1}{2}a^2-\frac{\kappa}{4}a^4
 \end{align}
and we have
\begin{align}
 \label{real23} f'^2 = \frac{1}{\gamma}\Bigl(2e -f^2 +\frac{\kappa}{2}f^4\Bigr).
\end{align}
Since the square of $f'$ is a polynomial of degree four in $f$ we conclude that $f(\bar{z})$ is an elliptic function in the complex $\bar{z}$-plane, i.e. a meromorphic function with two complex periods. We only need $f(\bar{z})$ along the real line here. Using the ansatz $ f(\bar{z}) = a\ g(\sigma \bar{z}/a)$ with $g(x)$ such that $g(0)=0, g'(0)=1$, we find
\begin{equation}\label{real24}
\begin{aligned}
 g'^2 &= \frac{1}{\gamma\sigma^2}\Bigl(2e-a^2g^2+\frac{\kappa a^4}{2}g^4\Bigr)\\
 &= 1 - \frac{a^2}{\gamma \sigma^2} g^2 +\frac{\kappa a^4}{2\gamma \sigma^2}g^4\\
 &= 1 - (k^2+1) g^2 +k^2g^4 \\
 &= (1-g^2)(1-k^2g^2).
\end{aligned}
\end{equation}
The unique solution to this equation is $g(x)=\text{sn}(x,k^2)$. The function $\text{sn}(x,k^2)$ has two complex periods $4K$ and $2\rmi K'$  such that
\begin{align}
 \label{real25} \text{sn}(x+4K,k^2) = \text{sn}(x+2\rmi K',k^2) = \text{sn}(x,k^2)
\end{align} 
for all $x\in \mathbb{C}$, where $K=K(k^2)$ and $K'=K(1-k^2)$. We are interested in the real period $4K$ that we match to $2\bar{D}$ through
\begin{align}
 \label{real26} f(\bar{z}+2\bar{D}) = f(\bar{z})\ \Rightarrow\ \frac{2\sigma \bar{D}}{a} = 4K(k^2).
\end{align}
This completes the proof. $\square$

\section{Closed-form expression for $\bar{F}_0(\bar{D})$} \label{AppClosedEnergy}
In this appendix, we derive a closed form expression for the integral
\begin{align}
 \label{fnot1} \bar{F}_0(\bar{D}) = \int_0^{\bar{D}}\mbox{d}\bar{z}\ \Bigl(\frac{1}{2}f'^2-\frac{1}{2}f^2+\frac{1}{4}f^4\Bigr),
\end{align}
where $f(\bar{z})$ solves the equation $0=-f''-f+f^3$ with either specular or maximally pair-breaking boundary conditions at $\bar{z}=0$ and $\bar{z}=\bar{D}$.

Since $ff'$ vanishes at the boundaries of the integration domain for both maximally pair-breaking and specular cases, we obtain from Eq. (\ref{fnot1}) via partial integration
\begin{align}
 \label{fnot2} \bar{F}_0(\bar{D}) &= \int_0^{\bar{D}}\mbox{d}\bar{z}\ \Bigl(-\frac{1}{2}ff''-\frac{1}{2}f^2+\frac{1}{4}f^4\Bigr)\\
 \label{fnot3} &= -\frac{1}{4}\int_0^{\bar{D}}\mbox{d}\bar{z}\ f^4.
\end{align}
Crucially, we used that $f$ solves the equation of motion. For specular boundary conditions, $f=1$ and we simply have
\begin{align}
 \label{fnot2b} \bar{F}_0(\bar{D}) = -\frac{1}{4}\bar{D}\ \text{(specular)}.
\end{align}
For maximally pair-breaking boundaries, we parametrize the solution $f(\bar{z})$ through
\begin{align}
 \label{fnot6} &f(\bar{z}) = a\ \text{sn}\Bigl(\frac{\sigma\bar{z}}{a},k^2\Bigr),\ a=\sqrt{1-\sqrt{1-2\sigma^2}},\\
 \label{fnot7} &k^2=\frac{a^4}{2\sigma^2},\ k^2+1=\frac{a^2}{\sigma^2},\ \frac{\sigma \bar{D}}{a} =2K(k^2)
\end{align}
and use the identity
\begin{align}
 \nonumber \int_0^{2K}\mbox{d}x\ \text{sn}(x,k^2)^4 ={}& \frac{2}{3k^4}\Bigl((k^2+2)K(k^2)\\
 \label{fnot9} &-2(k^2+1)E(k^2)\Bigr).
\end{align}
Here $K=K(k^2)$ and $E(k^2)$ are the elliptic integrals of the first and second kind, respectively, given by
\begin{align}
 \label{fnot10} K(k^2) &=\int_0^1\mbox{d}t\ \frac{1}{\sqrt{(1-t^2)(1-k^2t^2)}},\\
 \label{fnot11} E(k^2) &=\int_0^1\mbox{d}t\ \sqrt{\frac{1-k^2t^2}{1-t^2}}.
\end{align}
We then arrive at
\begin{align}
 \nonumber \bar{F}_0(\bar{D}) &=-\frac{a^5}{4\sigma} \int_0^{2K}\mbox{d}x\ \text{sn}(x,k^2)^4\\
 \nonumber  &=-\frac{\sigma a k^2}{2} \int_0^{2K}\mbox{d}x\ \text{sn}(x,k^2)^4\\
 \label{fnot12} &=-\frac{\sigma a}{3k^2}\Bigl((k^2+2)K(k^2)-2(k^2+1)E(k^2)\Bigr).
\end{align}
This is the desired closed-form expression for $\bar{F}_0(\bar{D})$ for maximally pair-breaking boundary conditions. The parameters $\sigma,a ,k^2$ are all determined solely by $\bar{D}$ through Eqs. (\ref{fnot6}) and (\ref{fnot7}).

\begin{figure}[t!]
\centering
\includegraphics[width=8.6cm]{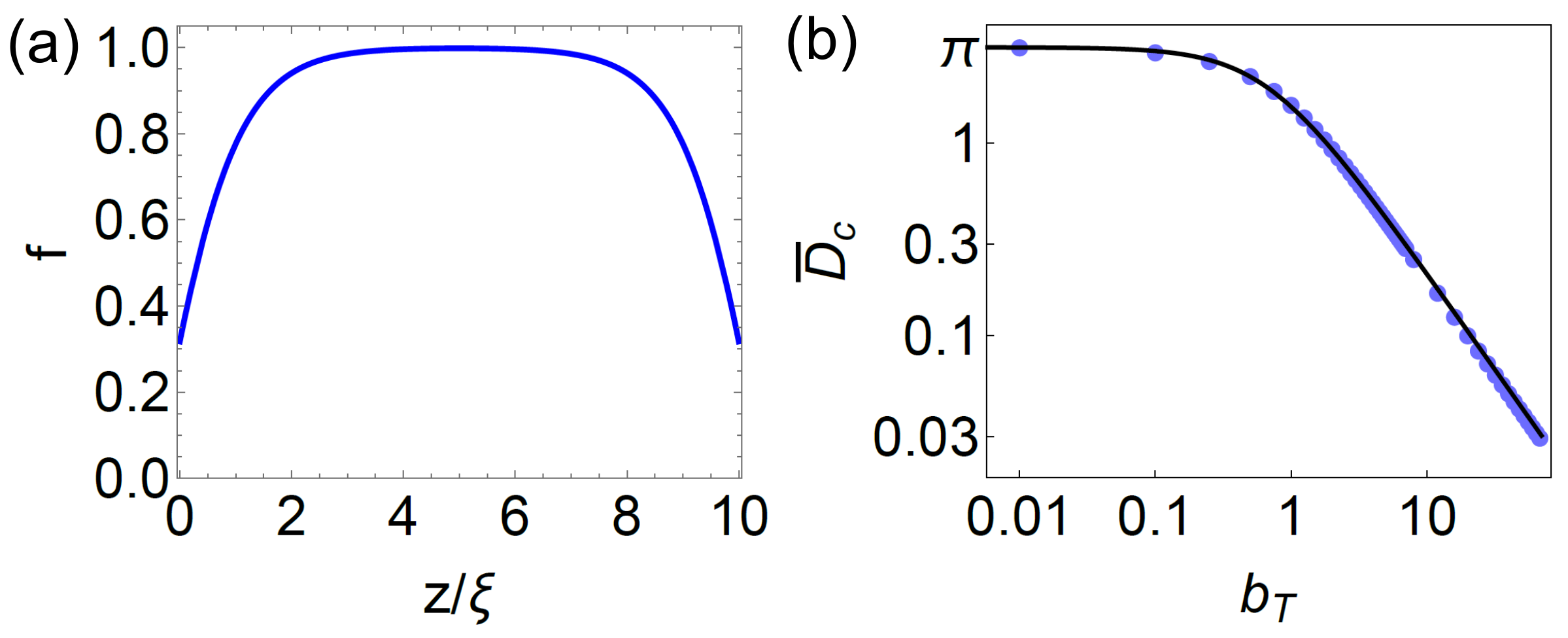}
\caption{Solutions $f(\bar{z})$ to Eq. (\ref{real13}) for diffusive boundary conditions (\ref{fnot13}) with parameter $0<b_{\rm T}<\infty$. \emph{Panel (a)}. We present the solution for $\bar{D}=10$ and the experimentally relevant case of $b_{\rm T}=0.5$. Note that neither the function nor its derivative vanish at the container walls $\bar{z}=0$ and $\bar{z}=\bar{D}$. The boundary conditions constrain the function such that if we linearly extrapolate the function into the region $\bar{z}<0$ by using the slope $f'(0)$, then this extrapolated function would vanish at $\bar{z}=-b_{\rm T}$, and similarly for the boundary at $\bar{z}=\bar{D}$. This also shows that the maximally pair-breaking and specular cases are recovered smoothly for $b_{\rm T}\to 0$ and $b_{\rm T}\to \infty$. \emph{Panel (b).} Solutions to Eq. (\ref{real13}) with diffusive boundaries exists for $\bar{D}>\bar{D}_{\rm c}(b_{\rm T})$. The blue points correspond to numerically determined values of $\bar{D}_{\rm c}(b_{\rm T})$, whereas the black curve is the empirical function in Eq. (\ref{pt5}). Equation (\ref{pt5}) is an ansatz with fitted parameters and does not derive from any underlying theory, while the true curve $\bar{D}_{\rm c}(b_{\rm T})$ likely corresponds to some properties of Jacobi elliptic functions.}
\label{FigAppbT}
\end{figure}

Note that the simplification from Eq. (\ref{fnot1}) to (\ref{fnot2}) is not valid for diffusive boundary condition, where we have
\begin{equation}\label{fnot13}
 \begin{aligned}
  f'(0) &= \frac{1}{b_{\rm T}}f(0),\\
f'(\bar{D}) &= -\frac{1}{b_{\rm T}}f(\bar{D}).
 \end{aligned}
\end{equation}
In this case, $ff'$ does not vanish at $\bar{z}=0$ and $\bar{z}=\bar{D}$. An example solution $f(\bar{z})$ for $b_{\rm T}=0.5$ is shown in Fig. \ref{FigAppbT}a. Solutions exists for $\bar{D}>\bar{D}_{\rm c}(b_{\rm T})$ shown in Fig. \ref{FigAppbT}b, with $\bar{D}_{\rm c}(0.5)=2.2143$. We then have
\begin{align}
 \label{fnot15} \bar{F}_0(\bar{D}) &= \frac{1}{2}\Bigl(f(\bar{D})f'(\bar{D})-f(0)f'(0)\Bigr)-\frac{1}{4}\int_0^{\bar{D}}\mbox{d}z\ f^4.
\end{align}
Now using the boundary conditions and the fact that $f(0)=f(\bar{D})$, because the function is symmetric about $\bar{D}/2$, we arrive at
\begin{align}
 \label{fnot16} \bar{F}_0(\bar{D}) &= -\frac{f(0)^2}{b_{\rm T}}-\frac{1}{4}\int_0^{\bar{D}}\mbox{d}z\ f^4.
\end{align}

\section{Derivation of Eqs. (\ref{real32b}) and (\ref{real41})}\label{AppReal}

In this appendix, we derive Eqs. (\ref{real32b}) and (\ref{real41}) for the phase boundaries of real orders in terms of $\bar{D}_{\rm c}$ and $\bar{D}_{\rm P}(\gamma,\zeta_{12})$.

Let us first consider the specific case of $\zeta_{12}=0$. The two components $f_1$ and $f_3$ then decouple and are described by the equations
\begin{equation}\label{real33}
 \begin{aligned}
 0 &=-f_1''-f_1 + f_1^3,\\
 0 &=-\gamma f_3'' -f_3+f_3^3.
 \end{aligned}
\end{equation} 
Clearly, the solution from Eq. (\ref{proof2}) applies to both equations and we find
\begin{equation}\label{real34}
 \begin{aligned}
 f_1(\bar{z}) &= a_1\text{sn}\Bigl(\frac{\sigma_1 \bar{z}}{a_1},k_1^2\Bigr),\\
 f_3(\bar{z}) &= a_3\text{sn}\Bigl(\frac{\sigma_3 \bar{z}}{a_3},k_3^2\Bigr).
\end{aligned}
\end{equation}
Herein,
\begin{equation}\label{real35}
 \begin{aligned}
 a_1 &=\sqrt{1-\sqrt{1-2\sigma_1^2}},\ a_3 =\sqrt{1-\sqrt{1-2\gamma \sigma_3^2}}
 \end{aligned}
\end{equation}
and $\sigma_{1,3}$ are determined to match $f_1(\bar{D})=f_3(\bar{D})=0$. The solution $f_3$ ceases to exist when $\sigma_3\to 0$, which corresponds to $k_3^2\to 0$. In this limit, $a_3 \to \sqrt{\gamma}\sigma_3$, and
\begin{align}
 \label{real36} \frac{f_3(\bar{z})}{a_3} \to \text{sn}\Bigl(\frac{\bar{z}}{\sqrt{\gamma}},0\Bigr) = \sin\Bigl(\frac{\bar{z}}{\sqrt{\gamma}}\Bigr).
\end{align}
The equation $f_3(\bar{D})=0$ is solved by $\bar{D}=\sqrt{\gamma}\pi$. Consequently, a solution with $f_3>0$ only exists for $\bar{D}>\bar{D}_{\rm P}(\gamma,0)=\sqrt{\gamma}\pi$. Similarly, since the equation for $f_1$ is identical to Eq. (\ref{real13}), we conclude that a solution $f_1>0$ only exists for $\bar{D}>\bar{D}_{\rm c} = \pi$.

To understand why the lower boundary of the P-phase in the phase diagram is $\bar{D}_{\rm c}=\pi$ for all $\zeta_{12}$, we note that Eq. (\ref{real30}) for $f_3=0$ yields
\begin{equation}\label{real37}
 \begin{aligned}
  0 &=-f_1''-f_1 +(1-\zeta_{12}) f_1^3\\
 \end{aligned}
\end{equation}
for the component $f_1$. Applying Eq. (\ref{proof2}) with $\gamma=1$ and $\kappa=1-\zeta_{12}$, we see that $a_1\to \sigma_1$
as $\sigma_1\to 0$. Hence $f_1(\bar{z})/a_1 \to \sin(\bar{z})$ and $\bar{D}_{\rm c}=\pi$ for all $\zeta_{12}$. 

As an alternative derivation of this result, by a method that will be applied below to find the P-pdB-boundary, assume that $\bar{D}$ is slightly above the critical value $\bar{D}_{\rm c}$. By approaching the phase boundary as closely as necessary, we can make $f_1$ so small that the term $(1-\zeta_{12})f_1^3$ in the equation of motion can be neglected. We then have $0\approx -f_1''-f_1$ for all $\zeta_{12}$, with the solution $f_1(\bar{z})\propto\sin(\bar{z})$. Hence $\sin(\bar{D}_{\rm c})=0$ yields $\bar{D}_{\rm c}=\pi$ as before.

To understand the origin of Eq. (\ref{real41}) for the boundary between the P- and pdB-phases, we recall some properties of the Lam\'{e} equation and its solutions. The Lam\'{e} equation reads
\begin{align}
 \label{real42} 0 = \psi''(x) + \Bigl( \mathcal{E}-\nu(\nu+1)k^2\text{sn}(x,k^2)^2\Bigr)\psi(x).
\end{align}
We may think of this as a stationary Schr\"{o}dinger equation for a particle moving in the one-dimensional periodic potential $V(x) \propto \text{sn}(x,k^2)^2$. Solutions to the Lam\'{e} equation exists if $\mathcal{E}$ has certain discrete values, called the Lam\'{e} eigenvalues. They are labelled by integers $m=0,1,2,\dots$ and depend on the boundary conditions and symmetries chosen. In our case, the Lam\'{e} eigenvalues read $\mathcal{E}_m=a_\nu^{(2m+1)}(k^2)$ with Lam\'{e} eigenfunctions $\psi_m(x)=\text{Ec}_\nu^{(2m+1)}(x,k^2)$. In Mathematica, they are included through the functions $\textsc{LameEigenvalueA}[\nu,2m+1,k^2]$ and $\textsc{LameC}[\nu,2m+1,x,k^2]$,

To derive Eq. (\ref{real41}), consider then a value of $\bar{D}$ slightly larger than $\bar{D}_{\rm P}$, such as point C in Fig. \ref{FigRealPD}. By approaching the phase boundary as closely as necessary, we can make $f_3$ arbitrarily small and thus neglect terms of order $f_3^2$ and $f_3^3$ in Eqs. (\ref{real30}). In this limit, the equations read
\begin{equation}\label{real43}
\begin{aligned}
 0 &\approx-f_1''-f_1 +(1-\zeta_{12}) f_1^3,\\
 0 &\approx -\gamma f_3'' -(1-2\zeta_{12}f_1^2)f_3.
\end{aligned}
\end{equation}
Crucially, the equation for $f_1$ is independent of $f_3$. It is challenging to rigorously justify approximations such as this for nonlinear differential equations; however, the numerical solution for $f_1$ shown in Fig. \ref{FigRealPD} C and D strongly suggests that $f_1$ is, indeed, independent of $f_3$ close to the boundary. The solution according to Eq. (\ref{proof2}) is given by $f_1(\bar{z})= a_1\text{sn}(\sigma_1\bar{z}/a_1,k_1^2)$ with $a_1,\sigma_1,k_1$ such that $k_1^2+1=a_1^2/\sigma_1^2$, $k_1^2=(1-\zeta_{12})a_1^4/2\sigma_1^2$, and $\sigma_1\bar{D}/a_1=2K(k_1^2)$. Write $f_3(\bar{z})=\psi(\sigma_1\bar{z}/a_1)$ to find that the equation for $f_3$ becomes
\begin{equation}
 \label{real44} 0 = \psi''(x) +\Biggl(\frac{k_1^2+1}{\gamma} - \frac{4\zeta_{12}}{\gamma(1-\zeta_{12})}k_1^2\text{sn}(x,k_1^2)^2\Biggr)\psi(x).
\end{equation}
This is identical to Lam\'{e}'s equation with $\mathcal{E}=(k_1^2+1)/\gamma$ and $\nu$ as in Eq. (\ref{real40}). As $\bar{D}$ approaches the phase boundary from above, the value of $k_1^2$ decreases. If $k_1^2$ is so small that $\mathcal{E}$ drops below the lowest possible eigenvalue $a_\nu^{(1)}(k_1^2)$ ($m=0$), then no solution for $\psi(x)$ exists. This determines the critical value $k_{1\rm c}^2>0$ through Eq. (\ref{real39}). The corresponding critical confinement follows from $\bar{D}_{\rm P} = \frac{2a_1}{\sigma_1}K(k_{1\rm c}^2)$, which gives Eq. (\ref{real41}) upon applying $k_1^2+1=a_1^2/\sigma_1^2$ once more. $\square$\\

\section{Stripe phase}\label{AppStripe}

The order parameter $O_{\rm S}(\bar{y},\bar{z})$ in the S-phase is given in Eq. (\ref{spec10}). Since it is a real order parameter, its equation of motion is given by the generalization of Eq. (\ref{real7}) to arbitrary spatial dependence, hence
\begin{equation}\label{str1}
\begin{aligned}
 0 ={}&-\bar{\nabla}^2O_{\mu i}-(\gamma-1)\bar{\partial}_i\bar{\partial}_j O_{\mu j}-O_{\mu i}\\
&+\zeta_{12}\mbox{tr}(OO^T)O_{\mu i}+\zeta_{345}(OO^TO)_{\mu i}.
\end{aligned}
\end{equation}
For the individual components this reads
\begin{widetext}
\begin{equation}\label{str2}
\begin{aligned}
0 ={}& -\begin{pmatrix} (\partial_{\bar{y}}^2+\partial_{\bar{z}}^2)O_{1} & & \\ & (\gamma\partial_{\bar{y}}^2+\partial_{\bar{z}}^2)O_{2} & (\partial_{\bar{y}}^2+\gamma\partial_{\bar{z}}^2)O_{23} \\ & (\gamma \partial_{\bar{y}}^2+\partial_{\bar{z}}^2)O_{32} & (\partial_{\bar{y}}^2+\gamma\partial_{\bar{z}}^2)O_{3} \end{pmatrix} - (\gamma-1) \begin{pmatrix} 0 & 0 & 0 \\0 & \partial_{\bar{y}}\partial_{\bar{z}} O_{23} & \partial_{\bar{y}}\partial_{\bar{z}} O_{2} \\ 0 & \partial_{\bar{y}}\partial_{\bar{z}} O_{3} & \partial_{\bar{y}}\partial_{\bar{z}} O_{32} \end{pmatrix} \\
&- \begin{pmatrix} O_1 & & \\ & O_2 & O_{23} \\ & O_{32} & O_3 \end{pmatrix}+\zeta_{12} (O_1^2+O_2^2+O_{23}^2+O_{32}^2+O_3^2)\begin{pmatrix} O_1 & & \\ & O_2 & O_{23} \\ & O_{32} & O_3 \end{pmatrix}\\
&+\zeta_{345}\begin{pmatrix} O_1^3 & & \\ & O_2(O_2^2+O_{23}^2+O_{32}^2)+O_{23}O_{32}O_3 & O_{23}(O_{23}^2+O_2^2+O_3^2)+O_2O_{32}O_3\\ & O_{32}(O_{32}^2+O_2^2+O_3^2)+O_2O_{23}O_3& O_3(O_3^2+O_{23}^2+O_{32}^2)+O_2O_{23}O_{32} \end{pmatrix}.
\end{aligned}
\end{equation}
\end{widetext}
The free energy that generalizes Eq. (\ref{real12b}) is
\begin{equation}
 \label{str2b} \bar{F}[O_{\rm S}(\bar{y},\bar{z})] =\frac{1}{\bar{L}_{\rm S}}\int_0^{\bar{L}_{\rm S}}\mbox{d}\bar{y} \int_0^{\bar{D}}\mbox{d}\bar{z} \Bigl( \bar{\text{f}}_{\rm S,kin}+\bar{\text{f}}_{\rm S,pot}\Bigr),
\end{equation} 
where the $\bar{y}$-integration is evaluated over one half-period in y-direction of the solution $O_{\rm S}$ to the equations of motion. In principle, one could integrate the free energy density over any sufficiently long interval $\bar{y}\in[0,\bar{L}_{\rm y}]$ and normalize by $\bar{L}_{\rm y}$, but since the solution $O_{\rm S}(\bar{y},\bar{z})$ needs to be determined numerically, this would lead to a visible finite-size effect in practice. The kinetic and potential free energy densities read
\begin{widetext}
\begin{align}
 \label{str3} \bar{\text{f}}_{\rm S,kin} ={}&\frac{1}{2}\mbox{tr}(\bar{\nabla} O\cdot \bar{\nabla}O^T) +\frac{\gamma-1}{2}(\bar{\partial}_iO_{\mu i})(\bar{\partial}_j O_{\mu j})\\
 \nonumber ={}&\frac{1}{2}\Bigl[(\partial_{\bar{y}}O_1)^2+(\partial_{\bar{z}}O_1)^2+(\partial_{\bar{y}}O_2)^2+(\partial_{\bar{z}}O_2)^2+(\partial_{\bar{y}}O_{23})^2+(\partial_{\bar{z}}O_{23})^2+(\partial_{\bar{y}}O_{32})^2+(\partial_{\bar{z}}O_{32})^2\\
 \label{str4} &+(\partial_{\bar{y}}O_3)^2+(\partial_{\bar{z}}O_3)^2\Bigr]+\frac{\gamma-1}{2}\Bigl[(\partial_{\bar{y}} O_{2}+\partial_{\bar{z}} O_{23})^2+(\partial_{\bar{y}} O_{32}+\partial_{\bar{z}} O_{3})^2\Bigr]\\
 \nonumber ={}&\frac{1}{2}\Bigl[(\partial_{\bar{y}}O_1)^2+(\partial_{\bar{z}}O_1)^2+\gamma (\partial_{\bar{y}}O_2)^2+(\partial_{\bar{z}}O_2)^2+(\partial_{\bar{y}}O_{23})^2+\gamma(\partial_{\bar{z}}O_{23})^2+\gamma(\partial_{\bar{y}}O_{32})^2+(\partial_{\bar{z}}O_{32})^2\\
 \label{str5} &+(\partial_{\bar{y}}O_3)^2+\gamma(\partial_{\bar{z}}O_3)^2\Bigr]+(\gamma-1)\Bigl[(\partial_{\bar{y}}O_2)(\partial_{\bar{z}}O_{23})+(\partial_{\bar{y}}O_{32})(\partial_{\bar{z}}O_3)\Bigr]
\end{align}
and
\begin{align}
\label{str6} \bar{\text{f}}_{\rm S,pot} ={}& -\frac{1}{2}\mbox{tr}(OO^T) + \frac{\zeta_{12}}{4}[\mbox{tr}(OO^T)]^2+\frac{\zeta_{345}}{4}\mbox{tr}(OO^TOO^T)\\
 \nonumber ={}&-\frac{1}{2}(O_1^2+O_2^2+O_{23}^2+O_{32}^2+O_3^2)+\frac{\zeta_{12}}{4}(O_1^2+O_2^2+O_{23}^2+O_{32}^2+O_3^2)^2\\
 \label{str7} &+\frac{\zeta_{345}}{4}\Bigl[O_1^4+O_2^4+O_{23}^4+O_{32}^4+O_3^4+2(O_2^2+O_3^2)(O_{23}^2+O_{32}^2)+4O_2O_{23}O_{32}O_3\Bigr].
\end{align}
\end{widetext}
In all of these expression, we have $\zeta_{345}=1-3\zeta_{12}$ due to Eq. (\ref{real8}). Note that the parameter $\gamma>1$, through the term in the first square brackets in Eq. (\ref{str5}), leads to an increased kinetic energy cost for modulations of the third column of the order parameter in the z-direction, and modulations of the second column in the y-direction. However, through the term in the second square brackets multiplying $(\gamma-1)$, the mutual interplay of these modulations can actually reduce the kinetic energy, as is explained in the main text.

\begin{figure}[t!]
\centering
\includegraphics[width=8.6cm]{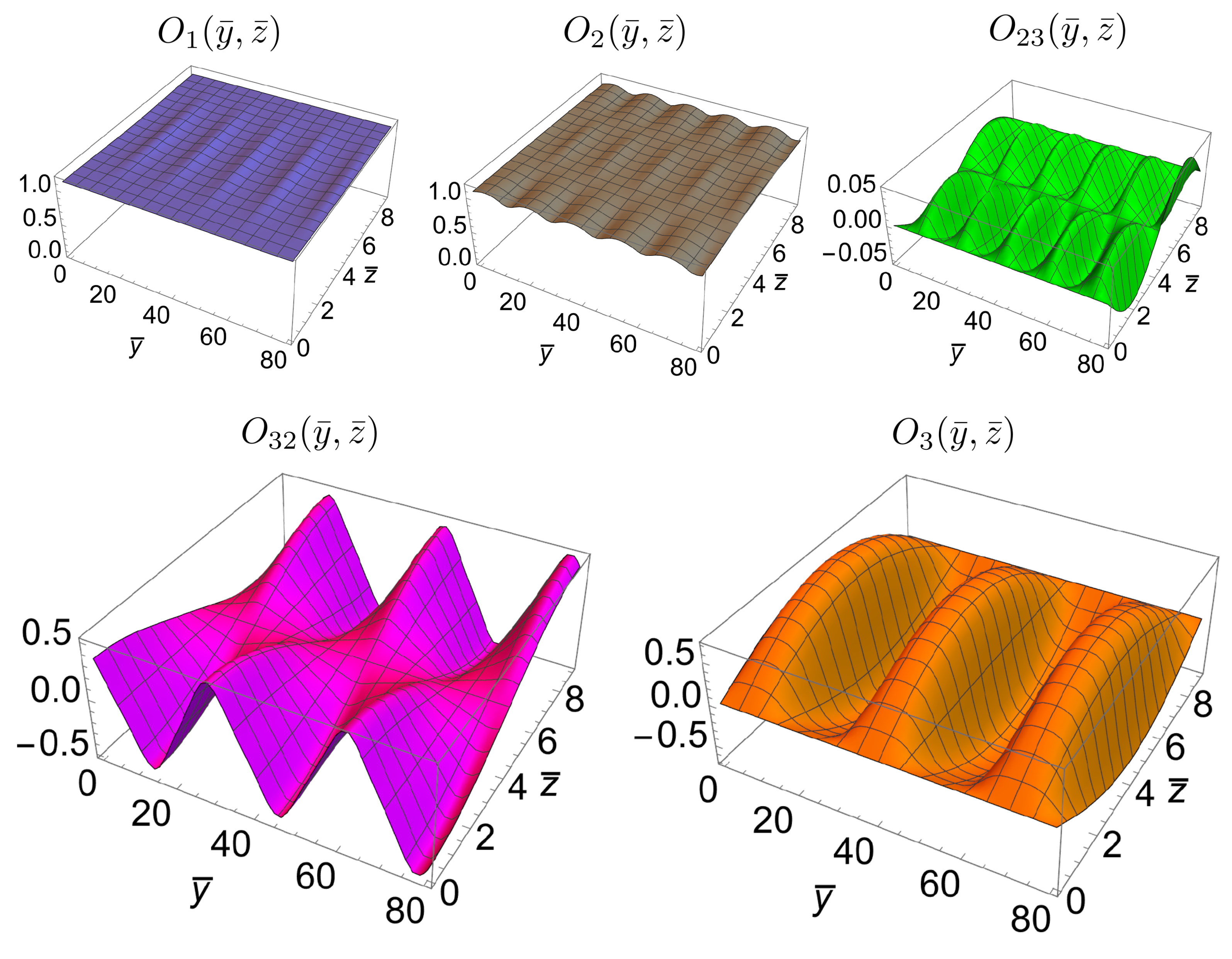}
\caption{Stripe solution to the equations of motion (\ref{str2}) obtained numerically with the finite-element method. We choose weak-coupling parameters $\zeta_{12}=0.2$ and $\gamma=3$. The simulation box is $(\bar{y},\bar{z})\in(\bar{L}_{\rm y},\bar{D})$ with $\bar{D}=9$, $\bar{L}_{\rm y}=L_{\rm y}/\xi=80$, and specular boundary conditions in both y- and z-direction. Note the following characteristics: (i) The five components approximately follow the form in Eq. (\ref{spec12}), with $O_1$ and $O_2$ close to unity and $O_{23}\approx 0$. (ii) The components $O_{23}$ and $O_3$ oscillate in an opposite manner. In particular, if $O_3$ vanishes at the wall at a position $(\bar{y},\bar{z}=0)$, say, then $O_{23}$ is maximal that position, and vice versa. In this manner, the kinetic free energy can be decreased. However, mechanism fails for maximally pair-breaking boundaries, where both $O_{23}$ and $O_3$ need to vanish at the surfaces. Hence the stripe solution does not exist for maximally pair-breaking boundary conditions.}
\label{FigAppInhom}
\end{figure}

We numerically solve the equations of motion (\ref{str2}) using the finite-element method for either maximally pair-breaking, specular, or diffusive (with $b_{\rm T}=0.5$) boundaries at $\bar{z}=0$ and $\bar{z}=\bar{D}$. We confine the system in the y-direction to a box of sufficiently large size $\bar{L}_{\rm y}\gg\bar{D}$ and choose specular boundary conditions in the y-direction. An example stripe solution for the five components of the order parameters in dependence of $\bar{z}$ and $\bar{y}$ is shown in Fig. \ref{FigAppInhom}.

\begin{figure*}[t!]
\centering
\includegraphics[width=17.5cm]{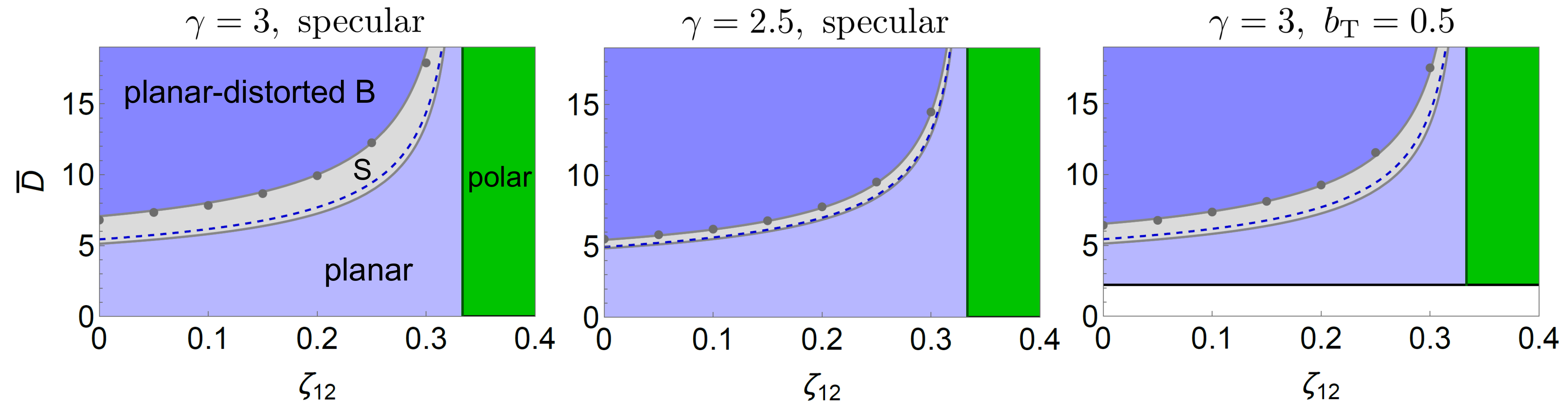}
\caption{Phase diagram of real order parameters including the S-phase. The left and central plot are for specular boundary conditions ($b_{\rm T}=\infty$) in the z-direction, whereas the right plot shows the result for $b_{\rm T}=0.5$. Furthermore, we vary the parameter $\gamma$ as indicated. The curve $\bar{D}_{\rm P}(\zeta_{12})=\pi \sqrt{\frac{\gamma(1-\zeta_{12})}{1-3\zeta_{12}}}$ is shown by the dashed-blue line. We observe that the S-phase is restricted to a narrow strip $\bar{D}_{\rm S1}<\bar{D}<\bar{D}_{\rm S2}$. Solving the equations of motion (\ref{str2}) numerically, we confirm excellent agreement of the lower boundary with the formula $\bar{D}_{\rm S1} = \frac{2\sqrt{\gamma-1}}{\gamma}\bar{D}_{\rm P} < \bar{D}_{\rm P}$ (lower gray line). The gray circles show transition points determined numerically along the upper transition line. We find that the numerical result is well-captured by the approximate formula $\bar{D}_{\rm S2} \simeq(1.3, 1.1, 1.2)\bar{D}_{\rm P}$ for the left, central, and right plot, respectively (upper gray line). The stripe solution does not exist for maximally pair-breaking boundary conditions. This is further corroborated by the fact that the region of stability becomes very narrow as $b_{\rm T}\to 0$. We also see that only a sufficiently large $\gamma>2$ can stabilize the stripe phase for specular boundaries.}
\label{FigAppRealSpecular}
\end{figure*}

By comparing the free energy $\bar{F}[O_{\rm S}(\bar{y},\bar{z})]$ from Eq. (\ref{str2b}) with the free energies of the competing real orders, we obtain the phase diagrams for real order shown in Fig. \ref{FigAppRealSpecular}. We observe that the onset of the S-phase preempts the P-pdB-transition at a critical value $\bar{D}_{\rm S1}<\bar{D}_{\rm P}$. However, the S-phase is restricted to a narrow strip of values $\bar{D}_{\rm S1}<\bar{D}<\bar{D}_{\rm S2}$. The strip becomes thinner as $\gamma\to2$ or $b_{\rm T}\to 0$. For maximally pair-breaking boundary conditions, we did not find a stripe-solution to the equations of motion. For $\bar{D}\to \infty$, the stability region of the S-phase shrinks to a point (similar to the P-phase), so that the S-phase does not appear in the bulk 3D phase diagram.

\section{Derivation of Eqs. (\ref{spec15}) and (\ref{spec16})}\label{AppDS1}

We repeat the calculation of Ref. \cite{wiman2016strong} to derive Eq. (\ref{spec15}) for the behavior of the S-phase close to $\bar{D}_{\rm S1}$ under specular boundary conditions. This also leads to the analytical expression for $\bar{D}_{\rm S1}$ in Eq. (\ref{spec16}). (While this is technically only an upper bound for $\bar{D}_{\rm S1}$, we find that, within the numerical errors, it agrees very well with the value determined by the numerical solution of the equations of motion (\ref{str2}), and we will assume that both expressions for $\bar{D}_{\rm S1}$ are equal.)

We start from the variational ansatz
\begin{align}
 O_1(\bar{y},\bar{z}) &= O_2(\bar{y},\bar{z}) = O_1,\\
 O_{23}(\bar{y},\bar{z}) &=0,\\
 O_{32}(\bar{y},\bar{z}) &=\mathcal{A}_{32} \cos\Bigl(\frac{\pi \bar{y}}{\bar{L}_{\rm S}}\Bigr) \cos\Bigl(\frac{\pi \bar{z}}{\bar{D}}\Bigr),\\
 O_{3}(\bar{y},\bar{z}) &=\mathcal{A}_{3} \sin\Bigl(\frac{\pi \bar{y}}{\bar{L}_{\rm S}}\Bigr) \sin\Bigl(\frac{\pi \bar{z}}{\bar{D}}\Bigr).
\end{align}
We assume that the amplitudes $\mathcal{A}_{32},\mathcal{A}_3>0$ are small compared to $O_1>0$. The free energy density $\bar{\text{f}}_{\rm S}=\bar{\text{f}}_{\rm S,kin}+\bar{\text{f}}_{\rm S,pot}$ for this order parameter reads
\begin{align}
 \nonumber \bar{\text{f}}_{\rm S} &= \frac{1}{2}\Bigl[\gamma \frac{\pi^2}{\bar{L}_{\rm S}^2}O_{32}^2+\frac{\pi^2}{\bar{D}^2}O_{32}^2 + \frac{\pi^2}{\bar{L}_{\rm S}^2}O_3^2+\gamma \frac{\pi^2}{\bar{D}^2}O_3^2\Bigr]\\
 \nonumber &-(\gamma-1) \frac{\pi^2}{\bar{L}_{\rm S}\bar{D}}\mathcal{A}_{32} \mathcal{A}_{3} \sin^2\Bigl(\frac{\pi \bar{y}}{\bar{L}_{\rm S}}\Bigr) \cos^2\Bigl(\frac{\pi \bar{z}}{\bar{D}}\Bigr)\\
 \nonumber &-\frac{1}{2}(2O_1^2+O_{32}^2+O_3^2) + \frac{\zeta_{12}}{4}(2O_1^2+O_{32}^2+O_3^2)^2\\
 &+\frac{1-3\zeta_{12}}{4}\Bigl[ 2O_1^4+2O_1^2O_{32}^2+(O_{32}^2+O_3^2)^2\Bigr].
\end{align}
Performing the $\bar{y}$- and $\bar{z}$-integrations, we arrive at
\begin{align}
 \nonumber \bar{F}_{\rm S} &= \frac{\bar{D}}{2} \Biggl(\Bigl[\gamma \frac{\pi^2}{\bar{L}_{\rm S}^2}+\frac{\pi^2}{\bar{D}^2}-1+(1-\zeta_{12})O_1^2\Bigr]\frac{1}{4}\mathcal{A}_{32}^2\\
 \nonumber &+ \Bigl[\frac{\pi^2}{\bar{L}_{\rm S}^2}+\gamma \frac{\pi^2}{\bar{D}^2}-1+2\zeta_{12}O_1^2\Bigr]\frac{1}{4}\mathcal{A}_{3}^2\\
 \nonumber &-2(\gamma-1) \frac{\pi^2}{\bar{L}_{\rm S}\bar{D}}\frac{1}{4}\mathcal{A}_{32} \mathcal{A}_{3}-2O_1^2+(1-\zeta_{12})O_1^4\\
 & +\frac{9(1-2\zeta_{12})}{128}(\mathcal{A}_{32}^4+\mathcal{A}_3^4)+\frac{1-2\zeta_{12}}{64}\mathcal{A}_{32}^2\mathcal{A}_3^2\Biggr).
\end{align}
Note that the kinetic term proportional to $\gamma-1$ is negative for $\gamma>1$. Hence the term $(\gamma-1)(\partial_{\bar{y}} O_{32})(\partial_{\bar{z}}O_3)$ is responsible for lowering the free energy. We minimize this expression with respect to $O_1^2$ by solving $\frac{\partial \bar{F}_S}{\partial O_1^2}=0$, which yields
\begin{align}
 O_1^2 &= \frac{1}{1-\zeta_{12}} -\frac{1}{8}\mathcal{A}_{32}^2-\frac{\zeta_{12}}{4(1-\zeta_{12})}\mathcal{A}_3^2.
\end{align}
We insert this value for $O_1^2$ into the free energy and obtain
\begin{align}
 \nonumber &\bar{F}_{\rm S} = \bar{F}_{\rm P} + \frac{\bar{D}}{8}\Biggl( \Bigl[\gamma \frac{\pi^2}{\bar{L}_{\rm S}^2}+\frac{\pi^2}{\bar{D}^2}\Bigr]\mathcal{A}_{32}^2\\
 &+\Bigl[\frac{\pi^2}{\bar{L}_{\rm S}^2}+\gamma \frac{\pi^2}{\bar{D}^2}-\frac{1-3\zeta_{12}}{1-\zeta_{12}}\Bigr]\mathcal{A}_3^2-2(\gamma-1) \frac{\pi^2}{\bar{L}_{\rm S}\bar{D}}\mathcal{A}_{32} \mathcal{A}_{3}\Biggr),
\end{align}
where we dropped terms of quartic order in $\mathcal{A}_3$ and $\mathcal{A}_{32}$, and identified
\begin{align}
\bar{F}_{\rm P} = -\frac{\bar{D}}{2(1-\zeta_{12})}
\end{align}
as the free energy of the planar phase for specular boundary conditions. 

The free energy $\bar{F}_{\rm S}$ depends on three variational parameters, $\mathcal{A}_{32}$, $\mathcal{A}_3$, and $\bar{L}_{\rm S}$.
The planar phase corresponds to $\mathcal{A}_{32}=\mathcal{A}_3=0$. If it is favorable to have nonzero amplitudes, then the stripe phase wins. We introduce the ratios $R=\mathcal{A}_{32}/\mathcal{A}_3$ and $L=\bar{L}_{\rm S}/\bar{D}$ to write 
\begin{align}
 \nonumber \bar{F} &= \bar{F}_{\rm P} - \frac{\bar{D}}{8}\mathcal{A}_{3}^2 \frac{1-3\zeta_{12}}{1-\zeta_{12}}\\
 &+\frac{\pi^2}{8\bar{D}}\mathcal{A}_{3}^2\Bigl(\Bigl[\frac{\gamma}{L^2}+1\Bigr]R^2+\Bigl[\frac{1}{L^2}+\gamma \Bigr]-2(\gamma-1) \frac{R}{L}\Bigr).
\end{align}
To determine the phase boundary between P and S, it is then sufficient to minimize $\bar{F}_{\rm S}$ with respect to the two parameters $R$ and $L$. The real solution to the set of equations
\begin{align}
0 &= \frac{\partial \bar{F}_{\rm S}}{\partial R}\Bigr|_{L} \propto 2\Bigl[\frac{\gamma}{L^2}+1\Bigr]R-2(\gamma-1)\frac{1}{L},\\
 0 &= \frac{\partial \bar{F}_{\rm S}}{\partial L}\Bigr|_{R} \propto \frac{2}{L^3}\Bigl(-1-\gamma R^2+(\gamma-1)RL\Bigr)
\end{align}
exists for $\gamma>2$ and is given by
\begin{align}
 R =\sqrt{\frac{\gamma-2}{\gamma}},\ L = \sqrt{\frac{\gamma}{\gamma-2}},
\end{align}
which proves Eqs. (\ref{spec15}). Inserting these values into the free energy, we obtain
\begin{align}
 \bar{F}_0 &= \bar{F}_{\rm P} +\frac{\bar{D}}{8}\mathcal{A}_{3}^2 \Biggl( -\frac{1-3\zeta_{12}}{1-\zeta_{12}}+\frac{4(\gamma-1)\pi^2}{\gamma \bar{D}^2}\Biggr).
\end{align}
We have $\bar{F}_{\rm S}<\bar{F}_{\rm P}$ when the term in brackets is negative. This happens for
\begin{align}
\bar{D}>\bar{D}_{\rm S1} = 2\pi \sqrt{\frac{\gamma-1}{\gamma} \frac{1-\zeta_{12}}{1-3\zeta_{12}}} = \frac{2\sqrt{\gamma-1}}{\gamma}\bar{D}_{\rm P},
\end{align}
which proves Eq. (\ref{spec16}).

\section{Thermodynamic data}\label{AppThermo}

\begin{figure}[t!]
\centering
\includegraphics[width=8.6cm]{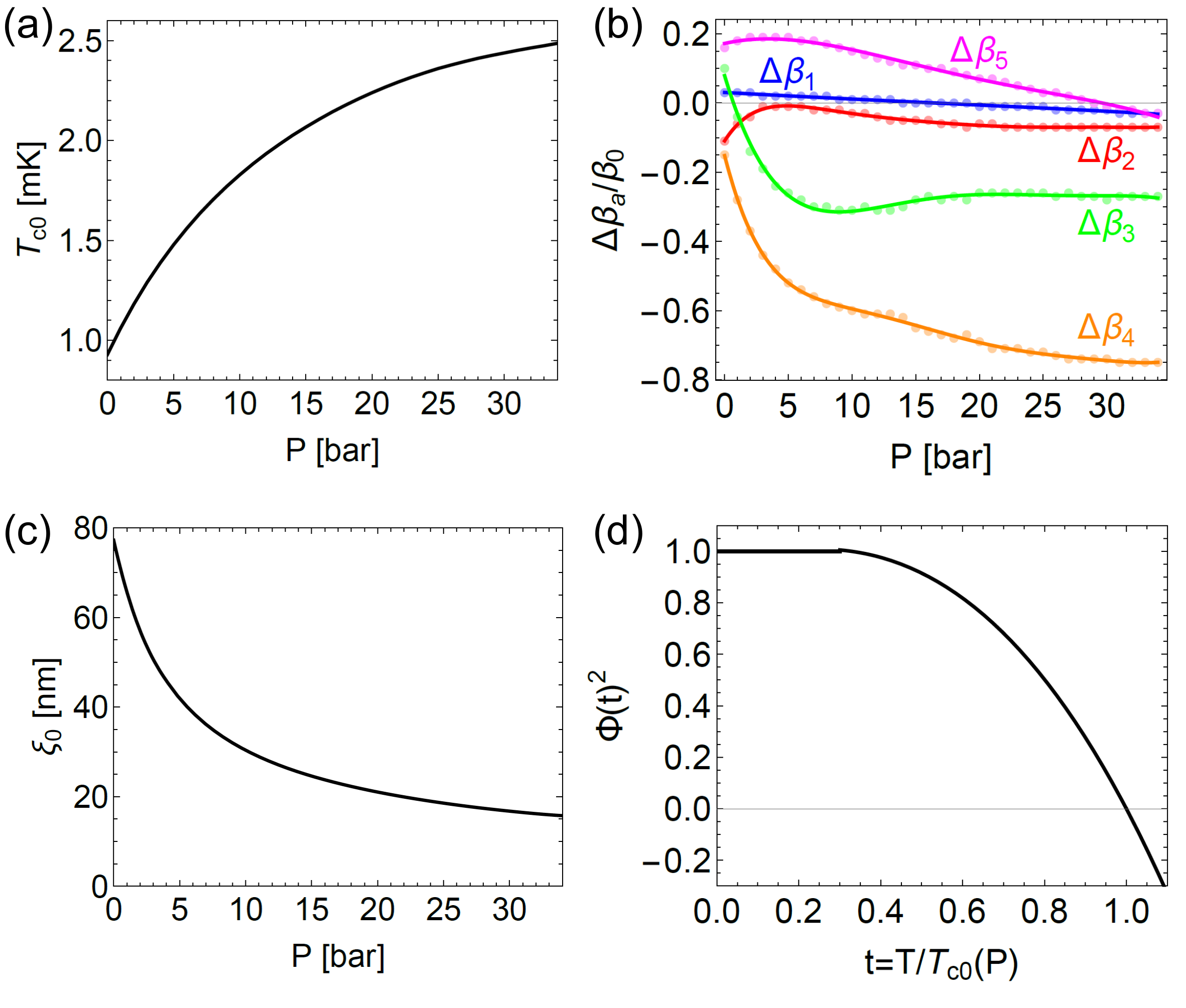}
\caption{Thermodynamic data of $^3$He used to compute the $P$-$T$ phase diagrams. \emph{Panel (a).} The critical temperature in vanishing magnetic field, $T_{\rm c0}(P)$, is obtained from the measurements by Greywall \cite{PhysRevB.33.7520}. \emph{Panel (b).} The strong-coupling coefficients $\beta_a(T,P)$ follow Eq. (\ref{pt9}) with $\Delta \beta_a(P)$ shown here and $\beta_0$ from Eq. (\ref{weak4}). The data for $\Delta\beta_a$ (points) is taken from Ref. \cite{PhysRevB.75.174503}, whereas the continuous lines constitute polynomial interpolations to smoothen the data, as described in the text. \emph{Panel (c).} The zero-temperature correlation length $\xi_0(P)$ enters Eq. (\ref{pt7}) and is approximated here by the data from Table I of Ref. \cite{PhysRevB.92.144515}. \emph{Panel (d).} The function $\Phi(t)$ from Eq. (\ref{pt8}) parametrizes the temperature dependence of $\xi(T,P)$ and thus of $\bar{D}=D/\xi(T,P)$. For $H=0$, it is restricted to the region $t = T/T_{\rm c0}\leq 1$. For $H>0$, we analytically continue $\Phi(t)^2$ into the region $t>1$, where the it is negative and approximately linear.}
\label{FigAppThermo}
\end{figure}

The functions $T_{\rm c0}(P)$ and $\xi_0(P)$ used in the computation of the $P$-$T$ phase diagrams are shown in Fig. \ref{FigAppThermo}, together with the function $\Phi(t)^2$ from Eq. (\ref{pt8}). In order to use the strong-coupling corrections $\Delta \beta_a(P)$ in Eq. (\ref{pt9}) from the values given in Choi et al. \cite{PhysRevB.75.174503}, we apply a polynomial fit to the data points $\Delta \beta_a(P)$ to create a smooth function. For this purpose, since the complexity of the five coefficient varies, we approximate $\Delta\beta_1(P),\Delta\beta_2(P),\Delta\beta_3(P),\Delta\beta_4(P),\Delta\beta_5(P)$ by a 4th, 7th, 5th, 7th, 4th order polynomial in $P$, respectively. These orders are chosen empirically to capture the features of the data in a smooth manner while avoiding overfitting. The data points and polynomial fits for the strong-coupling corrections are shown in Fig. \ref{FigAppThermo}b.

The parameter $\alpha_0(T)$ in weak-coupling or BCS theory is given by \cite{PhysRevB.108.144503}
\begin{align}
 \label{alpha1} \alpha_{0}(T) &\simeq \frac{1}{u} - \frac{N(0)}{3} \int_{-\Lambda}^\Lambda \mbox{d}\xi \frac{1-2n_{\rm F}(\xi)}{2\xi}\\
 \label{alpha2} &\simeq \frac{1}{u} - \frac{N(0)}{3} \log\Bigl(\frac{2 e^{\gamma_{\rm E}}}{\pi} \frac{\Lambda}{T}\Bigr)\\
 \label{alpha3}  &= \frac{N(0)}{3} \log\Bigl(\frac{T_{\rm c0}}{T}\Bigr),
\end{align}
with microscopic coupling $u>0$, ultraviolet cutoff $\Lambda$, $\xi=\frac{k^2}{2m^*}-\mu$, $n_{\rm F}(\xi)=(e^{\xi/T}+1)^{-1}$, $\gamma_{\rm E}$ Euler's constant, and weak-coupling critical temperature $T_{\rm c0} = \frac{2e^\gamma\Lambda}{\pi} e^{-3/[N(0)u]}$. For $T$ below and close to $T_{\rm c0}$, we find
\begin{align}
 \label{alpha4} \alpha_{0}(T) &\simeq \frac{N(0)}{3}\Biggl[\Bigl(1-\frac{T}{T_{\rm c0}}\Bigr)+\frac{1}{2}\Bigl(1-\frac{T}{T_{\rm c0}}\Bigr)^2+\dots\Biggr].
\end{align}
The  expression in Eq. (\ref{alpha3}) diverges for $T\to 0$, which expresses the fact that temperature $T>0$ is an infrared regulator for the logarithmic singularity of the integral on the Fermi surface ($\xi=0$ or $k=k_{\rm F}$). This does not mean, however, that the free energy is divergent at low temperatures. Instead, the system is superfluid at low temperatures and the superfluid gap regularizes the divergence. Indeed, defining $\alpha_0$ in the superfluid regime as the coefficient of $-\mbox{tr}(\delta A^\dagger\delta A)$ in an expansion of $\text{f}_{\rm pot}(A)$ around the global minimum at $A_0$ via $A=A_0+\delta A$, the expression in Eq. (\ref{alpha1}) assuming $A_0=0$ cannot be applied for $T\to 0$. To qualitatively estimate the so-defined $\alpha_0(T)$ for low temperatures, we can replace $\xi \to \sqrt{\xi^2+|\Delta(0)|^2}$ in Eq. (\ref{alpha1}), with $|\Delta(0)|$ the zero temperature gap. The resulting function $\alpha_0(T)$ remains finite as $T\to 0$. This highlights again that our free energy in Eq. (\ref{gl1b}), which relies on an expansion around $A_0=0$, is only valid close to $T_{\rm c}$, whereas applying it to low temperatures $T\to 0$ is an extrapolation.

\begin{figure}[t!]
\centering
\includegraphics[width=8.6cm]{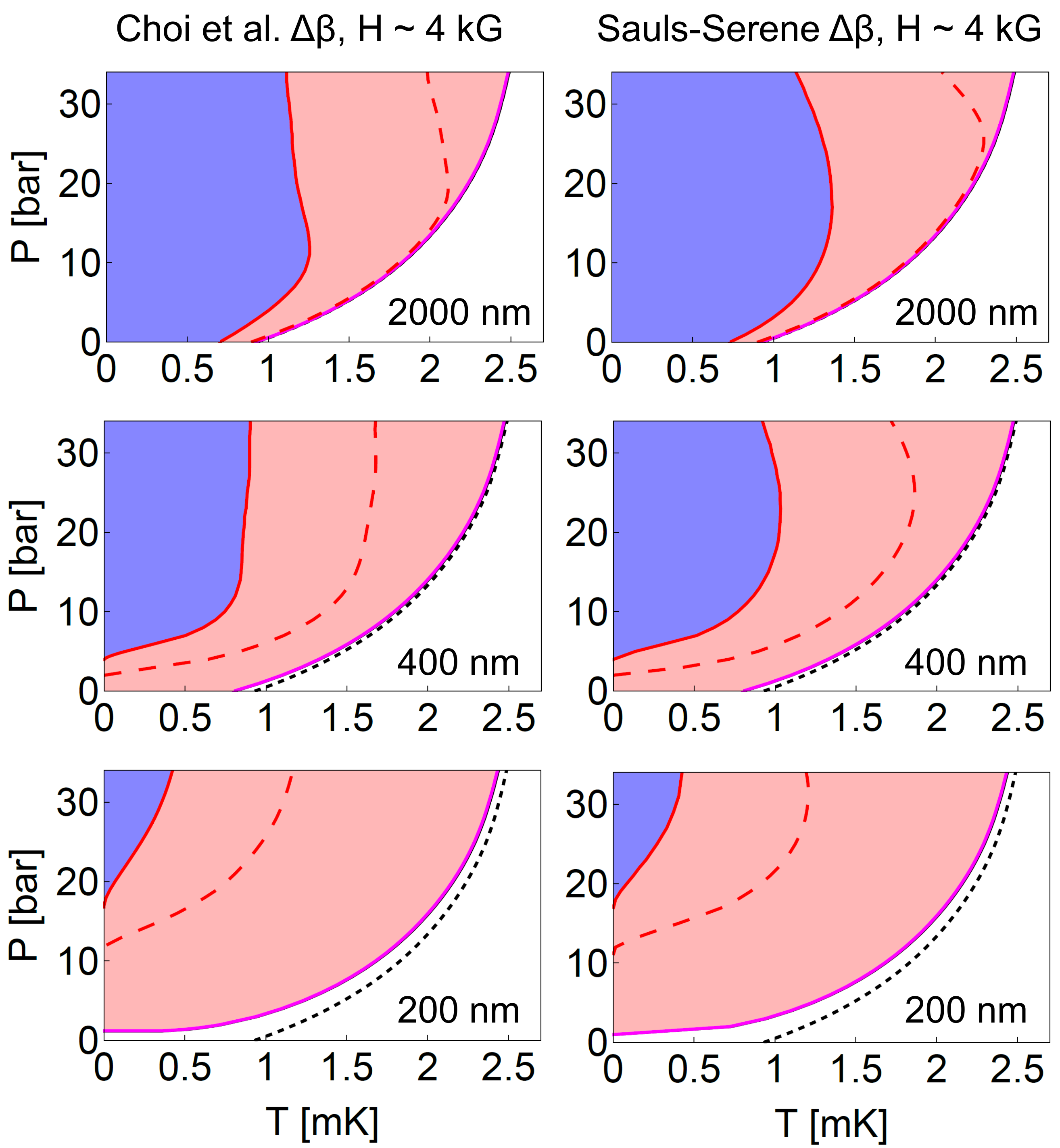}
\caption{Comparison of strong-coupling corrections $\Delta \beta_a$ to quartic coefficients $\beta_a$ in Eq. (\ref{pt9}), measured by Choi et al. \cite{PhysRevB.75.174503} (see Fig. \ref{FigAppThermo}, used in the maintext) and calculated by Sauls--Serene from quasiparticle potential scattering \cite{PhysRevB.24.183} (implemented as in Ref. \cite{PhysRevB.92.144515}).  We show the $P$-$T$ phase diagram for maximally pair-breaking boundaries in a finite magnetic field $H\sim 4\ \text{kG}$ as in Fig. \ref{FigPhasePTH}, with the zero-field AB transition indicated by the dashed red line. The quartic coefficients differ only at $T>0$, so that the resulting $T=0$ quantum phase diagram in both cases coincides with the weak-coupling result shown in \ref{FigQPD}.}
\label{FigSaulsSerene}
\end{figure}

A theoretical determination of the functions $\Delta \beta_a(P)$ that enter Eq. (\ref{pt9}) has been given by Sauls and Serene using quasiparticle potential-scattering models for $P\geq12\ \text{bar}$ \cite{PhysRevB.24.183}. To compare the difference in choice of quartic coefficients $\beta_a(P,T)$ on the phase diagram, we use here the implementation of the Sauls--Serene coefficients in Tables III and IV of Ref. \cite{PhysRevB.92.144515}, where the results are extrapolated to all $P\geq 0$ such that the functions $\Delta \beta_a(P)$ vanish at $-5\ \text{bar}$. The resulting $P$-$T$ phase diagram for $H\sim 4\ \text{kG}$ is shown in Fig. \ref{FigSaulsSerene} for both Choi et al. and Sauls--Serene quartic coefficients, together with the zero-field AB transition line (red dashed). While some quantitative difference are visible, the overall phase structure is robust and not affected by the choice of quartic coefficients.

By construction of Eq. (\ref{pt9}), both the strong-coupling corrections from Choi et al. and Sauls--Serene vanish at $T=0$. As a result, applying either choice results in the $T=0$ quantum phase diagram predicted by weak-coupling theory, shown in Fig. \ref{FigQPD}. Phenomenological improvements of the quartic coefficients in the zero-temperature regime have recently been proposed in Refs. \cite{zhelev2016observation,PhysRevLett.134.136001}. However, the Taylor expansion of the free energy in powers of the order parameter $A$ breaks down in this regime and a more sophisticated treatment of the functional is required, as outlined in Sec. \ref{SecConcl}.

\section{Coherence length}\label{AppCoherence}

In this appendix, we clarify how our definition of the coherence length compares to other conventions used in the literature. The magnetic field does not affect this discussion, so we set $H=0$ in the remainder of this section. We define the coherence length $\xi$ through Eq. (\ref{gl2}) as
\begin{align}
 \xi = \sqrt{\frac{K}{\alpha_0}}.
\end{align}
In the weak-coupling limit, $K$ and $\alpha_0$ are given by \cite{vollhardt2013superfluid}
\begin{align}
 K_{\rm wc} &= \frac{7 \zeta(3)}{20} \frac{N(0)}{3} \xi_{0,\rm wc}^2,\\
\alpha_{0,\rm wc} &= \frac{N(0)}{3} \Bigl(1-\frac{T}{T_{\rm c0}}\Bigr),
\end{align}
with $\xi_{0,\rm wc} = \frac{\hbar v_{\rm F}}{2\pi k_{\rm B}T_{\rm c0}}$ the weak-coupling prediction for the coherence length at zero temperature. The behavior of $\xi_0$ is shown in Fig. \ref{FigAppThermo}, taking values approximately between 70 nm at $P=0\ \text{bar}$ to 20 nm at $P=34\ \text{bar}$ \cite{PhysRevB.92.144515}. Inserting the weak-coupling values for $K$ and $\alpha_0$, we obtain
\begin{align}
\xi_{\rm wc} = \frac{\xi_0}{\sqrt{1-T/T_{\rm c0}}}.
\end{align} 
This equation should be compared to Eq. (\ref{pt7}), given by
\begin{align}
 \xi = \frac{\xi_0}{\Phi(t)},
\end{align}
where $\Phi(t)$ models a more realistic temperature dependence of the coherence length compared to the weak-coupling form $\sqrt{1-t}$, see Fig. \ref{FigAppThermo}. This reflects the fact that $K$ and $\alpha_0$ deviate from $K_{\rm wc}$ and $\alpha_{0,\rm wc}$. The explicit form of the weak-coupling value of $\xi$ is often referred to as Ginzburg--Landau coherence length $\xi_{\rm GL}$ in the literature (see e.g. \cite{wiman2016strong,PhysRevB.104.094520}), namely
\begin{align}
 \xi_{\rm GL} = \xi_{\rm wc}.
\end{align}

In the definition of the diffusive boundary conditions in Eqs. (\ref{diffBC}), we define the modifying denominator as $b_T \xi$. A more common definition evaluates the denominator as $b_T\xi_0$, see for instance Refs. \cite{PhysRevB.92.144515,wiman2016strong}. With our choice of $\Phi(t)$, both definitions agree for $T\leq 0.3 T_{\rm c0}$. At high temperatures, where $\xi > \xi_0$, our choice with $b_T\xi$ is effectively more specular than the definition employing $b_T\xi_0$. This, however, does not change the effect clearly visible in the phase diagrams in Figs. \ref{FigPDCrossover} and \ref{FigQPD} that the diffusive limit interpolates between the maximally pair-breaking and specular boundary conditions.

\section{Relevant order parameters in a magnetic field}\label{AppMagOrder}

In this appendix, we first discuss the relevant order parameters in a magnetic field without utilizing the matrix notation. We then discuss the spin-gap matrix formulation.

It is useful to characterize the orders in a magnetic by the use of two sets of triads, $(\vec{d},\vec{d}',\vec{d}'')$ for the spin part and $(\vec{m},\vec{n},\vec{\ell})$ for the orbital part, which consist of real, mutually orthogonal unit vectors. While they are arbitrary in principle, we define standard triads by
\begin{align}
  \label{mag6} (\vec{d},\vec{d}',\vec{d}'') = (\vec{m},\vec{n},\vec{\ell}) = (\vec{e}_{\rm x},\vec{e}_{\rm y},\vec{e}_{\rm z}),
\end{align}
which facilitate a matrix notation and are in accordance with the energetic constraints imposed by the magnetic field derived below. We also define $\vec{\hat{H}}=\vec{H}/H=\vec{e}_{\rm z}$.

The order parameter of the A$_2$-phase is given by
\begin{align}
 \label{mag7} A^{(\rm A_2)}_{\mu i} = (Bd_\mu-\rmi M d'_\mu)(m_i+\rmi n_i),
\end{align}
where $B>M>0$ as in Eq. (\ref{mag6b}). For this order parameter we have
\begin{align}
 \label{mag8} \text{f}_{\rm mag,A_2}^{(1)} &= -4 g_1 \vec{H}\cdot(\vec{d}\times\vec{d}')BM,\\
 \label{mag9} \text{f}_{\rm mag,A_2}^{(2)} &= 2g_2\Bigl(B^2(\vec{H}\cdot\vec{d})^2+M^2(\vec{H}\cdot\vec{d}')^2\Bigr)
\end{align}
for the linear and quadratic Zeeman terms. To lower the free energy, it is favorable to have both $\vec{d},\vec{d}'$ orthogonal to $\vec{H}$, and $\vec{d}\times \vec{d}'=\hat{\vec{H}}$, such that
\begin{align}
 \label{mag10} \text{f}_{\rm mag,A_2}^{(1)} &= -4  g_1 H BM,\\
 \label{mag11} \text{f}_{\rm mag,A_2}^{(2)} &=0.
\end{align}
All conditions are satisfied by the standard triads and yield the matrix representation in Eq. (\ref{mag12b}).

For the B$_2$-phase, we have
\begin{align}
 \label{mag16} A_{\mu i}^{(\rm B_2)} &=  B(d_\mu m_i+d'_\mu n_i) +\rmi M(d_\mu n_i-d'_\mu m_i) + C d''_\mu \ell_i,
\end{align}
with $B,M>0$ and $C\in\mathbb{R}$ from Eq. (\ref{mag6b}). The pdB-, planar, and polar phases are special cases with $M=0$. We have
\begin{align}
 \label{mag18} \text{f}_{\rm mag,B_2}^{(1)} &=-4 g_1  \vec{H}\cdot(\vec{d}\times \vec{d}')BM,\\
 \nonumber \text{f}_{\rm mag,B_2}^{(2)} &= g_2 \Bigl[(B^2+M^2) \Bigl((\vec{H}\cdot\vec{d})^2+(\vec{H}\cdot\vec{d}')^2\Bigr)\\
 \label{mag19} &+C^2(\vec{H}\cdot\vec{d}'')^2\Bigr].
\end{align}
Again, the quadratic Zeeman term is optimized by having both $\vec{d},\vec{d}'$ orthogonal to $\vec{H}$ so that $\vec{d}''= \pm\hat{\vec{H}}$, and the linear Zeeman term becomes negative for $\vec{d}\times\vec{d}'=\vec{\hat{H}}$. With these constraints we find
\begin{align}
 \label{mag20} \text{f}_{\rm mag,B_2}^{(1)} &=-4 g_1 H BM,\\
 \nonumber \text{f}_{\rm mag,B_2}^{(2)} &= g_2 H^2 C^2 .
\end{align}
Again, the standard triads satisfy these conditions. The associated matrix representation of the order parameter is given in Eq. (\ref{mag21}).

It is instructive to define the momentum-dependent spin gap matrix \cite{vollhardt2013superfluid} through
\begin{align}
 \label{mag25} \hat{\Delta}(\vec{k}) = \begin{pmatrix} \Delta_{\up}(\hat{\vec{k}}) & \Delta_0(\hat{\vec{k}}) \\ \Delta_0(\hat{\vec{k}})^* & \Delta_\down(\hat{\vec{k}}) \end{pmatrix} = A_{\mu i} \sigma_\mu (\rmi \sigma_2) \hat{k}_i,
\end{align}
with normalized momentum $\hat{\vec{k}} = \vec{k}/k_{\rm F}$ and $\sigma_\mu$ the Pauli matrices. We then have
\begin{equation}\label{mag26}
\begin{aligned}
 \hat{\Delta}_{\rm A_2}(\vec{k}) &= \begin{pmatrix} -(B-M) & 0 \\ 0 & B+M\end{pmatrix}(\hat{k}_1+\rmi \hat{k}_2)\\
  &= \begin{pmatrix} -\Delta_{\up\up} & 0 \\ 0 & \Delta_{\down\down}\end{pmatrix}(\hat{k}_1+\rmi \hat{k}_2)
\end{aligned}
\end{equation}
for the A$_2$-phase, and
\begin{align}
 \nonumber \hat{\Delta}_{\rm B_2}(\vec{k}) &= \begin{pmatrix} -(B-M)(\hat{k}_1-\rmi \hat{k}_2) & C\hat{k}_3 \\ C\hat{k}_3 & (B+M)(\hat{k}_1+\rmi \hat{k}_2)\end{pmatrix}\\
 \label{mag27} &= \begin{pmatrix} -\Delta_{\up\up}(\hat{k}_1-\rmi \hat{k}_2) & \Delta_{\up\down}\hat{k}_3 \\ \Delta_{\up\down}\hat{k}_3 & \Delta_{\down\down}(\hat{k}_1+\rmi \hat{k}_2)\end{pmatrix}
\end{align}
for the B$_2$-phase. The quadratic part of the free energy density is given by
\begin{align}
 \nonumber \text{f}_2 ={}& -\frac{3}{2}\alpha_0 \langle \mbox{tr}(\hat{\Delta}\hat{\Delta}^\dagger)\rangle +\frac{3}{2} g_1 H \langle \mbox{tr}(\sigma_3 \hat{\Delta}\hat{\Delta}^\dagger)\rangle\\
 \label{mag28} &+3g_2H^2\langle|\Delta_0|^2\rangle\\
 \nonumber ={}& -\frac{3}{2}(\alpha_0+g_1 H) \langle |\Delta_\down|^2\rangle -\frac{3}{2}(\alpha_0-g_1 H)\langle|\Delta_\up|^2\rangle \\
 \label{mag29} &-3 (\alpha_0 -g_2 H^2)\langle|\Delta_0|^2\rangle,
\end{align}
where $\langle \dots \rangle = \int\frac{d\Omega}{4\pi}(\dots)$ denotes the angular average over $\hat{\vec{k}}$. This confirms that $g_1 H>0$ favors pairing of spin-down atoms. We also see that a fully polarized superfluid of spin-down atoms, i.e. the A$_1$-phase, can form even when $\alpha_0>-g_1 H$ is slightly negative.

\section{Bulk free energies in a magnetic field}\label{AppMag}

In this appendix, we derive Eqs. (\ref{mbulk3}) - (\ref{mbulk5}) for the free energy densities in the presence of a magnetic field.

We begin by considering the A$_2$-phase and its limitting cases, with the order parameter parametrized in Eq. (\ref{mag12b}). We assume that $B,M\in\mathbb{C}$ are arbitrary complex numbers and  arrive at the conclusion that $B>M\geq0$ can be chosen real. Inserting the A$_2$-ansatz into the free energy functional (\ref{gl1}) with the magnetic terms from Eq. (\ref{mag4}), we obtain the free energy density
\begin{align}
 \label{fs1} \text{f} &= -\alpha_0 \delta + \beta_{245} \delta^2 -\beta_5 p^2 -g_1 H p
\end{align}
with
\begin{align}
 \label{fs2} \delta &= 2\Bigl(|B|^2+|M|^2\Bigr),\\
 \label{fs3} p&= 2(BM^*+B^*M) = 4\ \text{Re}(BM).
\end{align}
The equations of motion read
\begin{align}
 \label{fs4} 0 = \begin{pmatrix} \partial_\delta \text{f} \\ \partial_p \text{f} \end{pmatrix} = \begin{pmatrix} - \alpha_0 + 2\beta_{245}\delta \\ -2\beta_5 p - g_1 H \end{pmatrix}.
\end{align}
Assuming $\beta_{245}>0$ and $\beta_5<0$, we observe that a nontrivial solution to the first equation with $\delta>0$ only exists for $\alpha_0>0$. The second equation can always be solved. The nontrivial solution is given by
\begin{align}
 \label{fs5} \delta &= \frac{\alpha_0}{2\beta_{245}},\ p = \frac{-g_1 H}{2\beta_5}.
\end{align}
The Hessian matrix evaluated at the solution,
\begin{align}
 \label{fs6} \text{Hess} = \begin{pmatrix} \partial_\delta^2 \text{f} & \partial_\delta\partial_p \text{f} \\ \partial_\delta \partial_p \text{f} & \partial_p^2 \text{f} \end{pmatrix} = \begin{pmatrix} 2\beta_{245} & 0 \\ 0 & -2\beta_5 \end{pmatrix},
\end{align}
is positive definite, thus the solution is a free energy minimum. The condition $\alpha_0>0$ for the solution to exist underestimates the critical value of $\alpha_{\rm cA}$, because Eqs. (\ref{fs2}) and (\ref{fs3}) put additional bounds on the possible values of $\delta$ and $p$. Indeed, for any two complex numbers $B,M$ we have
\begin{align}
 0 \leq 2|B\pm M|^2 = \delta \pm p,
\end{align}
hence $\delta\geq |p|$. We thus infer that Eqs. (\ref{fs5}) are only applicable for
\begin{align}
 \alpha_0> \alpha_{\rm cA} = \frac{\beta_{245}}{|\beta_5|} g_1H.
\end{align}
Inserting the solution into Eq. (\ref{fs1}) we then arrive at 
\begin{align}
 \label{fs7} \text{f}_{\rm A_2} = -\frac{\alpha_0^2}{4\beta_{245}}+\frac{(g_1H)^2}{4\beta_5}
\end{align}
for $\alpha_0>\alpha_{\rm cA}$, which is Eq. (\ref{mbulk4}). Choosing the overall phase of the order parameter such that $B>0$, we can satisfy Eqs. (\ref{fs5}) with a real $M>0$.

Let us comment on other choices than $M\in\mathbb{R}$ for the A$_2$-phase. The sign of $p$ is determined by the sign of $-g_1H/\beta_5$, which is positive in our convention, but changing the direction of the magnetic field would change the sign of $p$. The optimal value for $p$ requires a complex $M$ with a nonzero real part so that $p=4\ \text{Re}(BM)\neq 0$. Indeed, if $M$ were chosen entirely imaginary, say $M=\rmi m$ with $m\in\mathbb{R}$, then $p=0$. The best possible free energy density from Eq. (\ref{fs1}) with $p=0$ is
\begin{align}
 \label{fs8} \text{f}_{\rm A} = -\frac{\alpha_0^2}{4\beta_{245}}
\end{align}
for $\alpha_0>0$, which is the free energy density of the A-phase. In fact, the A$_2$ order parameter with $B=\Delta\cos\alpha>0$ positive and $M=\rmi B \sin \alpha$ imaginary is identical to the A-state, because
\begin{align}
 \nonumber \begin{pmatrix} B & \rmi B & 0 \\ -\rmi M  & M & 0 \\ 0 & 0 & 0 \end{pmatrix} &= \Delta \begin{pmatrix} \cos\alpha & \rmi \cos\alpha & 0 \\ \sin \alpha & \rmi \sin\alpha & 0 \\ 0 & 0 & 0 \end{pmatrix} \\
 \label{fs9} &= \Delta\begin{pmatrix} \cos \alpha & -\sin \alpha & 0 \\ \sin \alpha & \cos\alpha & 0 \\ 0 & 0 & 0 \end{pmatrix} \begin{pmatrix} 1 & \rmi & 0 \\ 0 & 0 & 0 \\ 0 & 0 & 0 \end{pmatrix}.
\end{align}
The rotation matrix is from $\text{SO}(2)_{\rm S}$, hence within the symmetry group of the free energy in a magnetic field. Equation (\ref{fs8}) also shows that the A-phase is always disfavored compared to the A$_2$-phase in a magnetic field.

In the A$_1$-phase with $M=B$ (thus $\delta=p=4|B|^2$), the free energy density simplifies to 
\begin{align}
 \label{fs10} \text{f} = -4(\alpha_0+g_1H)|B|^2+16\beta_{24}|B|^4.
\end{align}
Assuming $\beta_{24}>0$, this is minimized by
\begin{align}
 \label{fs11} |B|^2 = \frac{\alpha_0+g_1H}{8\beta_{24}}
\end{align}
for $\alpha_0+g_1H>0$, with the minimal free energy density
\begin{align}
 \label{fs12} \text{f}_{\rm A_1} = -\frac{(\alpha_0+g_1 H)^2}{4\beta_{24}},
\end{align}
which is Eq. (\ref{mbulk3}). For concreteness, we can again choose $B>0$.

For the P$_2$-state, we consider the order parameter given in Eq. (\ref{mag23}), which we extend to $B,M\in\mathbb{C}$. Inserting this parametrization into the free energy we obtain
\begin{align}
 \nonumber \text{f} ={}& -\alpha_0 \delta +\frac{1}{2}(2\beta_{12}+\beta_{345})\delta^2 \\
 \label{fs13} &-\frac{1}{2}(2\beta_{12}+\beta_{345}-2\beta_{24})p^2-g_1 Hp,
\end{align}
with $\delta $ and $p$ as in Eqs. (\ref{fs2}) and (\ref{fs3}). The equations of motion are
\begin{align}
 \label{fs14} 0 = \begin{pmatrix} \partial_\delta \text{f} \\ \partial_p \text{f} \end{pmatrix} =\begin{pmatrix} - \alpha_0+(2\beta_{12}+\beta_{345})\delta \\ -(2\beta_{12}+\beta_{345}-2\beta_{24})p-g_1H\end{pmatrix}.
\end{align}
A nontrivial solution exists for $\alpha_0>0$, given by
\begin{align}
 \label{fs15} \delta= \frac{\alpha_0}{2\beta_{12}+\beta_{345}},\ p= \frac{-g_1H}{2\beta_{12}+\beta_{345}-2\beta_{24}}
\end{align} 
with positive definite  Hessian matrix
\begin{align}
 \label{fs16} \text{Hess} = \begin{pmatrix} 2\beta_{12} +\beta_{345} & 0 \\ 0 & -(2\beta_{12}+\beta_{345}-2\beta_{24}) \end{pmatrix}.
\end{align}
Here we assume $\mathcal{C}_2=2\beta_{12}+\beta_{345}-2\beta_{24}<0$ as in Eq. (\ref{mbulk0b}). Note how the situation is analogous to the A$_2$-case, and so we can choose $B,M>0$ positive. Furthermore, the condition $\delta\geq \pm |p|$ yields the critical value
\begin{align}
 \alpha_{\rm cP} = \frac{2\beta_{12}+\beta_{345}}{|2\beta_{12}+\beta_{345}-2\beta_{24}|}g_1H.
\end{align}
The minimal free energy density is given by
\begin{align}
 \label{fs17} \text{f}_{\rm P_2} = -\frac{\alpha_0^2}{2(2\beta_{12}+\beta_{345})}+\frac{(g_1H)^2}{2(2\beta_1+\beta_{35}-\beta_4)}
\end{align}
for $\alpha_0>\alpha_{\rm cP}$, which is Eq. (\ref{mbulk5}). We have
\begin{align}
 \nonumber \alpha_{\rm cP} - \alpha_{\rm cA} &= \frac{2\beta_{12}+\beta_{345}}{|2\beta_{12}+\beta_{345}-2\beta_{24}|} - \frac{\beta_{245}}{|\beta_5|}\\
 \nonumber &= -\frac{2\beta_{12}+\beta_{345}}{2\beta_{12}+\beta_{345}-2\beta_{24}} + \frac{\beta_{245}}{\beta_5} \\
 \label{fs17b} &= -\frac{1-\zeta_{12}}{1-\zeta_{12}-2\zeta_{24}} + \frac{\zeta_{245}}{\zeta_5}.
\end{align}
Inserting Eqs. (\ref{mbulk0}) and (\ref{mbulk0b}), we find
\begin{align}
 \label{fs17c} \alpha_{\rm cP} - \alpha_{\rm cA} = \frac{\beta_{24}\mathcal{C}_1}{\beta_5\mathcal{C}_2}>0,
\end{align}
which yields Eq. (\ref{mbulk6e}).

In the bulk system, the free energy density $\text{f}=F/(L_{\rm x}L_{\rm y}D)$ is related to the dimensionless free energy $\bar{F}$ through 
\begin{align}
 \label{fs18} \bar{F} &= (3\beta_{12}+\beta_{345})\frac{\bar{D}}{\alpha_H^2}\text{f}.
\end{align}
(Here $\bar{D}$ is assumed to be large but finite, like $L_{\rm x}$ and $L_{\rm y}$.) For the A$_1$-phase we have
\begin{align}
 \label{fs19} \bar{F}_{\rm A_1} &= -\frac{1}{4\zeta_{24}}\bar{D}
\end{align}
for $\alpha_H>0$ or, equivalently, $h<\infty$. For the A$_2$-phase we have
\begin{align}
 \label{fs20} \text{f}_{\rm A_2} &= -\frac{\alpha_H^2}{4\beta_{245}}\Bigl(1-h-\frac{\zeta_{24}}{4\zeta_5}h^2\Bigr),\\
 \label{fs21} \bar{F}_{\rm A_2} &= - \frac{1}{4\zeta_{245}}\Bigl(1-h-\frac{\zeta_{24}}{4\zeta_5}h^2\Bigr)\bar{D}
\end{align} 
for $h<h_{\rm cA}$. For the P$_2$-state we have
\begin{align}
 \label{fs22} \text{f}_{\rm P_2} &= -\frac{\alpha_H^2}{2(2\beta_{12}+\beta_{345})}\Bigl(1-h-\frac{\zeta_{24}}{2(1-\zeta_{12}-2\zeta_{24})}h^2\Bigr),\\
  \label{fs23} \bar{F}_{\rm P_2} &= - \frac{1}{2(1-\zeta_{12})}\Bigl(1-h-\frac{\zeta_{24}}{2(1-\zeta_{12}-2\zeta_{24})}h^2\Bigr)\bar{D}
\end{align}
for $h<h_{\rm cP}$. Expressions (\ref{fs21}) and (\ref{fs23}) agree with the specular-boundary expression from Eq. (\ref{thin16}) upon inserting the values of $h_{\rm cA}$ and $h_{\rm cP}$ from Eqs. (\ref{thin9}) and (\ref{thin10}).

For the B$_2$-phase parametrized in Eq. (\ref{mag21}) with $B,C,M\in\mathbb{R}$, we have
\begin{align}
 \nonumber \text{f}_{\rm B_2}={}& -\alpha_0[2(B^2+M^2)+C^2]+\beta_1[2(B^2-M^2)+C^2]^2\\
 \nonumber &+\beta_2[2(B^2+M^2)+C^2]^2+\beta_{35}[2(B^2-M^2)^2+C^4]\\
 \nonumber &+\beta_4[2(B^2+M^2)^2+8B^2M^2+C^4] \\
 \label{fs24} &-4g_1 H BM+g_2H^2C^2.
\end{align}
The corresponding equations of motion read
\begin{align}
 \nonumber 0 ={}& -4\alpha_0 B +8[\beta_{12}C^2-(2\beta_1-2\beta_2+\beta_{35}-3\beta_4)M^2]B\\
 \label{fs25} &+8(2\beta_{12}+\beta_{345})B^3-4g_1 H M,\\
 \nonumber 0 ={}& -4\alpha_0 M+8(2\beta_{12}+\beta_{345})M^3-4g_1 H B \\
 \label{fs26} &-8[(\beta_1-\beta_2)C^2+(2\beta_1-2\beta_2+\beta_{35}-3\beta_4)B^2]M,\\
 \nonumber 0 ={}& -2\alpha_0 C +8[\beta_{12}B^2-(\beta_1-\beta_2)M^2]C\\
 \label{fs27} &+4(\beta_{12}+\beta_{345})C^3+2g_2 H^2 C.
\end{align}
We solve these equations numerically for a given set of parameters $\alpha_0$, $\beta_a$, finding the solution with $C>0$ that minimizes $\text{f}_{\rm B_2}$. The equations of motion expressed here in terms of $B,M,C$ are equivalent to Eqs. (\ref{b2})-(\ref{b4}) for the dimensionless order parameter upon neglecting the derivative terms.

The B$_2$-solution to the equations of motion with $C>0$ only exists for sufficiently large $\alpha_0>\alpha_{\rm cB}$. To estimate the value of $\alpha_{\rm cB}$ to leading order, we neglect all terms of order $B^2C$, $M^2C$ and $C^3$ in Eq. (\ref{fs27}) to get $\alpha_0 \simeq \alpha_{\rm cB}$ with
\begin{align}
 \label{fs28} \alpha_{\rm cB} \simeq g_2H^2.
\end{align}
We obtain the same estimate from the last line of Eq. (\ref{mag29}), where $\alpha_0=g_2H^2$ is the point where the quadratic term multiplying $\langle|\Delta_0|^2\rangle$ changes sign. To compute the precise value of $\alpha_{\rm cB}$, which comprises the transition from the P$_2$- to the B$_2$-phase, we follow the logic that lead to Eqs. (\ref{b8}) and (\ref{b9}) in the main text, which were used there to estimate $\bar{D}_{\rm A_2B_2}$. For this, we assume the bulk system to be a confined system with specular boundaries in the limit $D\to \infty$. We solve the equations of motion for $B$ and $M$ setting $C=0$, which yields the bulk or specular P$_2$-solution
\begin{align}
 2(B_0^2+M_0^2) &= \delta_0= \frac{\alpha_0}{2\beta_{12}+\beta_{345}},\\
 4B_0M_0 &= p_0= \frac{-g_1H}{2\beta_{12}+\beta_{345}-2\beta_{24}}.
\end{align} 
We insert this into the equation to $C$ under confinement, which we linearize in $C$, assuming $C$ to be small. This yields
\begin{align}
 \nonumber 0 \simeq{}& K\gamma \partial_z^2C +\Bigl[2\alpha_0-2g_2H^2\\
  \label{fs29} &-8\beta_1(B_0^2-M_0^2)-8\beta_2(B_0^2+M_0^2)\Bigr]C.
\end{align}
Using $B_0^2-M_0^2=\frac{1}{2}\sqrt{\delta_0^2-p_0^2}$ we arrive at
\begin{align}
  \label{fs30}  0 \simeq{}& K\gamma \partial_z^2C +2\bar{\omega}_{\rm H}^2C,
\end{align}
with
\begin{align}
  \label{fs31} \bar{\omega}_{\rm H}^2 = \alpha_0-g_2H^2-2\beta_1\sqrt{\delta_0^2-p_0^2}-2\beta_2\delta_0.
\end{align}
The existence of a solution $C>0$ is bound to the criterion $\bar{\omega}_{\rm H}^2\geq 0$, which determines the critical value $\alpha_0=\alpha_{\rm cB}$ through $\bar{\omega}_{\rm H}=0$. Approximating $p_0\ll \delta_0$ we obtain
\begin{align}
  \label{fs32} \alpha_{\rm cB} \simeq \Bigl(1-\frac{2(\beta_1-\beta_2)}{2\beta_{12}+\beta_{345}}\Bigr)g_2H^2.
\end{align}

\end{appendix}

\bibliography{refs_he3}

\end{document}